\documentclass[pdftex]{nolta}

\usepackage{float}
\usepackage{graphicx}
\usepackage{caption}
\usepackage{subcaption}
\usepackage{multirow}
\usepackage{amsmath}
\usepackage{bm}
\usepackage[switch]{lineno}
\usepackage[a4paper,top=15mm,bottom=15mm]{geometry}

\Vol{2}%
\No{3}%
\Year{2011}%
\Month{10}%

\title{Limited Improvement of Connectivity in Scale-Free Networks by Increasing 
\\ the Power-Law Exponent 
\textcolor{red}{}}

\AUTHOR*{%
\author*{Yingzhou MOU}{1}\orcid{0009-0003-0761-2747}, 
\author{Yukio HAYASHI}{1}\orcid{00}
}

\AFFILIATE{%
\affiliate{Division of Transdisciplinary Sciences, Japan Advanced Institute of Science and Technology, 1-1 Asahidai, Nomi, Ishikawa 923-1292, Japan}{1}
}

\begin{document}

\begin{abstract} 
It has been well-known that many real networks are scale-free (SF) but extremely vulnerable against attacks. We investigate the robustness of connectivity and the lengths of the shortest loops in randomized SF networks with realistic exponents $2.0 < \gamma \leq 4.0$. We show that smaller variance of degree distributions leads to stronger robustness and longer average length of the shortest loops, which means the existing of large holes. These results will provide important insights toward enhancing the robustness by changing degree distributions.

\end{abstract}


\maketitle 

\linenumbers
\section{Introduction}
\label{sec:introduction}
Many real networks of technological (e.g., World Wide Web, Internet, and power grids), social (e.g., actor collaborations, and citation networks) and biological (e.g., protein–interaction, and metabolic networks) systems have common scale-free (SF) structure \cite{barabasi1999emergence, amaral2000classes} generated by preferential attachment rule \cite{Barabasi1999meanfield}. In the power-law degree distributions $P(k) \sim k^{-\gamma}$ with a heavy-tail, \(k\) denotes the number of links connected to a node, where the exponent is $2<\gamma\leq4$ in many cases, such as WWW with $\gamma \approx 2.1$ and the Western US power grid with $\gamma \approx 4.0$ \cite{barabasi1999emergence, amaral2000classes, Barabasi1999meanfield, goh2001universal, goh2002classification, nguyen2021new, bellingeri2023forecasting}. Such networks are known to be extremely vulnerable against degrees attacks on hubs \cite{albert2000error, callaway2000network, cohen2003efficient}. The percolation theory has shown that the critical threshold $p_c$ depends on the power-law exponent $\gamma$ in SF networks \cite{gallos2005stability, gallos2007scale}. Beyond degrees attacks, betweenness centralities attacks have also been investigated for a synthetic SF network of Barabási–Albert (BA) model with $\gamma = 3.0$ and real SF networks: scientific collaboration networks with $\gamma \approx 2.5 \sim 3.0$, Internet with $\gamma \approx 2.1$ \cite{holme2002attack}, peer-to-peer with $\gamma \approx 2.1$ \cite{adamic2001search}, and protein-protein interaction networks with $\gamma \approx 2.0 \sim 3.0$ \cite{alvarez2015proteins}. Moreover, the belief propagation (BP) attacks is known as the approximately worst attacks by destructing loops for any networks \cite{mugisha2016identifying}. Note that dismantling and decycling problems are equivalent \cite{alfredo2016network}, and corresponds to the worst attacks and eliminating loops, respectively.

To enhance the robustness of connectivity against malicious attacks in networks beyond vulnerable SF structure, several approaches have been proposed. Recent studies suggest that reducing variance $\sigma^2 = \langle k^2 \rangle - \langle k \rangle^2$ of degree distributions $P(k)$ is crucial for improving the robustness in rewiring by heuristically enhancing loops \cite{Chujyo2021LoopEnhancement}. Here, $\langle k^2 \rangle$ and $\langle k \rangle$ denote the averages of square degrees and degrees, respectively, with respect to $P(k)$. In addition, large loops are more important than small loops such as triangles in adding links \cite{chujyo2022adding}. Recently, it has been revealed that smaller variances $\sigma^2$ give higher robustness of connectivity in the wide class of randomized networks with continuously changing degree distributions which include SF networks ($\gamma=3.0$), Erdős–Rényi (ER) random graph, and regular networks \cite{Chujyo2023OptimalRobustness}. Meanwhile, smaller variances $\sigma^2$ also give longer average length $\langle l \rangle$ of the shortest loops. Since the inside of the shortest loop is empty, such a loop represents a hole \cite{kawato2025larger}. Thus, large holes enhance the robustness of connectivity in any randomized networks.
 
\begin{figure}[H]
  \centering
  % 第一行
    \includegraphics[width=\textwidth]{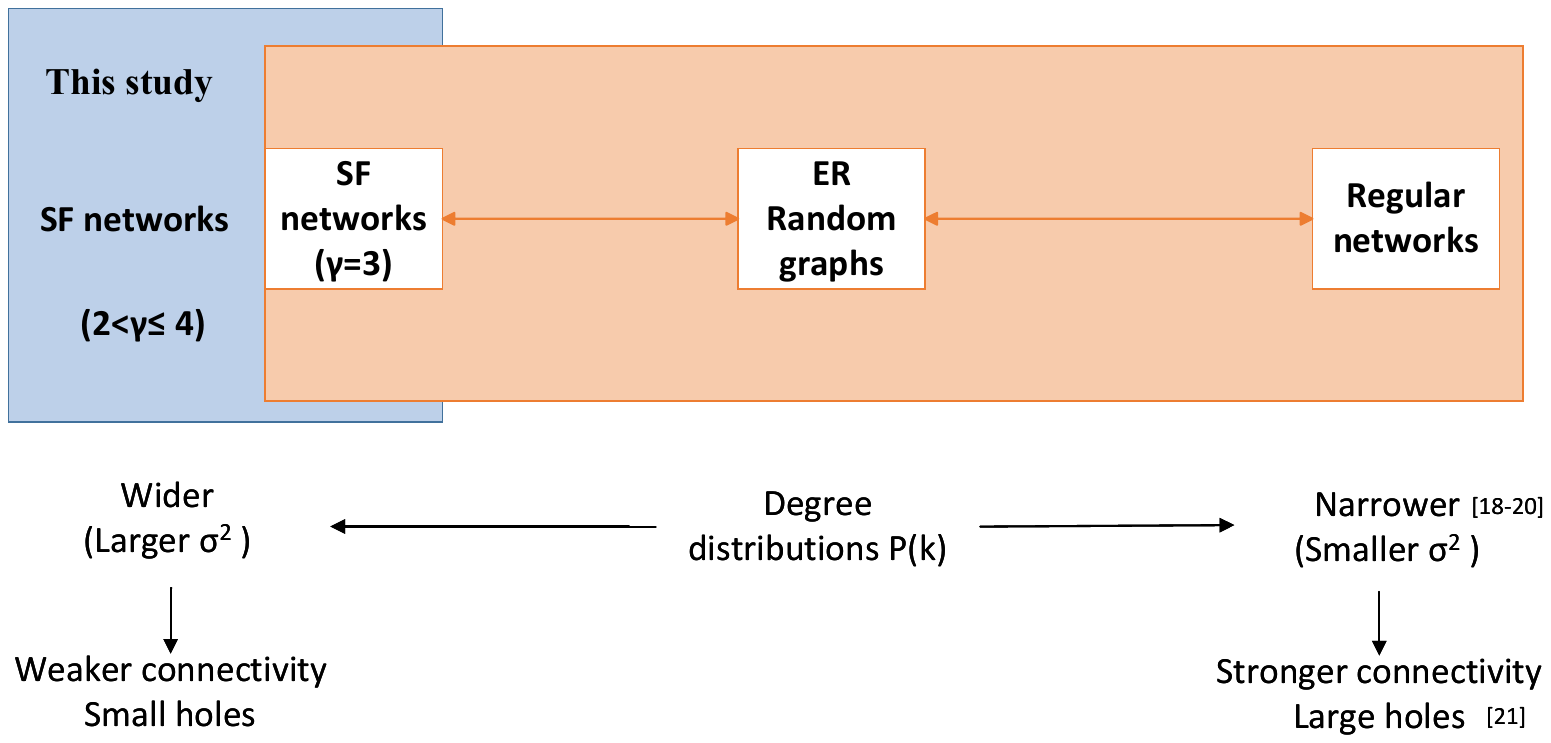}
    \caption{Schematic illustration of related works. The blue region indicates this study, which focuses on SF networks with tunable exponents $ 2 < \gamma \leq 4$. The orange region represents the previous research in the wide class of randomized networks. It has been revealed that the robustness against malicious attacks becomes stronger with large holes as $P(k)$ is narrower \cite{Chujyo2021LoopEnhancement, chujyo2022adding, Chujyo2023OptimalRobustness, kawato2025larger}.}
    \label{fig:relationship}
\end{figure}

Since many real networks have SF structures with power-law exponents $2 < \gamma < 3$, we extend the previous studies (orange area in Figure \ref{fig:relationship}) \cite{Chujyo2023OptimalRobustness, kawato2025larger} to randomized SF networks with tunable exponents (blue area in Figure \ref{fig:relationship}). The remainder of this paper is organized as follows. In section \ref{sec:methodology}, we describe the calculation methods for the robustness and the lengths of the shortest loops. In section \ref{sec:results}, we show the numerical results on the relation among the power-law exponent $\gamma$, the lengths of the shortest loops, and the robustness of connectivity. In section \ref{sec:conclusion}, we conclude that our findings for randomized SF networks with tunable exponents $2<\gamma\leq4$ are consistent with previous results in the wide class of randomized networks \cite{Chujyo2023OptimalRobustness, kawato2025larger}, as summarized in Figure ~\ref{fig:relationship}. We also provide a brief discussion for future research directions.

\section{Calculation methods for the robustness and the lengths of the shortest loops in SF networks}
\label{sec:methodology} 
We introduce the power-law exponents in SF networks, the robustness of connectivity against attacks, and the lengths of the shortest loops. Subsection~\ref{sec:network_generation} shows how the networks are generated by using DSM model \cite{dorogovtsev2000structure}. Subsection~\ref{sec:attack_robustness} explains three typical targeted node removals which are recalculated degrees, betweenness centralities, and BP attacks. We explain the robustness index $R$ based on the largest connected component (LCC). Subsection~\ref{sec:loop_structure_analysis} presents the calculation method for the average length $\langle l \rangle$ of the shortest loops.

\subsection{SF networks with tunable power-law exponents}
\label{sec:network_generation}
We consider SF networks, whose degree distributions $P(k)$ follow a power-law with the exponent $\gamma$.
If its form is
\begin{equation}
P(k) = C k^{-\gamma}, \quad k \geq k_{\min},
\label{eqn:p_k}
\end{equation}
then average degree is given by
\begin{equation}
   \langle k \rangle = \sum_{k} kP(k) \approx \int_{k_{\min}}^{\infty} k P(k) \, dk = \frac{\gamma - 1}{\gamma - 2} k_{\min}.
   \label{eqn:avg_degree}
\end{equation}
where $k$ denotes a degree, and the normalization constant is $C = (\gamma - 1)k_{\min}^{\gamma - 1}$. However, Eq.(\ref{eqn:avg_degree}) shows that $\langle k \rangle$ is not constant in depending on the value of $\gamma$, even when the minimum degree $k_{min}$ is constant.

Thus, to investigate the pure effect of the power-law exponent $\gamma$ on the robustness under the
condition of a constant average degree $\langle k \rangle \approx 2m$, we use the DSM model \cite{dorogovtsev2000structure} to generate scale-free networks. Here, $m = k_{\min}$ denotes the number of links attached to existing nodes from a new node at every time step. The degree distribution is given by
\begin{equation}
    P(k) \sim k^{-(3+c/m)},\;\;\;\; c > -m,
    \label{eqn:pk_price}
\end{equation}
where $c$ is a tunable constant. In the growth process, the probability of attachment to a node $i$ with degree $k_i$ is proportional to $k_i+c$. By adjusting $c/m$, different exponents $\gamma=3+c/m$ can be realized. However, networks generated by this growth process may have degree-degree or higher correlations. For example, degree–degree correlation refers to the tendency for a node’s degree to correlate with the degrees of its connected neighbors, such as high-degree nodes connect to other high-degree nodes in assortative mixing \cite{newman2002assortative, newman2003mixing}, or to low-degree nodes in disassortative mixing, while higher–correlations refer to characteristic connection patterns of degrees formed by more than two nodes. To reduce these correlations, we randomize these networks by using the configuration model \cite{catanzaro2005generation} to eliminate such correlations, and to investigate the pure effect of $\gamma$ on the robustness as shown later. The randomization process is as follows. First, after generating a networks by DSM model, each link is cut into two free-ends. Then there are $k_i$ free-ends emanated from a node $i$. Next, a pair of free-ends is randomly chosen and connected in prohibiting self-loops at a node and multi-links between nodes. We repeat them until all free-ends are connected in a LCC. Since these processes do not add or remove any links, the degree of each node is preserved for a given $P(k)$ in the network. Unless otherwise specified, all figures and results are obtained for SF networks with $N = 10^3$ and $m = 2$, while the exponent $\gamma$ ranges from $2.1$ to $4.0$ in steps of $0.1$. This setting allows us to investigate the pure effect of $\gamma$ under a fixed average degree $\langle k \rangle \approx 2m$. To examine whether our conclusions remain unchanged for denser or larger networks, additional experiments were performed for $N=10^3, m=3~and~4$ as denser networks, and $N=10^4, m=2,3~and~4$ as larger networks. These results are provided in the Supplementary Information and briefly summarized in Section~\ref{sec:conclusion} to confirm our conclusions. Following results are averaged over 100 realizations of the probabilistically generated networks.

\subsection{Robustness index $R$ against attacks}
\label{sec:attack_robustness}
To evaluate the robustness of connectivity against malcious attacks, we consider typical node removals of recalculated degrees (hub) \cite{albert2000error}, betweenness centralities \cite{holme2002attack}, and belief-propagation (BP) attacks \cite{mugisha2016identifying}. In recalculated degrees attacks, the node with the highest degree (know as hub node) is iteratively removed. Recalculated betweenness centralities attacks iteratively remove the node with the highest betweenness centrality, while recalculated BP attacks iteratively remove the node that is most likely to belong to the minimum Feedback Vertex Set (FVS) \cite{Karp1972reducibility, zhou2013spin}. The removal procedures of BP attacks are as follows. At each iteration, belief–propagation equations are solved on the 2-core of network to estimate node's belonging probability to the minimum FVS. Here, 2-core is a subgraph obtained after peeling all nodes of degree 0 or 1 recursively. Then, the node with the highest probability is selected and removed. After the removals, remaining nodes in dangling subtrees are selected as the removal targets in the decreasing order of degrees.

For investigating the robustness, we apply the usual measure of robustness index defined as follows.
\begin{equation}
  R \;=\; \frac{1}{N}\sum_{q=1/N}^{1} S(q), \;\;\;\; q=\frac{1}{N}, \frac{2}{N},..., \frac{N-1}{N}, \frac{N}{N},
  \label{eqn:R_discrete}
\end{equation}
where the relative size of the LCC is denoted by $S(q)/N$, $S(q)$ is the number of nodes in the LCC after a fraction $q$ of node removals \cite{schneider2011mitigation}. A larger value of $R$ indicates the stronger connectivity that the whole connectivity remains even after many nodes are removed.

\subsection{Calculation of the average length $\langle l \rangle$ of the shortest loops}
\label{sec:loop_structure_analysis}
Based on the approach \cite{kawato2025larger}, we compute the shortest loops associated with each link in a network as follows. For a given link $e_{ij}$ between nodes $i$ and $j$, we temporarily remove $e_{ij}$ and calculate the length of the shortest path between nodes $i$ and $j$ except of $e_{ij}$ itself. The length $l$ of the shortest loops is given by the length of the shortest path plus one (the length of $e_{ij}$).

Then, we restore the link $e_{ij}$. Repeat them for all links in the network. The average length $\langle l \rangle$ of the shortest loops is obtained by 
\begin{equation}
   \langle l \rangle = \sum_{l} l P(l),
   \label{eqn:shortest_loops}
\end{equation}
where $P(l)$ denotes the length distribution of the shortest loops.

Remember that the shortest loop represents a hole. In the next section, we show that large holes contribute to be robust connectivity against attacks. Although this phenomenon seems contradictory, the truth has been already revealed in the wide class of randomized networks, which includes SF networks with $\gamma=3.0$, ER random graph and regular networks \cite{Chujyo2023OptimalRobustness, kawato2025larger}. We extend the previous studies to SF networks with $2.1 \leq \gamma \leq 4.0$.

\section{Effects of the exponent $\gamma$ on the robustness and the shortest loops}
\label{sec:results}
We investigate the robustness in SF networks with various power-law exponents $\gamma=2.1 \sim 4.0$. Subsection \ref{sec:degree_distributions} numerically shows that the generated networks have the tails of power-law in degree distributions. Subsection \ref{sec:robustness_results} shows that the robustness index $R$ becomes larger as the exponent $\gamma$ increases with smaller variance $\sigma^2$ of $P(k)$. Subsection \ref{sec:loop_results} shows that the average length $\langle l \rangle$ of the shortest loops becomes larger as the exponent $\gamma$ increases. The obtained results are consistent with the previous ones in the wide class of randomized networks \cite{Chujyo2023OptimalRobustness, kawato2025larger} (see Figure \ref{fig:relationship} again). However, we also find that there is a limitation in SF networks.

\subsection{Tunable power-law exponents in generated SF networks}
\label{sec:degree_distributions}
\begin{figure}[H]
  \centering
  % 第一行
  \begin{subfigure}[b]{0.42\textwidth}
    \includegraphics[width=\textwidth]{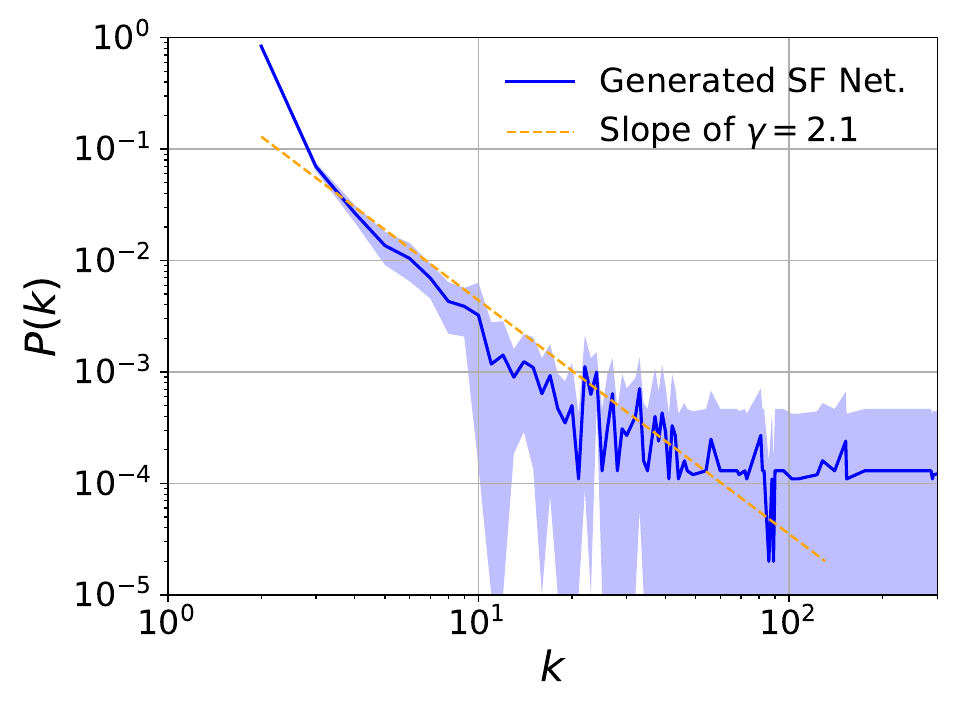}
    \caption{$\gamma=2.1$}
  \end{subfigure}
  \hfill
  \begin{subfigure}[b]{0.42\textwidth}
    \includegraphics[width=\textwidth]{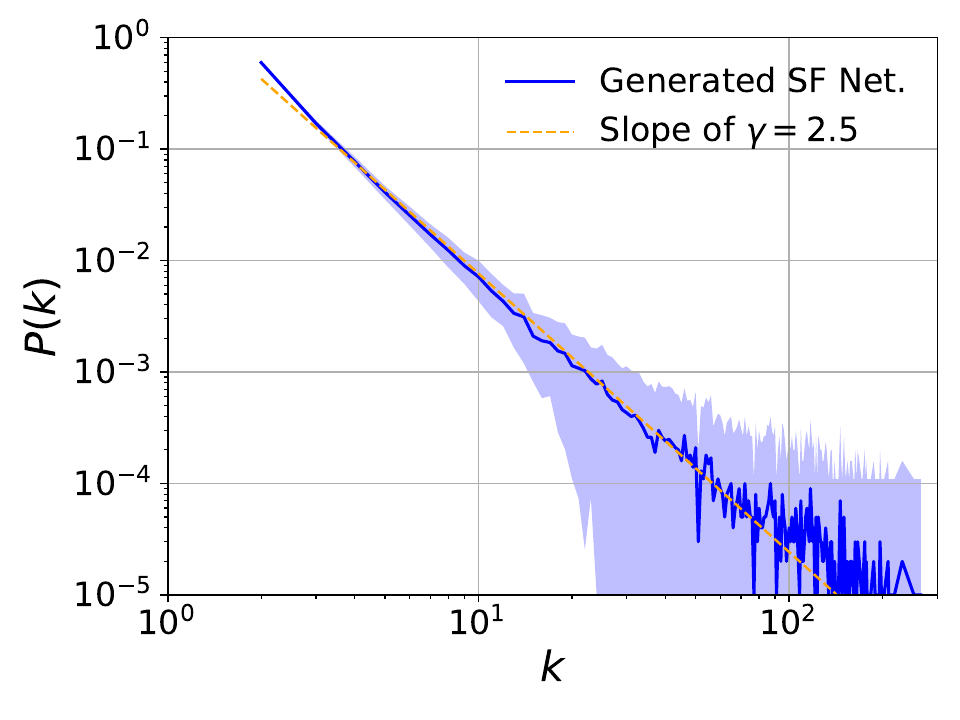}
    \caption{$\gamma=2.5$}
  \end{subfigure}

  \vspace{1em} % 两行之间的垂直间距
  
  \begin{subfigure}[b]{0.42\textwidth}
    \includegraphics[width=\textwidth]{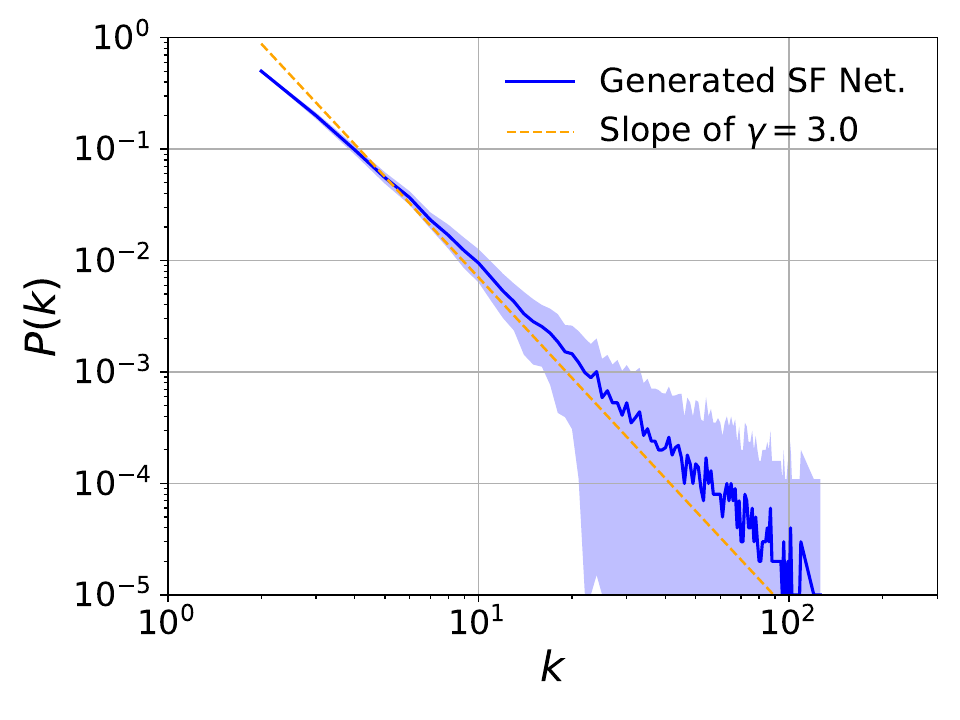}
    \caption{$\gamma=3.0$}
  \end{subfigure}
  \hfill
  \begin{subfigure}[b]{0.42\textwidth}
    \includegraphics[width=\textwidth]{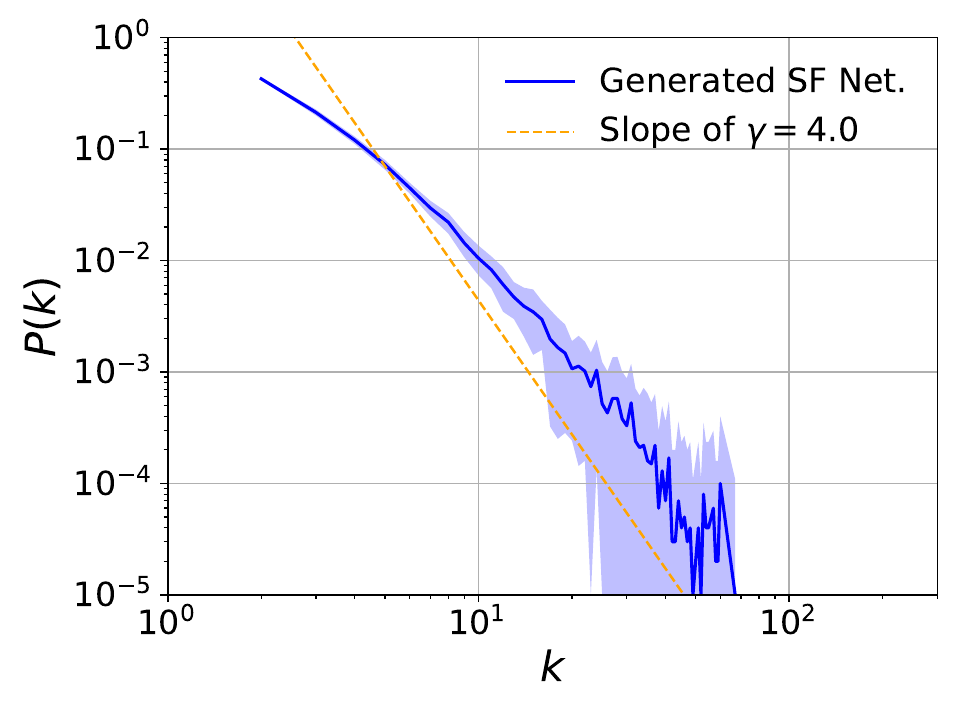}
    \caption{$\gamma=4.0$}
  \end{subfigure}
    \caption{Degree distributions $P(k) \sim k^{-\gamma}$ in generated SF networks with power-law exponents (a) $\gamma=2.1$, (b) $\gamma=2.5$, (c) $\gamma=3.0$, and (d) $\gamma=4.0$ for $N=10^3$ and $m=2$. Dashed lines guide the slope of power-law exponent $\gamma$ in the log-log plot. The shaded areas show the standard deviations in log-log scales.}
    \label{fig:degree_distributions}
\end{figure}

The generated SF networks by DSM models \cite{dorogovtsev2000structure} follow the tails of power-law in degree distributions with exponents $\gamma$ from $\gamma=2.1$ to $\gamma=4.0$ as shown in Figure~\ref{fig:degree_distributions}(a)-(d). We visualize the standard deviations shown as blue shaded areas in log-log plot, because the fluctuations of $P(k)$ are extremely small and hard to be observed in linear-scale plots. Since the fluctuations of $P(k)$ over 100 realizations are very small, the results are statistically stable. Detailed variance values are provided in the Supplementary Information.

Figure~\ref{fig:kmax_variance_gamma}(a)(b) show that both the maximum degree $k_{\max}$ and the variance $\sigma^2$ of the degree distribution decrease monotonically as exponent $\gamma$ increases. Remember the definition $\sigma^2 = \langle k^2 \rangle - \langle k \rangle^2$. Since the slope becomes steeper for larger $\gamma$ in the log-log plot of $P(k)$ versus $k$ (see the orange lines in Figure \ref{fig:degree_distributions}), the width of $P(k)$ is narrower with smaller $k_{max}$. However, hubs still exist even for larger $\gamma$ because of the convergence of large $k_{max} \approx 50 > \langle k \rangle$ in Figure \ref{fig:kmax_variance_gamma}(a). Moreover, we remark that the variance $\sigma^2$ of $P(k)$ is also convergent to a non-zero value for $\gamma>3.0$ in Figure \ref{fig:kmax_variance_gamma}(b). These convergences affect on the robustness index $R$ and the average length $\langle l \rangle$ of the shortest loops as shown later.

\begin{figure}[H]
  \centering
  % 第一行
  \begin{subfigure}[b]{0.42\textwidth}
    \includegraphics[width=\textwidth]{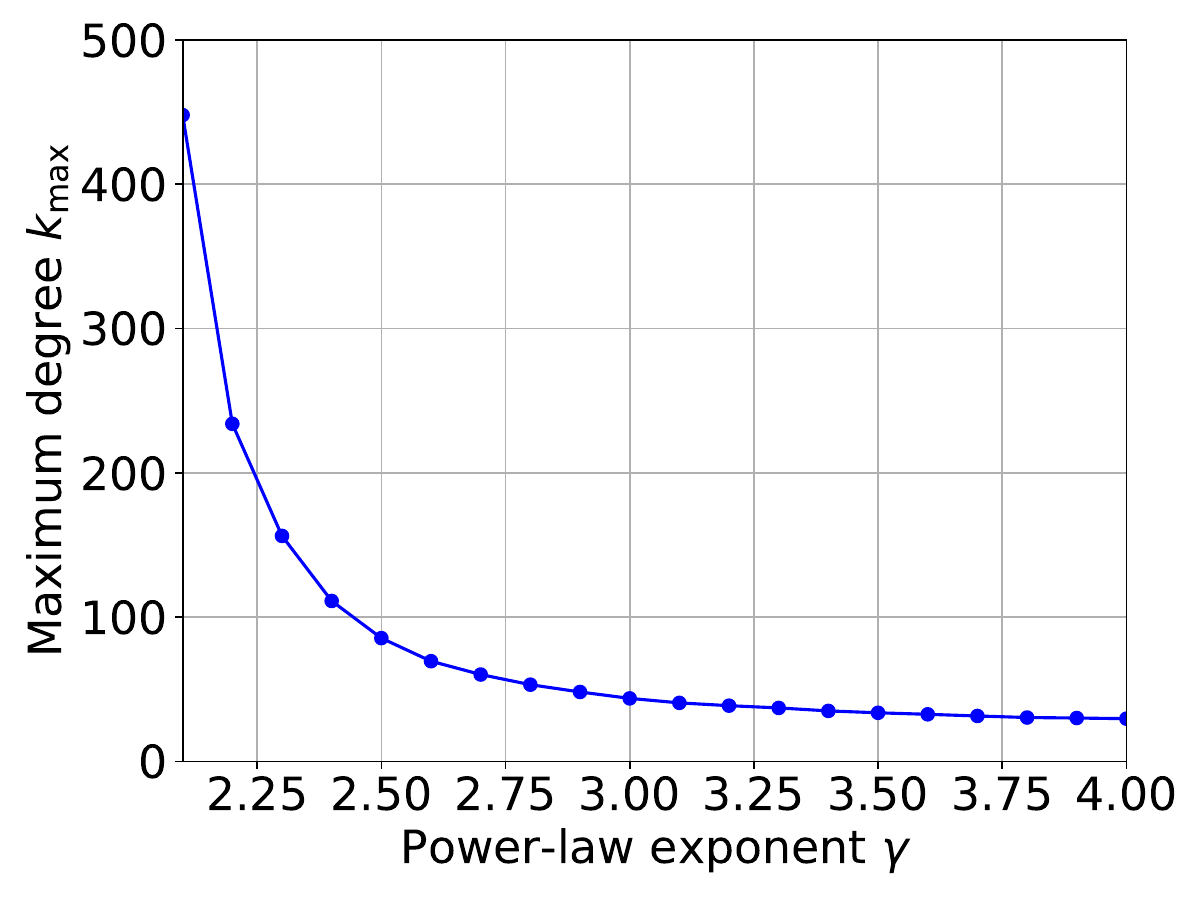}
    \caption{}
  \end{subfigure}
  \hfill
  \begin{subfigure}[b]{0.42\textwidth}
    \includegraphics[width=\textwidth]{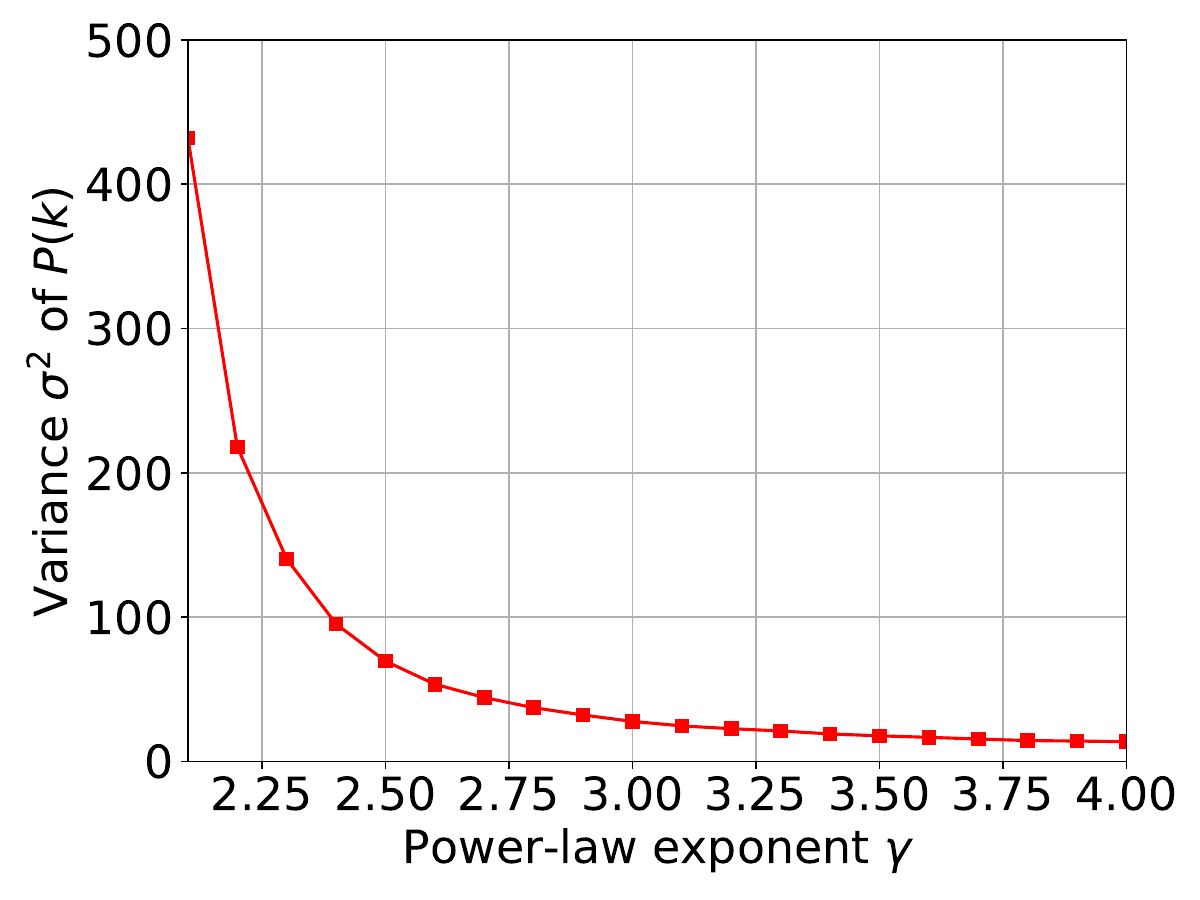}
    \caption{}
  \end{subfigure}
    \caption{Monotone decreasing of (a) the maximum degree $k_{max}$ and (b) the variance $\sigma^2$ of degree distribution $P(k)$ with the power-law exponent $\gamma$ for $N=10^3$ and $m=2$.}
    \label{fig:kmax_variance_gamma}
\end{figure}

\subsection{Robustness of connectivity against attacks}
\label{sec:robustness_results}
\begin{figure}
\centering   
\hfill
\begin{subfigure}[b]{0.42\textwidth}
\includegraphics[width=\textwidth]{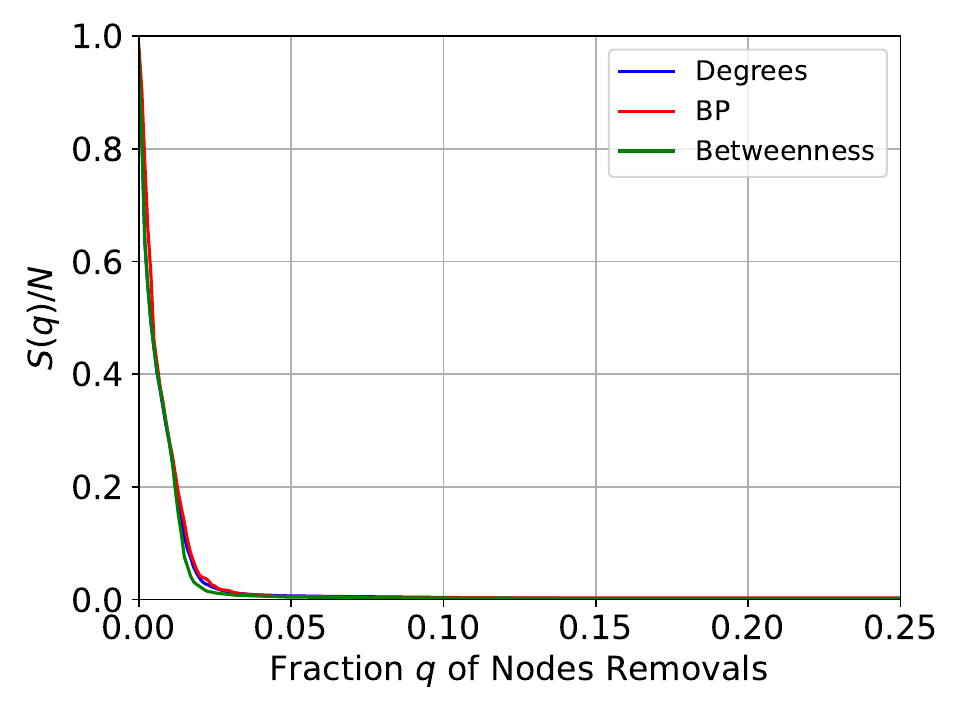}
\caption{$\gamma=2.1$}
\end{subfigure}
\hfill
\begin{subfigure}[b]{0.42\textwidth}
\includegraphics[width=\textwidth]{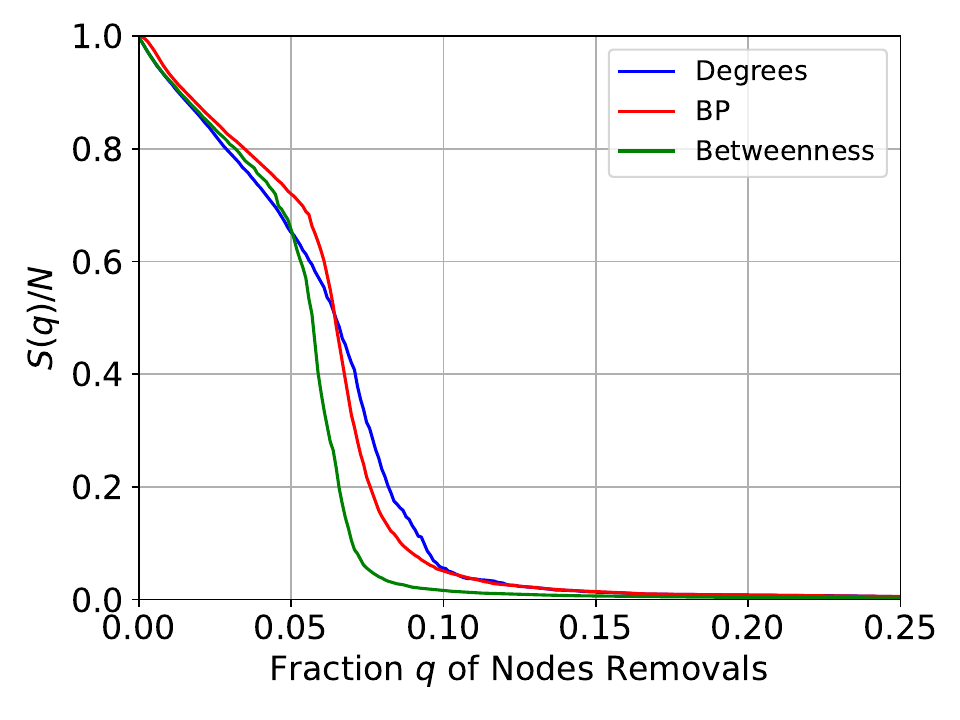}
\caption{$\gamma=2.5$}
\end{subfigure}

\vspace{1em}
\hfill
\begin{subfigure}[b]{0.42\textwidth}
\includegraphics[width=\textwidth]{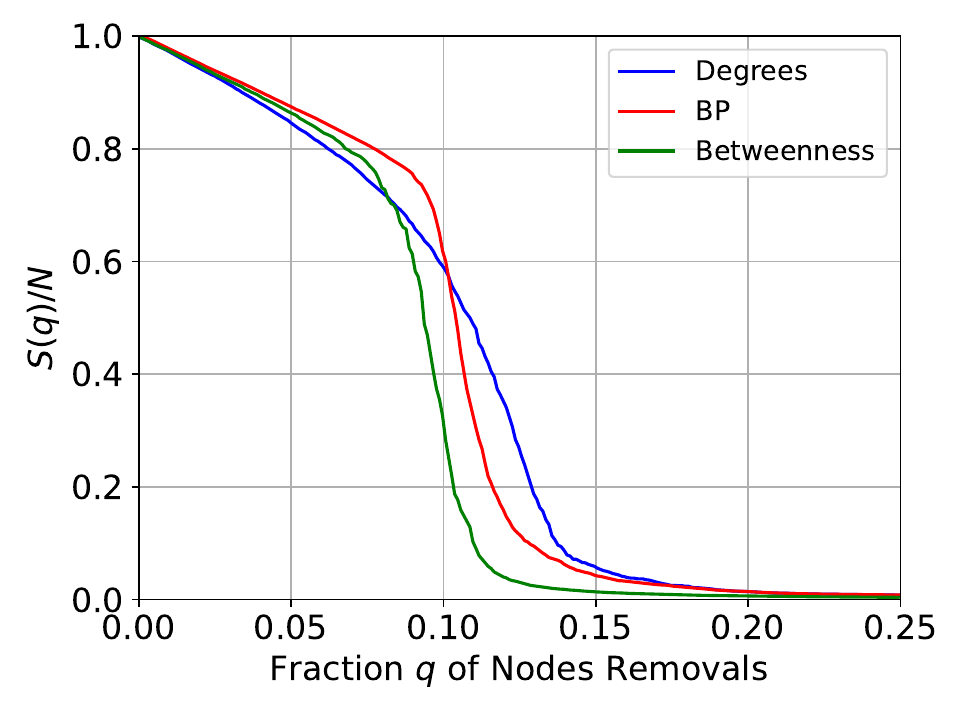}
\caption{$\gamma=3.0$}
\end{subfigure}
\hfill
\begin{subfigure}[b]{0.42\textwidth}
\includegraphics[width=\textwidth]{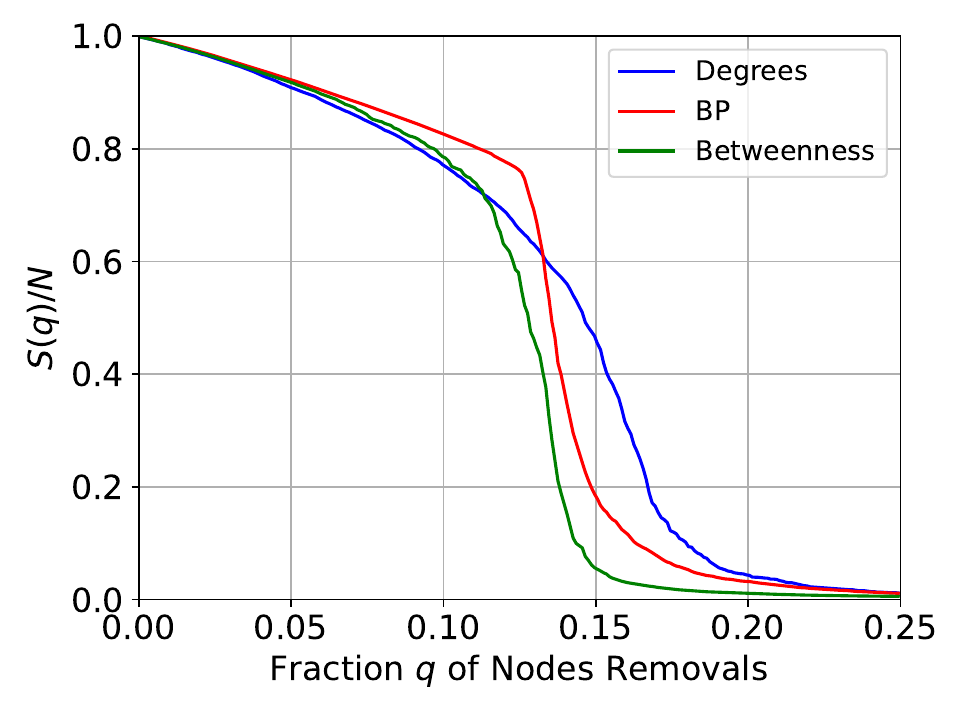}
\caption{$\gamma=4.0$}
\end{subfigure}
\caption{The relative size $S(q)/N$ of the largest connected component (LCC) against different attacks in randomized SF networks with the power-law exponents (a) $\gamma = 2.1$, (b) $\gamma = 2.5$, (c) $\gamma = 3.0$, and (d) $\gamma = 4.0$ for $N=10^3$ and $m=2$. Blue, red, and green curves correspond to recalculated degrees, betweenness centralities, and BP attacks, respectively. In comparing the areas under curves, red (BP attacks) and green (betweenness centralities) curves show more destructive with smaller areas than blue curves (degrees attacks).}
\label{fig:r_by_attacks}
\end{figure}

We show that the robustness of connectivity is slightly improved as the power-law exponent $\gamma$ increases from $\gamma=2.1 \sim 4.0$. Figure~\ref{fig:r_by_attacks}(a)-(d) show that the area under curves become larger as the exponent $\gamma$ increases, which means the robustness become stronger for a larger $\gamma$ from Figure \ref{fig:r_by_attacks} (a) to (d). The variances of $S(q)/N$ over 100 realizations are extremely small on the order of $10^{-3} \sim 10^{-5}$, and therefore difficult to be observed directly. We show the detailed variances in Table \ref{tab:SqN_variance_1000_m2} in the Supplementary Information rather than displaying them in Fig.~\ref{fig:r_by_attacks}(a)–(d). Note that both BP and betweenness centralities attacks are more destructive than degrees (hub) attacks. Because BP attacks approximately give the worst case of node removals from the equivalence of dismantling and decycling problems \cite{alfredo2016network}. The selected (removed) nodes as targets are belonging to the candidates of feedback vertex set, which are necessary to form loops in the network. Betweenness centralities attacks remove the nodes that are critical as bottlenecks on essential paths between different modules. Removing such nodes disconnects inter-module bridges and lead to rapid fragmentation of the largest connected component. However, in contrast, degrees attacks remove high-degree hubs, but hubs do not necessarily coincide with such structural bottlenecks.

\begin{figure}[H]
\centering
% 第一行
\hfill
\begin{subfigure}[b]{0.42\textwidth}
\includegraphics[width=\textwidth]{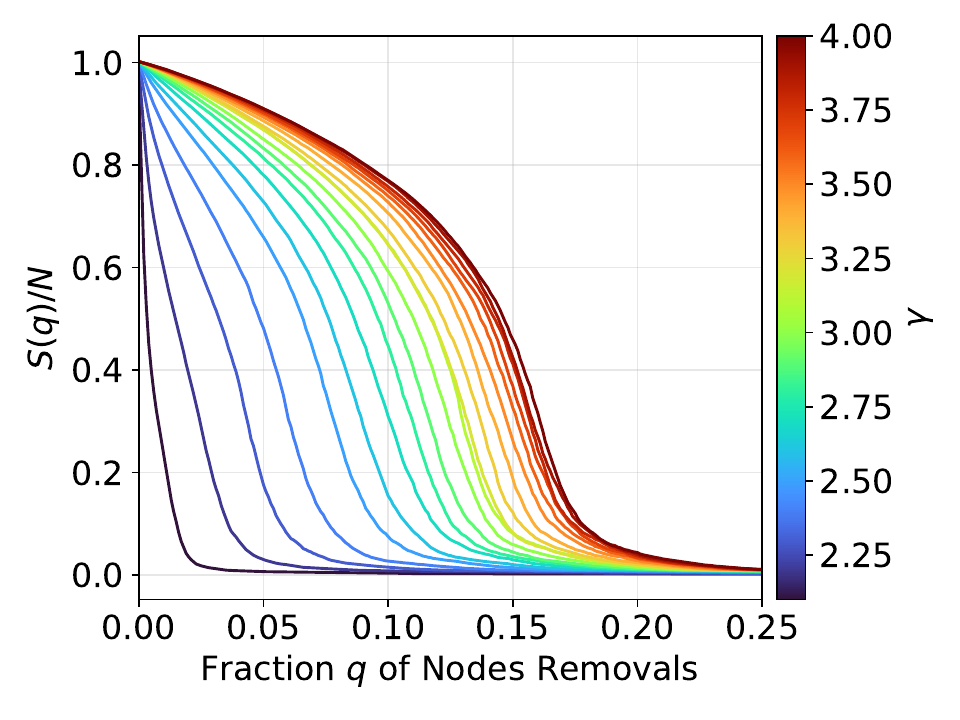}
\caption{Degrees attacks}
\end{subfigure}
\hfill
  \begin{subfigure}[b]{0.42\textwidth}
    \includegraphics[width=\textwidth]{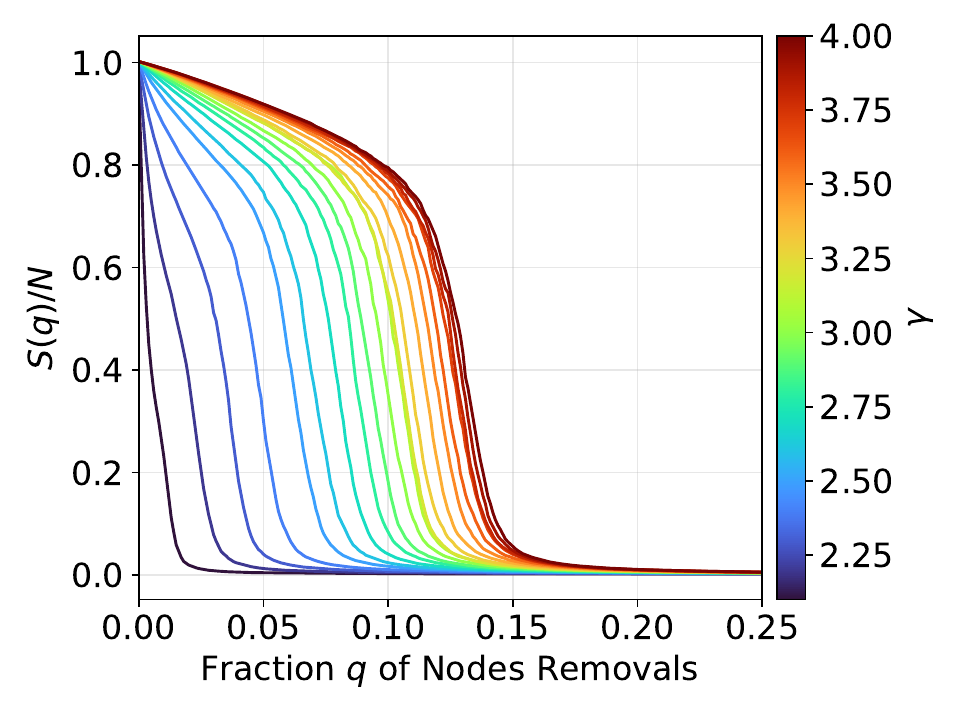}
    \caption{Betweenness centralities attacks}
  \end{subfigure}

\vspace{1em}
\hfill
\begin{subfigure}[b]{0.42\textwidth}
\includegraphics[width=\textwidth]{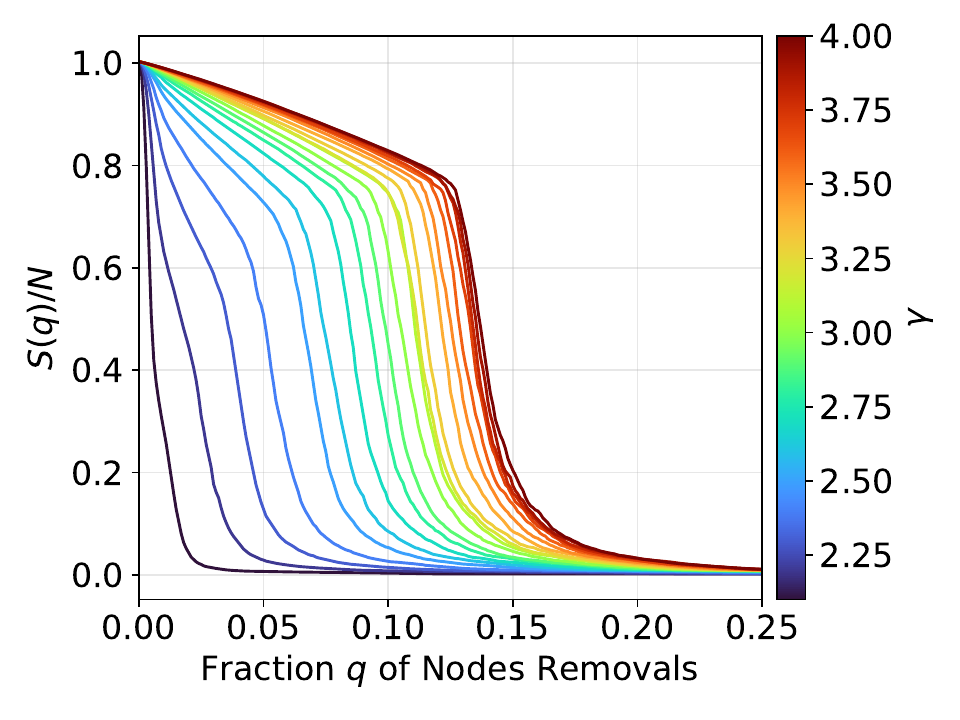}
\caption{Belief Propagation attacks}
\end{subfigure}
\caption{More detailed results for the robustness against recalculated (a) degrees, (b) betweenness centralities, and (c) belief propagation (BP) attacks for $N=10^3$ and $m=2$. The areas under colored curves represent the robustness index $R$ in SF networks with power-law exponents from $\gamma = 2.1$ (dark purple) to $\gamma = 4.0$ (red). As $\gamma$ increases, the areas under curves become larger from dark purple to red lines.}
\label{fig:r_by_gamma}
\end{figure}

We show more detailed results for the effect of $\gamma$ on the robustness against recalculated (a) degrees, (b) betweenness centralities, and (c) BP attacks in Figure~\ref{fig:r_by_gamma}. For all three attacks, the curves shift to right as $\gamma$ increases from 2.1 (dark purple curves) to 4.0 (red curves). However, the amount of this rightward shift decreases with increasing $\gamma$. For $\gamma > 3.0$, the curves nearly overlap with the convergence of the variance $\sigma^2$ of the degree distribution in Figure~\ref{fig:kmax_variance_gamma}(b). Since the areas under the curves from dark purple to red become larger, SF networks with larger $\gamma$ are more robust against these attacks. Thus, even in SF networks known as extremely vulnerable, the robustness becomes slightly stronger, as degree distributions are narrower (see Figures \ref{fig:degree_distributions} and \ref{fig:kmax_variance_gamma}). This extended results for SF networks with various exponent $\gamma$ are consistent with the previous results in the wide class of randomized networks with continuously changing $P(k)$, which include SF networks with $\gamma=3$, ER random graph, and regular networks \cite{Chujyo2023OptimalRobustness}.

\begin{figure}
\centering   
\hfill
\begin{subfigure}[b]{0.42\textwidth}
\includegraphics[width=\textwidth]{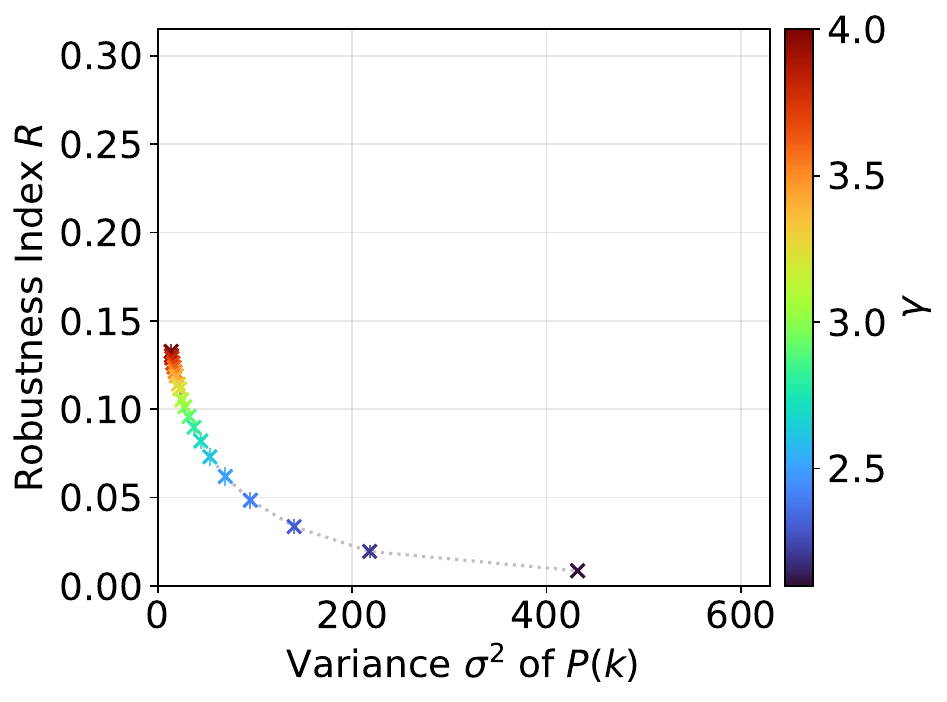}
\caption{Degrees attacks}
\end{subfigure}
\hfill
\begin{subfigure}[b]{0.42\textwidth}
\includegraphics[width=\textwidth]{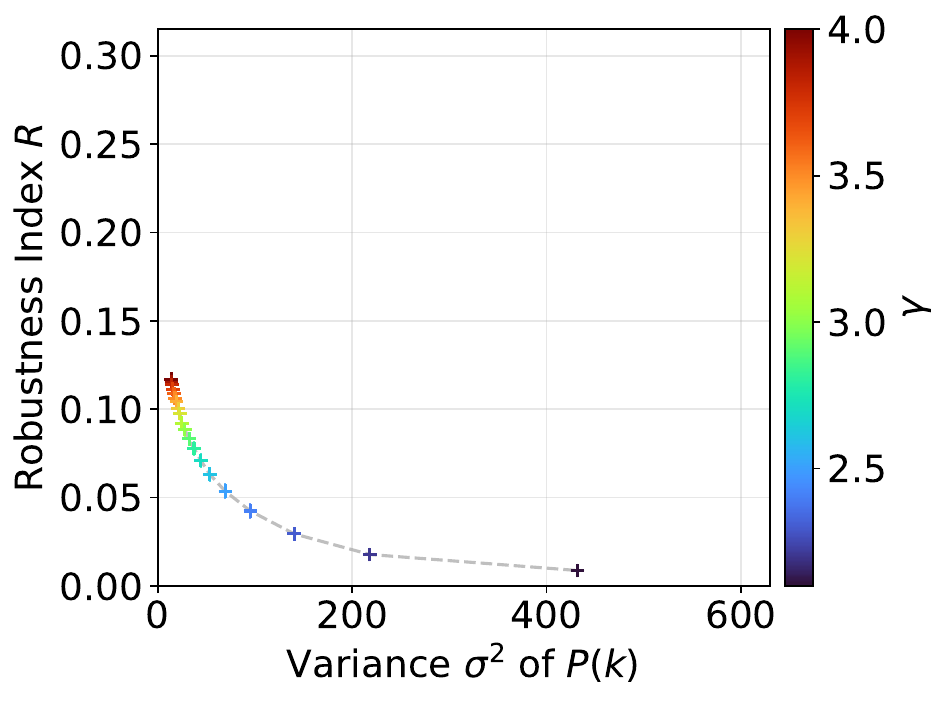}
\caption{Betweenness centralities attacks}
\end{subfigure}

\vspace{1em}
\hfill
\begin{subfigure}[b]{0.42\textwidth}
\includegraphics[width=\textwidth]{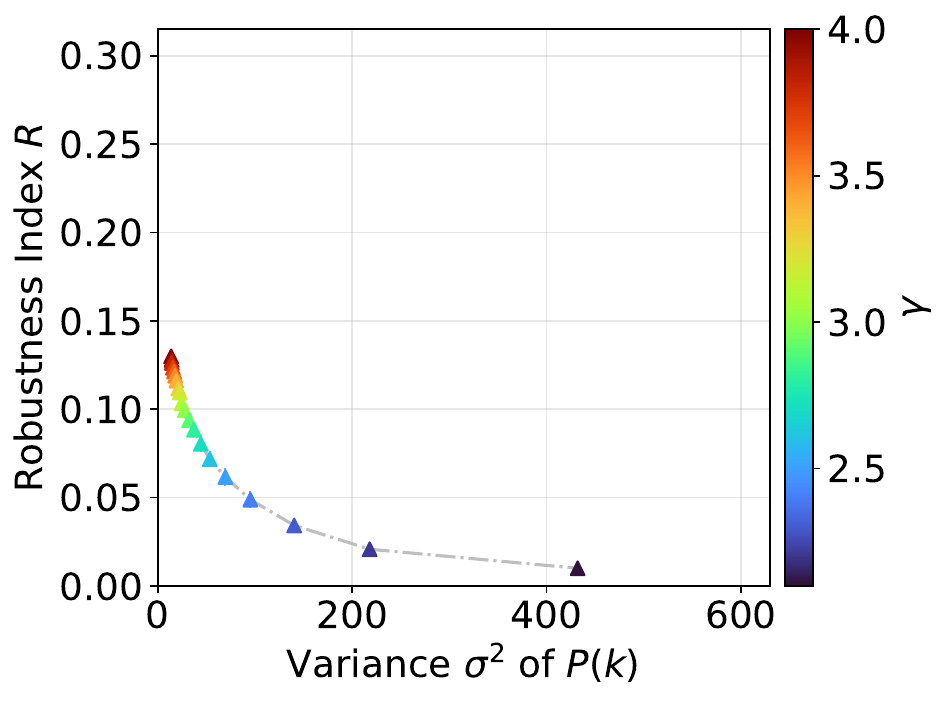}
\caption{Belief propagation attacks}
\end{subfigure}
\hfill
\begin{subfigure}[b]{0.42\textwidth}
\includegraphics[width=\textwidth]{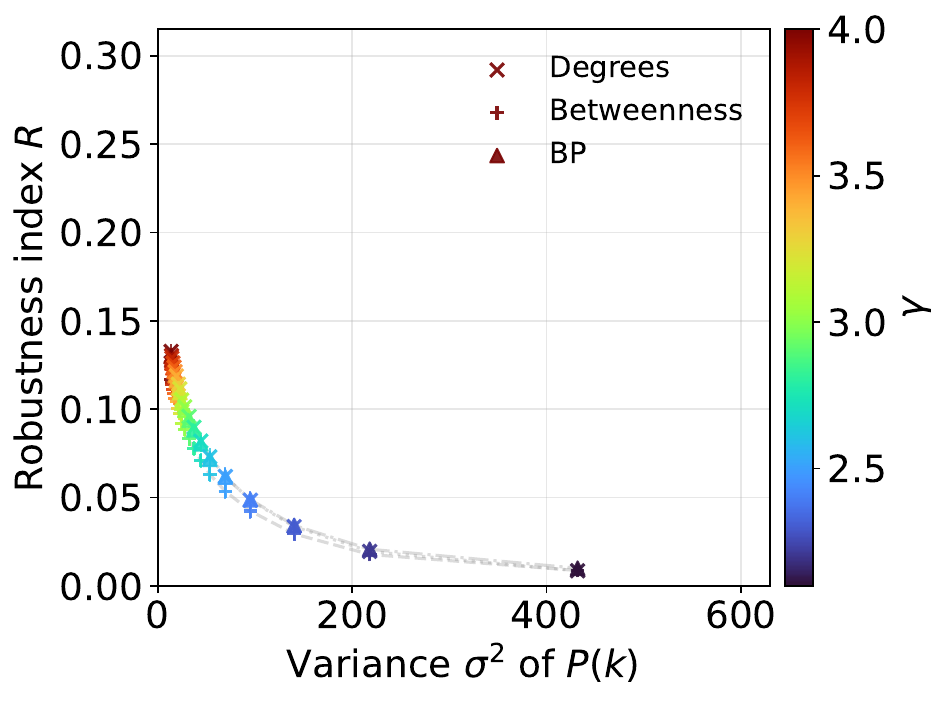}
\caption{Comparison of three attacks}
\end{subfigure}
\caption{Robustness index $R$ versus the variance $\sigma^2$ of degree distribution $P(k)$ in randomized SF networks against recalculated (a) degrees, (b) betweenness centralities, (c) belief propagation (BP) attacks, and (d) the comparison of robustness against these attacks for $N=10^3$ and $m=2$. Colored points represent the results for networks with power-law exponents $\gamma$ ranging from $\gamma = 2.1$ (dark purple points) to $\gamma = 4.0$ (red points). It is common that $R$ becomes larger as $\gamma$ increases. However, for $\gamma > 3$ (from green to red points), the improvement of $R$ is bounded.}
\label{fig:r_vs_var}
\end{figure}

Figure~\ref{fig:r_vs_var}(a)–(d) show clear relations between the robustness index $R$ and the variance $\sigma^2$ of $P(k)$ controlled by the values of exponent $\gamma$. As shown in Figure \ref{fig:r_vs_var}(d), three curves show almost coincident with colored points against recalculated (a) degrees, (b) betweenness centralities, and (c) BP attacks. It is common that $R$ becomes larger as $\gamma$ increases from 2.1 (dark purple) to 4.0 (red). In other words, the robustness against these attacks is determined by only the variance $\sigma^2$ of $P(k)$ and independent of nonlinear deviations in the heads of distributions (see Figure \ref{fig:degree_distributions}). Moreover, we emphasis that, even for a larger exponent $\gamma > 3.0$, SF networks are still vulnerable against these attacks. This limited improvement of robustness is related to the existing of hub nodes, since both $k_{max}$ and $\sigma^2$ converge to none-zero values (see Figure \ref{fig:kmax_variance_gamma}). 

In addition, the variances of $R$ are too small on the order of $10^{-4} \sim 10^{-7}$. Thus, we summary the variances of $R$ in Tables \ref{tab:variance_1000_m2} to \ref{tab:variance_10000_m4} in the Supplementary Information. Since these variances remain on the order of $10^{-4} \sim 10^{-7}$, the values of $R$ are not sensitive to random network generations.

\subsection{Similar trends in the average length $\langle l \rangle$ of the shortest loops and the robustness}
\label{sec:loop_results}
In the previous subsection, we have shown that the robustness index $R$ becomes larger, as the power-law exponent $\gamma$ increases in SF networks with smaller variance $\sigma^2$ of $P(k)$. A similar relation between the robustness of connectivity against attacks and the variance $\sigma^2$ of degree distributions has also been revealed in the wide class of randomized networks, including SF ($\gamma=3.0$) networks, ER random graph and regular networks \cite{Chujyo2023OptimalRobustness}. Moreover, the average length $\langle l \rangle$ of the shortest loops becomes longer as $\sigma^2$ decreases in this class \cite{kawato2025larger} (see Figure \ref{fig:relationship}).

\begin{figure}[H]
  \centering
  % 第一行
  \begin{subfigure}[b]{0.32\textwidth}
    \includegraphics[width=\textwidth]{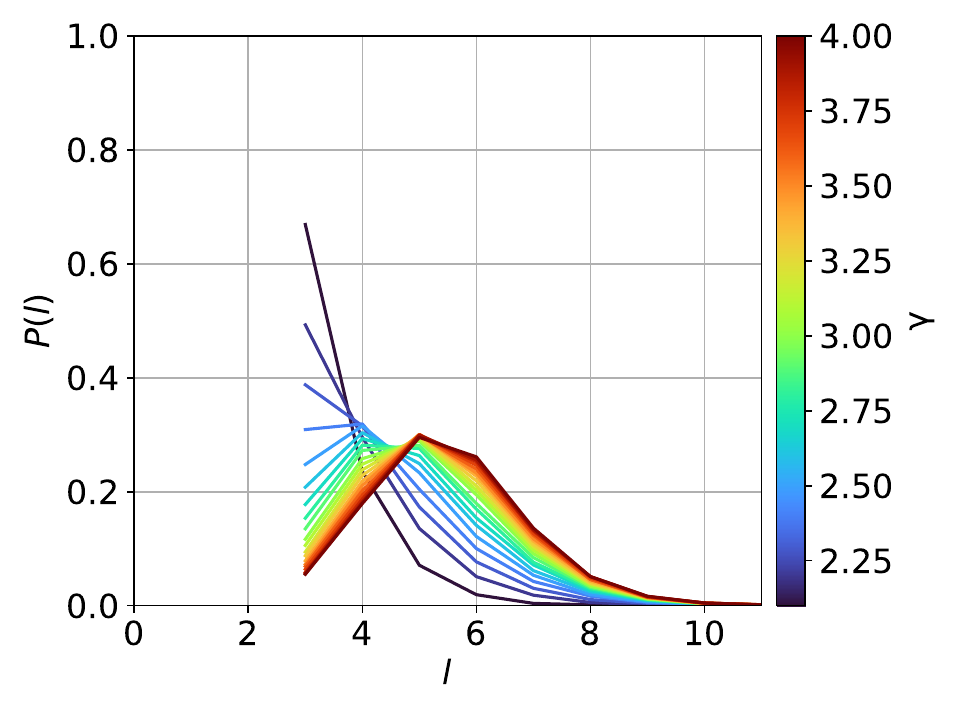}
    \caption{$N=10^3, ~m=2$}
  \end{subfigure}
  \hfill
  \begin{subfigure}[b]{0.32\textwidth}
    \includegraphics[width=\textwidth]{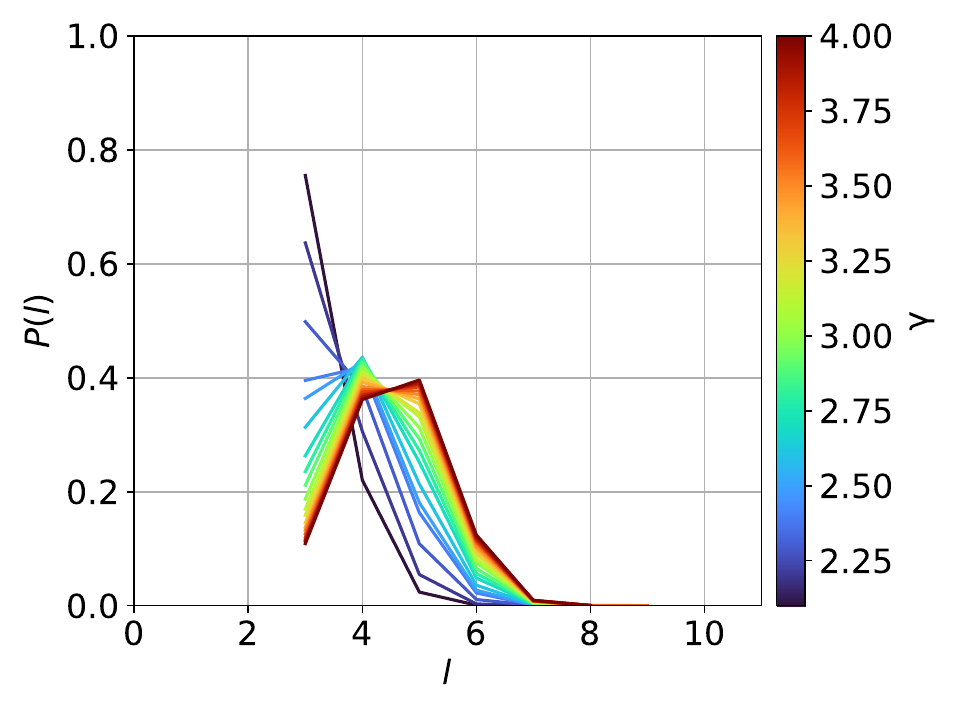}
    \caption{$N=10^3, ~m=3$}
  \end{subfigure}
  \begin{subfigure}[b]{0.32\textwidth}
    \includegraphics[width=\textwidth]{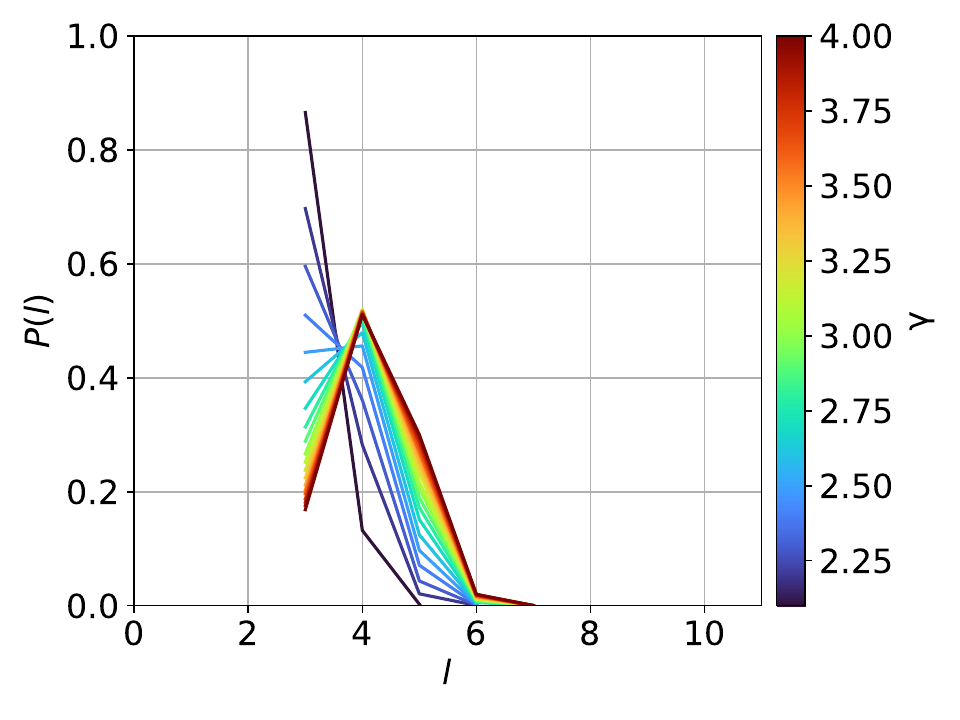}
    \caption{$N=10^3, ~m=4$}
  \end{subfigure}
  \vskip\baselineskip
    \hfill
  \begin{subfigure}[b]{0.32\textwidth}
    \includegraphics[width=\textwidth]{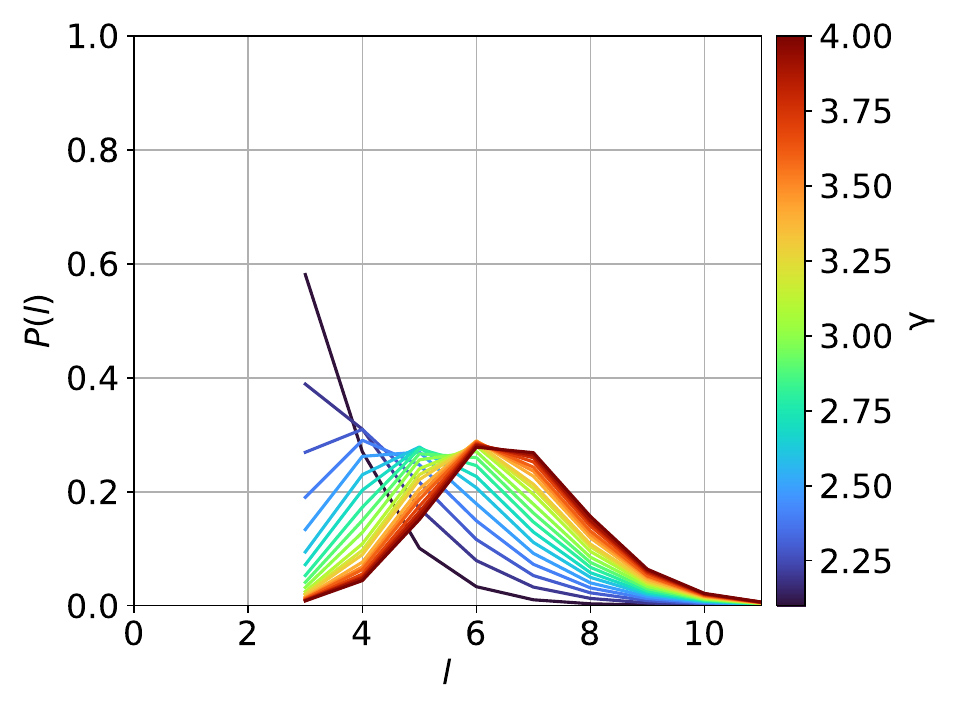}
    \caption{$N=10^4, ~m=2$}
  \end{subfigure}
      \hfill
  \begin{subfigure}[b]{0.32\textwidth}
    \includegraphics[width=\textwidth]{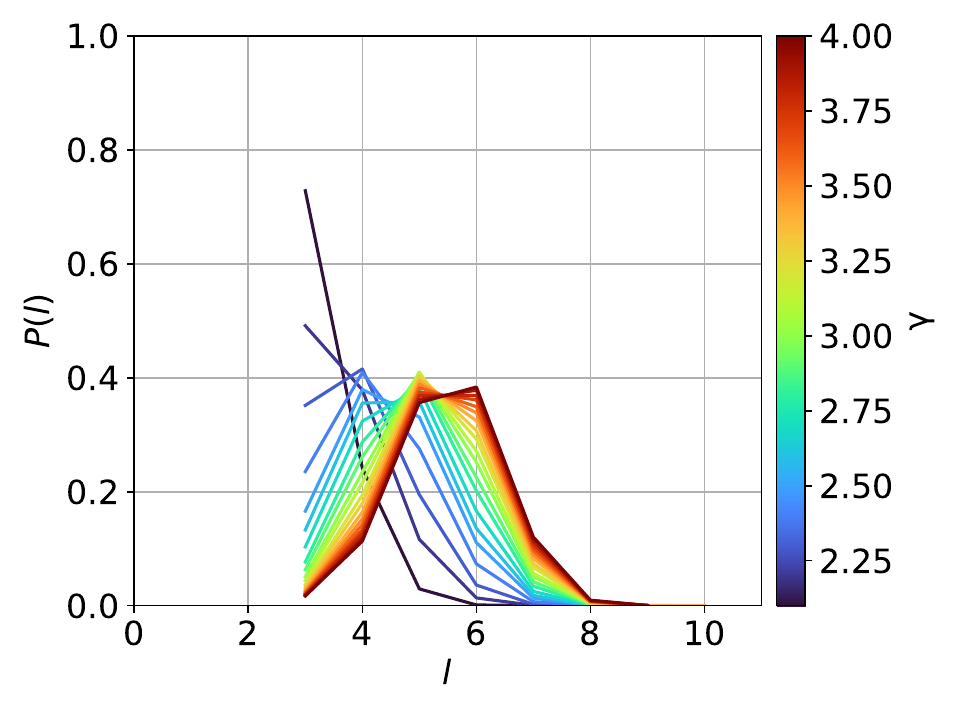}
    \caption{$N=10^4, ~m=3$}
  \end{subfigure}
      \hfill
  \begin{subfigure}[b]{0.32\textwidth}
    \includegraphics[width=\textwidth]{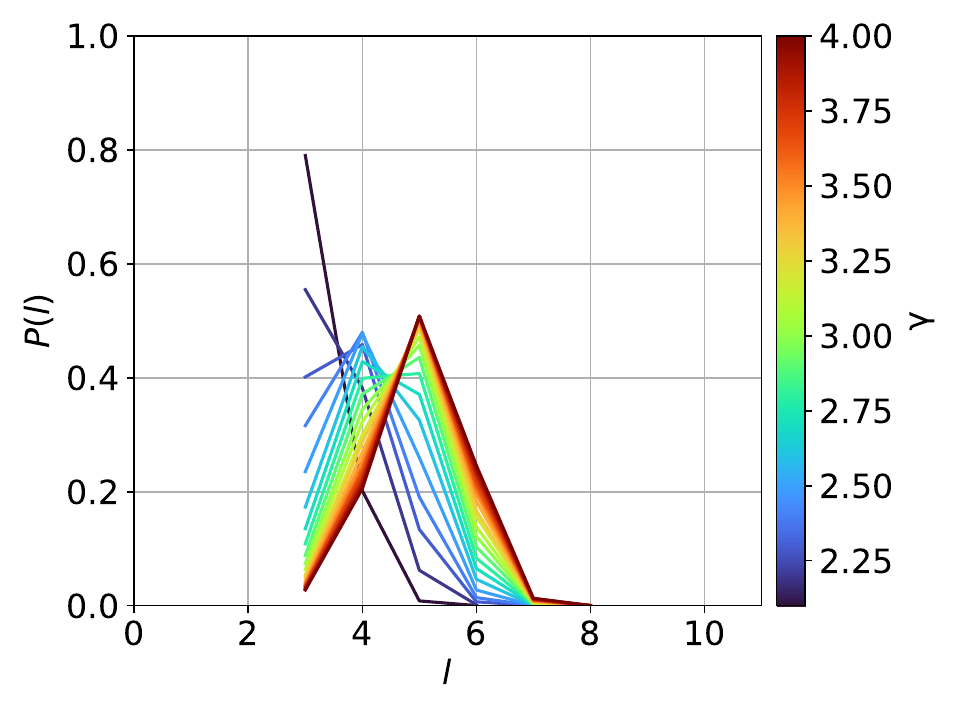}
    \caption{$N=10^4, ~m=4$}
  \end{subfigure}
    \caption{Length distributions $P(l)$ of the shortest loops in randomized SF networks with (a) $N=10^3, ~m=2$ (b) $N=10^3, ~m=3$ (c) $N=10^3, ~m=4$ (d) $N=10^4, ~m=2$ (e) $N=10^4, ~m=3$ and (f) $N=10^4, ~m=4$. Colored lines show the results for SF networks with power-law exponent $\gamma$ ranging from $\gamma=2.1$ (dark purple curves) to $\gamma=4.0$ (red curves). As $\gamma$ increases, $P(l)$ shifts to right, which means the existing of longer loops in SF networks with larger exponents $\gamma$.}
    \label{fig:loop_distribution}
\end{figure}

We further find a similar relation between the length distributions $P(l)$ of the shortest loops and the exponent $\gamma$. Figure~\ref{fig:loop_distribution} (a) shows the length distribution \(P(l)\) of the shortest loops over 100 realizations for each value of \(\gamma\). Each colored curve corresponds to a different value of \(\gamma\). The curves shift to right as $\gamma$ increases from $2.1$ (dark purple) to $4.0$ (red). This shifted change of $P(l)$ is associated with the convergence of the variance $\sigma^2$ of $P(k)$ as shown from green to red curves in Figure \ref{fig:kmax_variance_gamma}(b). Note that the rightward shifting is also observed in denser networks ($N=10^3, ~m=3$ and 4) and larger network ($N=10^4, ~m=2,3$ and 4), as shown in Figure~\ref{fig:loop_distribution}(b)-(f)). In addition, by comparing subfigures (a)–(c) or (d)–(f), we observe that for a fixed network size $N$, the length distribution $P(l)$ becomes narrower as the network becomes denser (from $m=2$ to $m=4$), indicating that the lengths of the shortest loops become more consistent in denser networks. Furthermore, comparisons between (a) and (d), (b) and (e), and (c) and (f) show that when $m$ is fixed but the network size increases (from $10^{3}$ to $10^{4}$), a slight rightward shift of $P(l)$ is observed, implying that larger networks tend to have longer shortest loops.

\begin{figure}[H]
  \centering
  % 第一行
  \begin{subfigure}[b]{0.32\textwidth}
    \includegraphics[width=\textwidth]{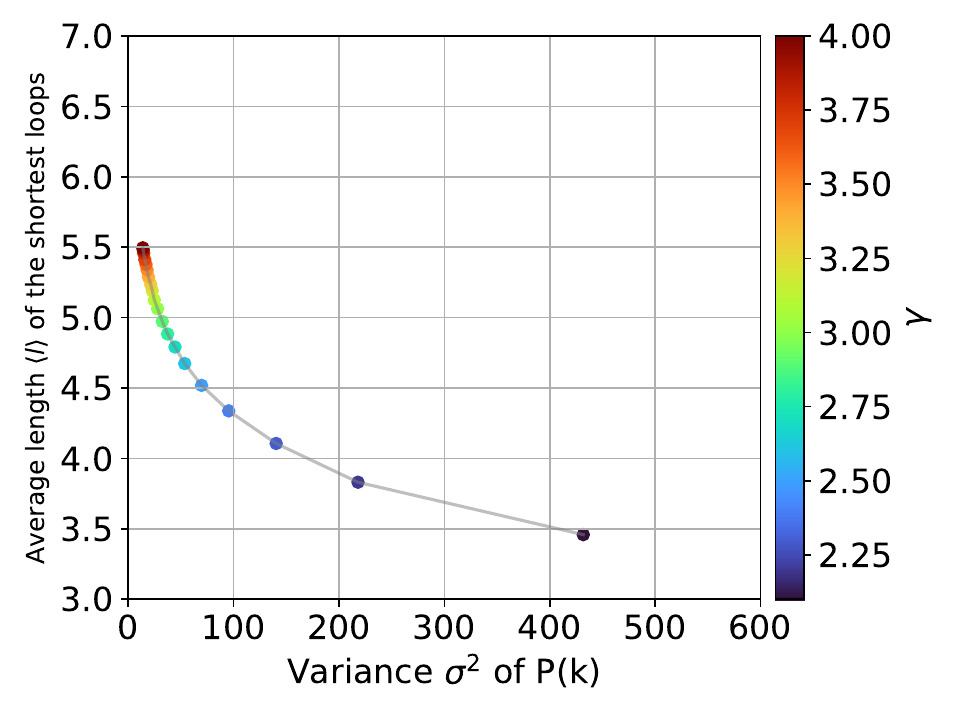}
    \caption{$N=10^3, ~m=2$}
  \end{subfigure}
  \hfill
  \begin{subfigure}[b]{0.32\textwidth}
    \includegraphics[width=\textwidth]{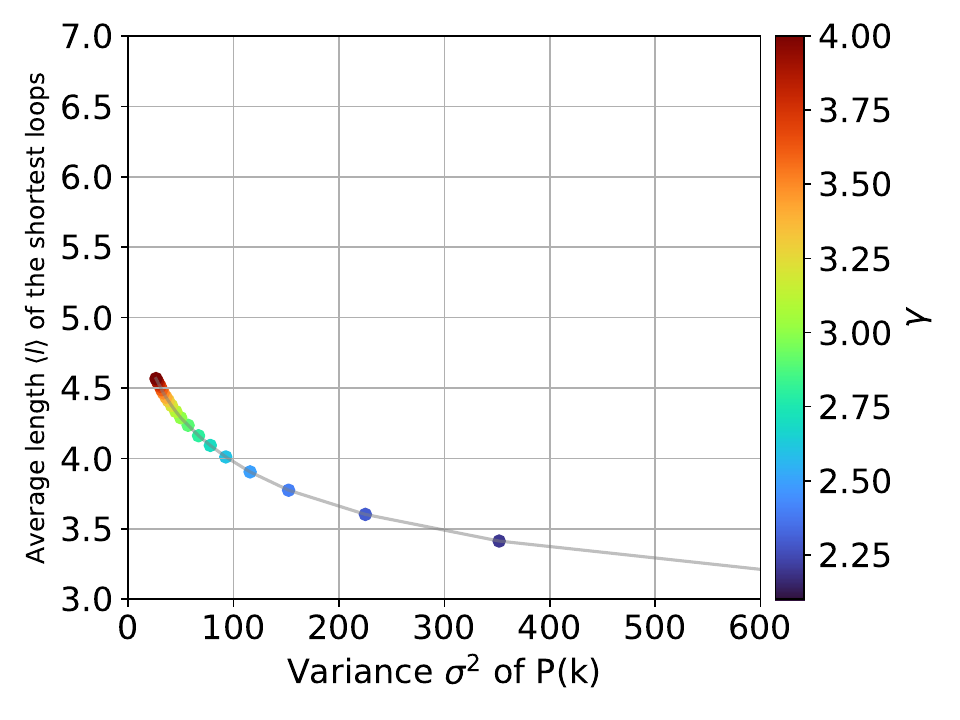}
    \caption{$N=10^3, ~m=3$}
  \end{subfigure}
  \hfill
  \begin{subfigure}[b]{0.32\textwidth}
    \includegraphics[width=\textwidth]{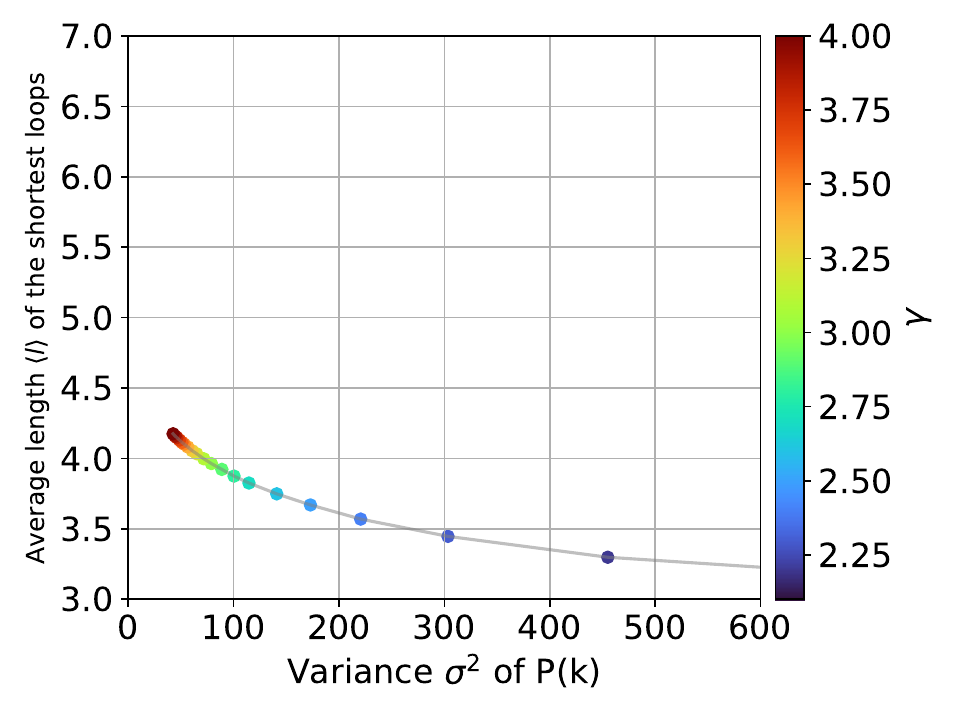}
    \caption{$N=10^3, ~m=4$}
  \end{subfigure}
\vskip\baselineskip
  \begin{subfigure}[b]{0.32\textwidth}
    \includegraphics[width=\textwidth]{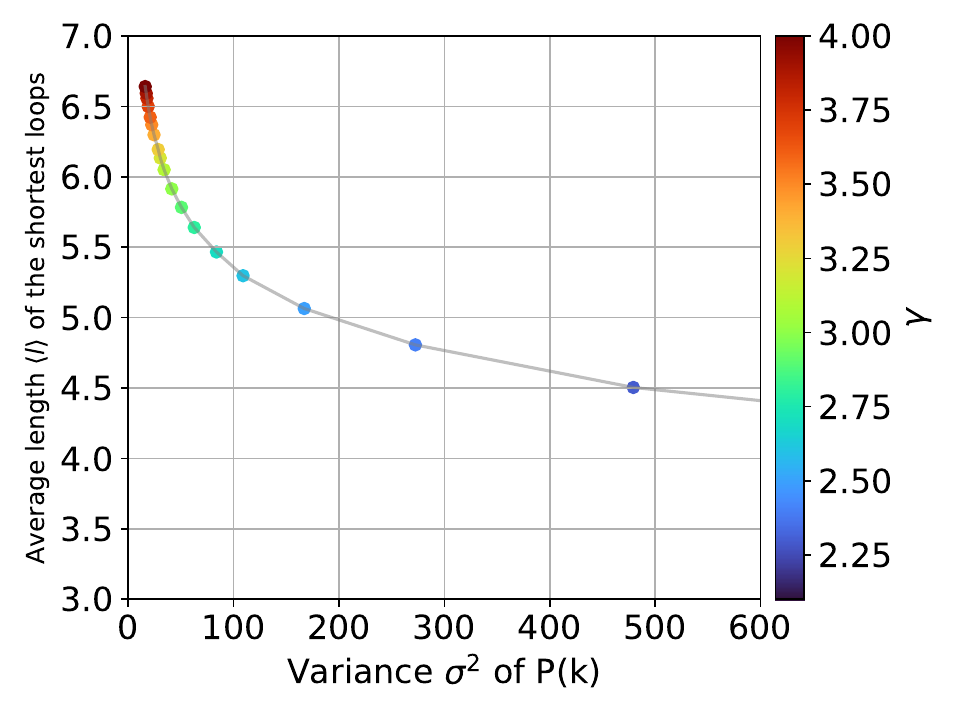}
    \caption{$N=10^4, ~m=2$}
  \end{subfigure}
    \begin{subfigure}[b]{0.32\textwidth}
    \includegraphics[width=\textwidth]{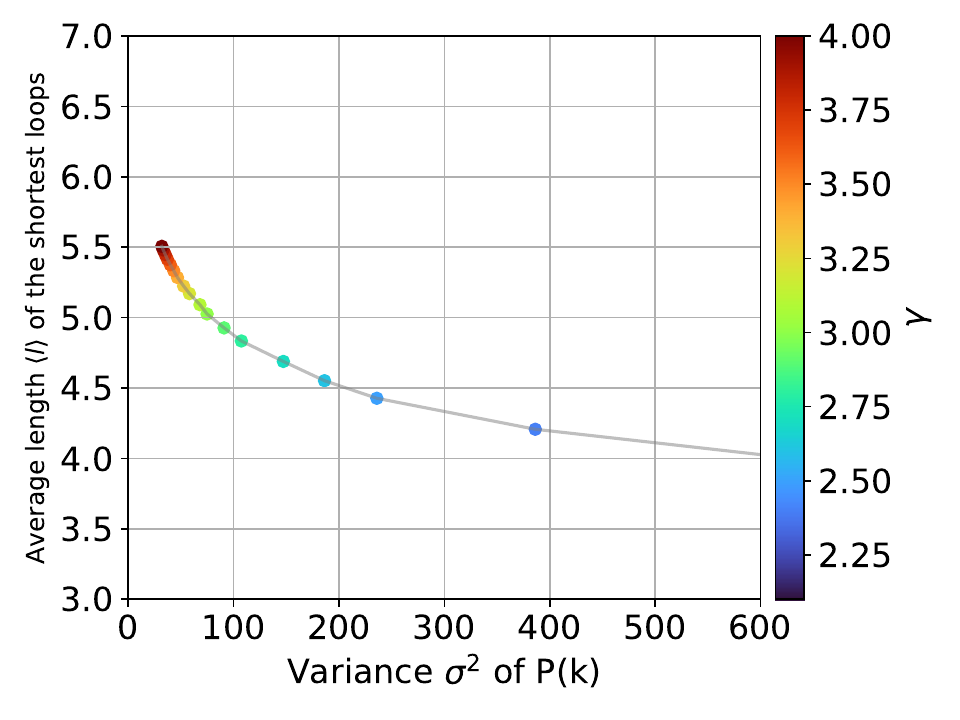}
    \caption{$N=10^4, ~m=3$}
  \end{subfigure}
    \begin{subfigure}[b]{0.32\textwidth}
    \includegraphics[width=\textwidth]{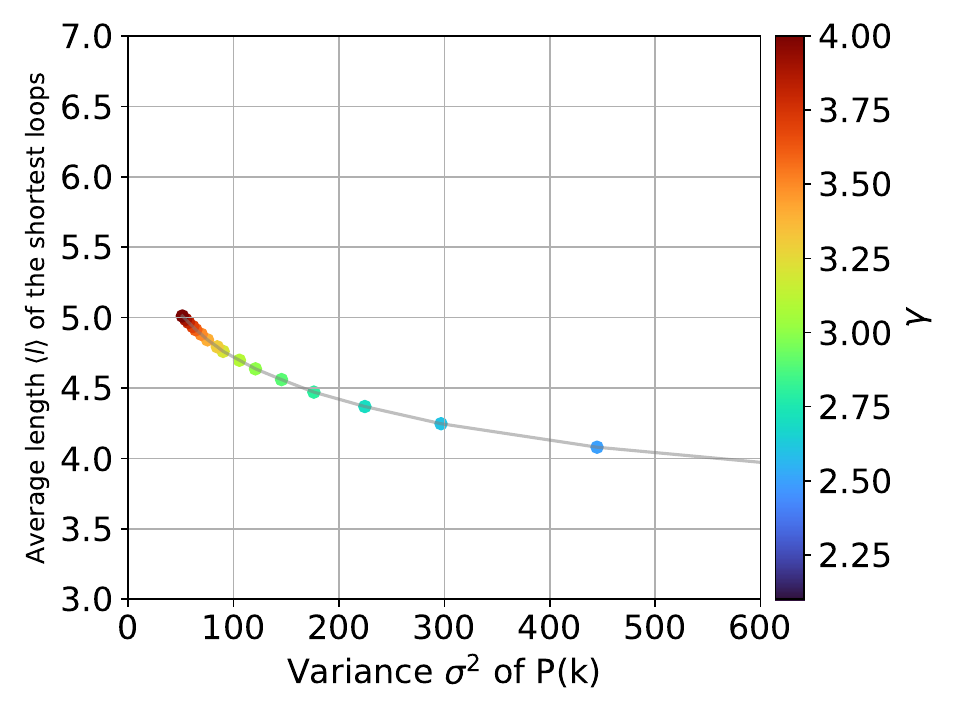}
    \caption{$N=10^4, ~m=4$}
  \end{subfigure}
   \caption{A monotone decreasing of the average length $\langle l \rangle$ of the shortest loops for the variance $\sigma^{2}$ of $P(k)$ in randomized SF networks. The colored points are corresponded to SF networks with power-law exponent $\gamma$ ranging from $\gamma=2.1$ (dark purple) to $\gamma=4.0$ (red). The average length $\langle l \rangle$ becomes larger, as $\gamma$ increases with smaller $\sigma^2$ in Figure \ref{fig:kmax_variance_gamma}(b).}
    \label{fig:loop_vs_var}
\end{figure}

Moreover, we observe a monotone decreasing between the average length $\langle l \rangle$ of the shortest loops and variance $\sigma^2$ of degree distributions $P(k)$. In Figures~\ref{fig:loop_vs_var}(a)-(f), each point represents the pair \((\sigma^{2}, \langle l \rangle)\) averaged over 100 realizations for a given \(\gamma\), (a) N=$10^3$, m=2, (b) N=$10^3$, m=3, (c) N=$10^3$, m=4, (d) N=$10^4$, m=2, (e) N=$10^4$, m=3, and (f) N=$10^4$, m=4. Dark purple to red points denotes increasing values of \(\gamma\), in showing how the average length of the shortest loops vary with the variance of degrees. However, from green to red points in Figures~\ref{fig:loop_vs_var} (a)-(f), $\langle l \rangle$ is bounded as $\gamma$ increases by the convergence of $\sigma^2$ in Figure \ref{fig:kmax_variance_gamma}(b). Similarly, the improvement of robustness for SF networks with $\gamma > 3.0$ is also bounded by the same convergence of $\sigma^2$, as discussed in subsection \ref{sec:robustness_results}. In addition, by comparing (a)–(c) or (d)–(f), we find that when $N$ is fixed, the curves shift downward as the network becomes denser (from $m=2$ to $m=4$), indicating that denser networks tend to have shorter shortest loops. Conversely, comparisons between (a) with (d), (b) with (e), and (c) with (f) show that when $m$ is fixed but the network size increases (from $10^{3}$ to $10^{4}$), the curves shift upward, implying that larger networks tend to have longer shortest loops. Since the variances of $\langle l \rangle$ are too small on the order of $10^{-3}\sim10^{-4}$, the detailed results of the variances are summarized for each $\gamma$ in Tables \ref{tab:variance_1000_m2} to \ref{tab:variance_10000_m4} in the Supplementary Information. Thus, the values of $\langle l \rangle$ are not sensitive to random network generations.

\begin{figure}[H]
  \centering
  % 第一行
  \begin{subfigure}[b]{0.32\textwidth}
    \includegraphics[width=\textwidth]{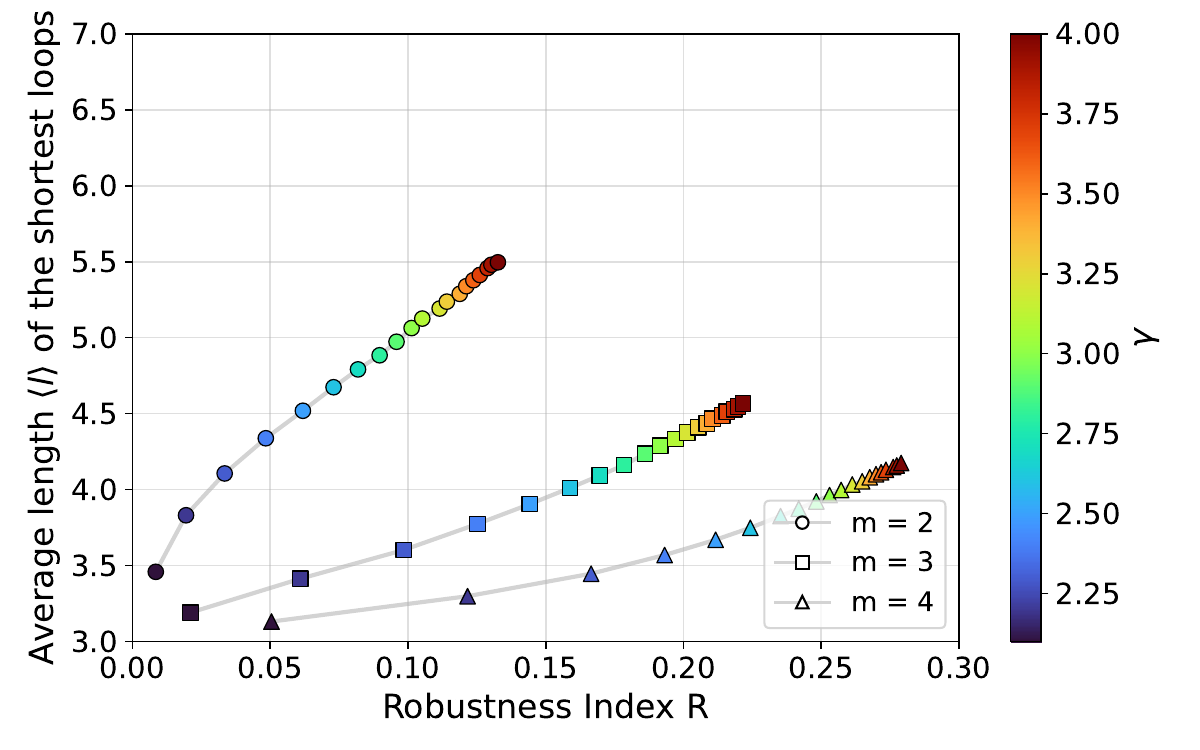}
    \caption{Degrees}
  \end{subfigure}
  \hfill
  \begin{subfigure}[b]{0.32\textwidth}
    \includegraphics[width=\textwidth]{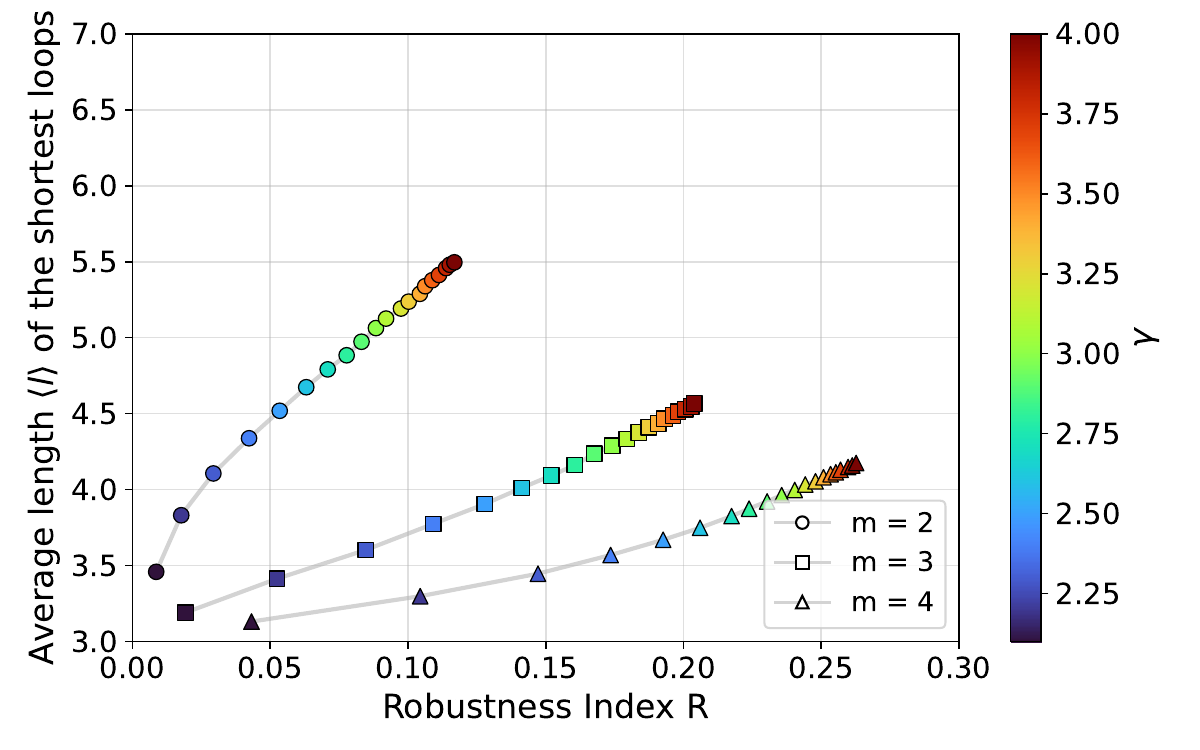}
    \caption{Betweenness}
  \end{subfigure}
  \hfill
  \begin{subfigure}[b]{0.32\textwidth}
    \includegraphics[width=\textwidth]{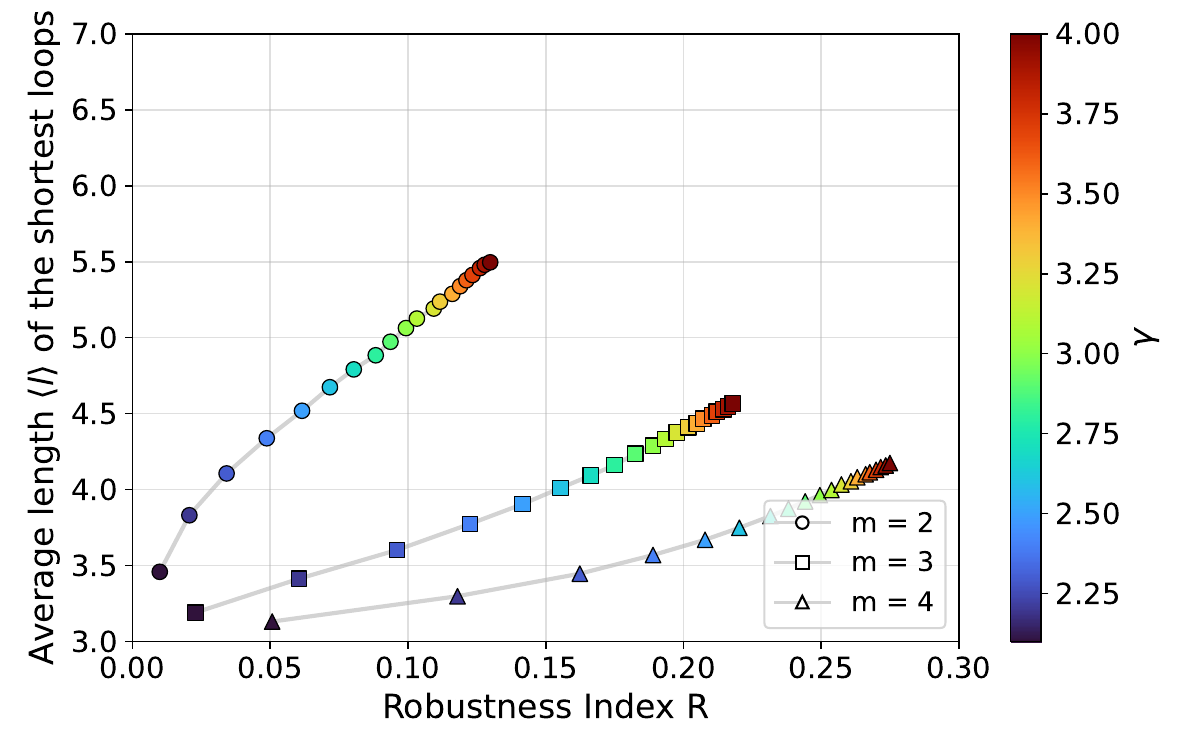}
    \caption{BP}
  \end{subfigure}
  \vskip\baselineskip
  \begin{subfigure}[b]{0.32\textwidth}
    \includegraphics[width=\textwidth]{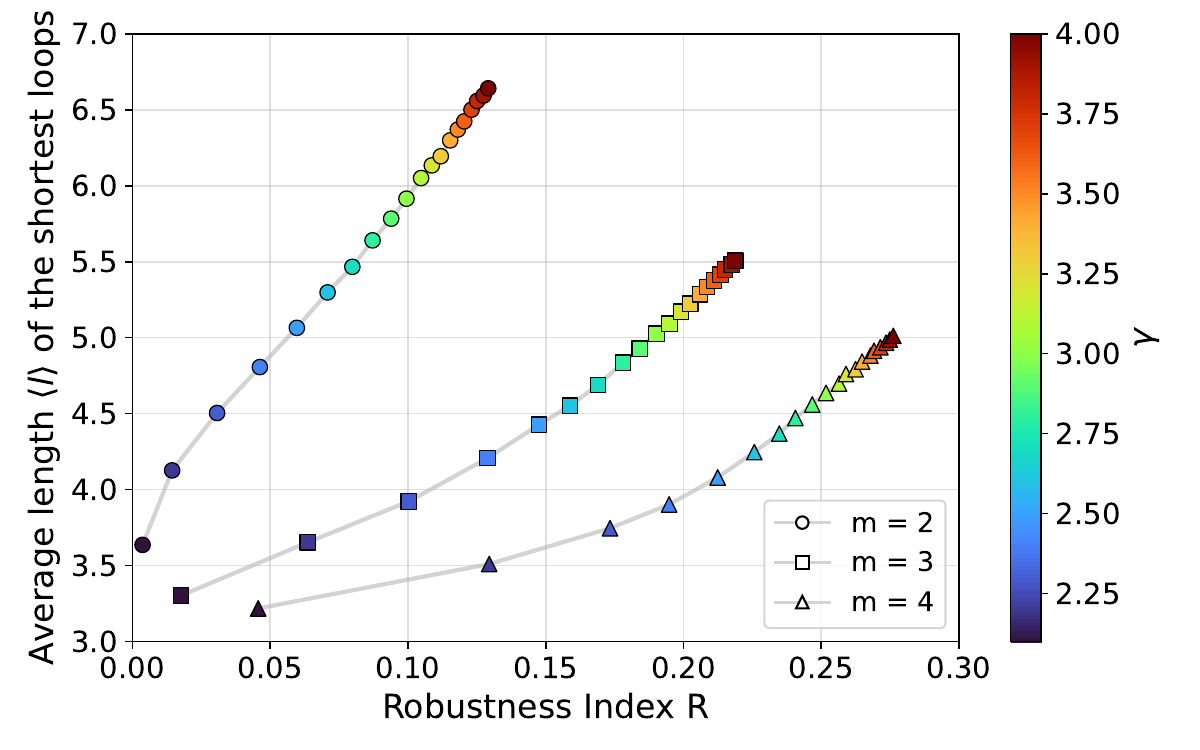}
    \caption{Degrees}
  \end{subfigure}
  \hfill
  \begin{subfigure}[b]{0.32\textwidth}
    \includegraphics[width=\textwidth]{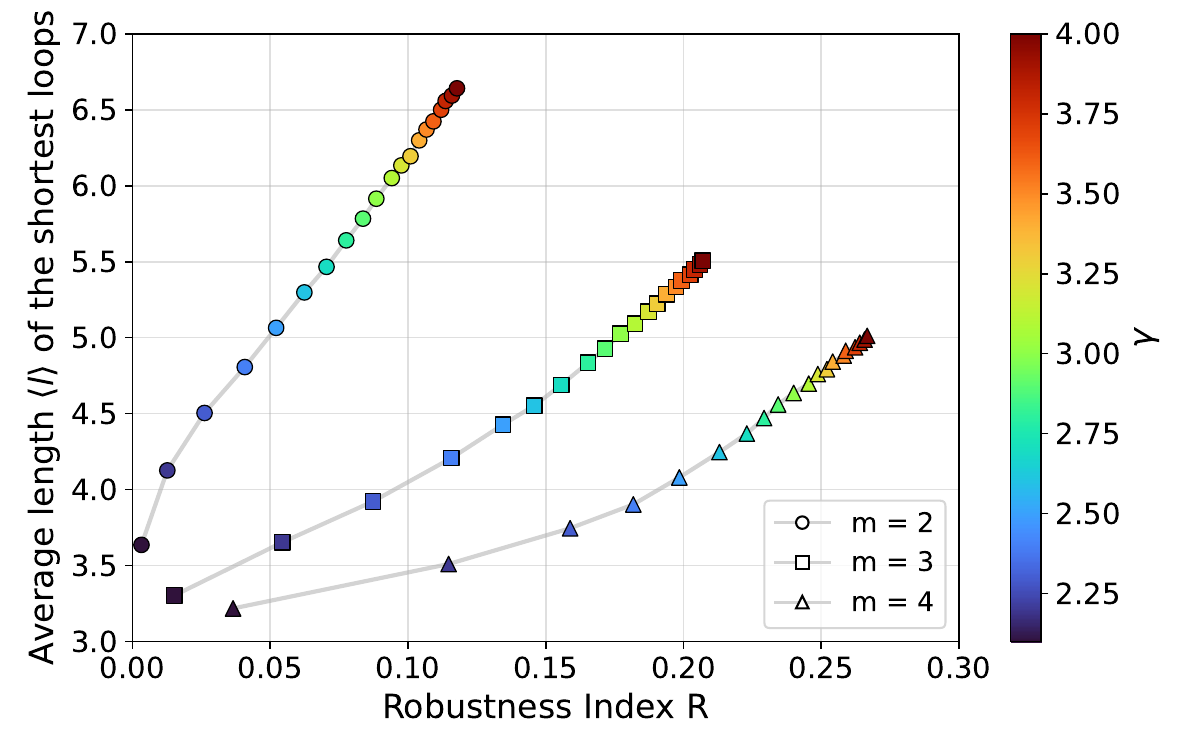}
    \caption{Betweenness}
  \end{subfigure}
  \hfill
  \begin{subfigure}[b]{0.32\textwidth}
    \includegraphics[width=\textwidth]{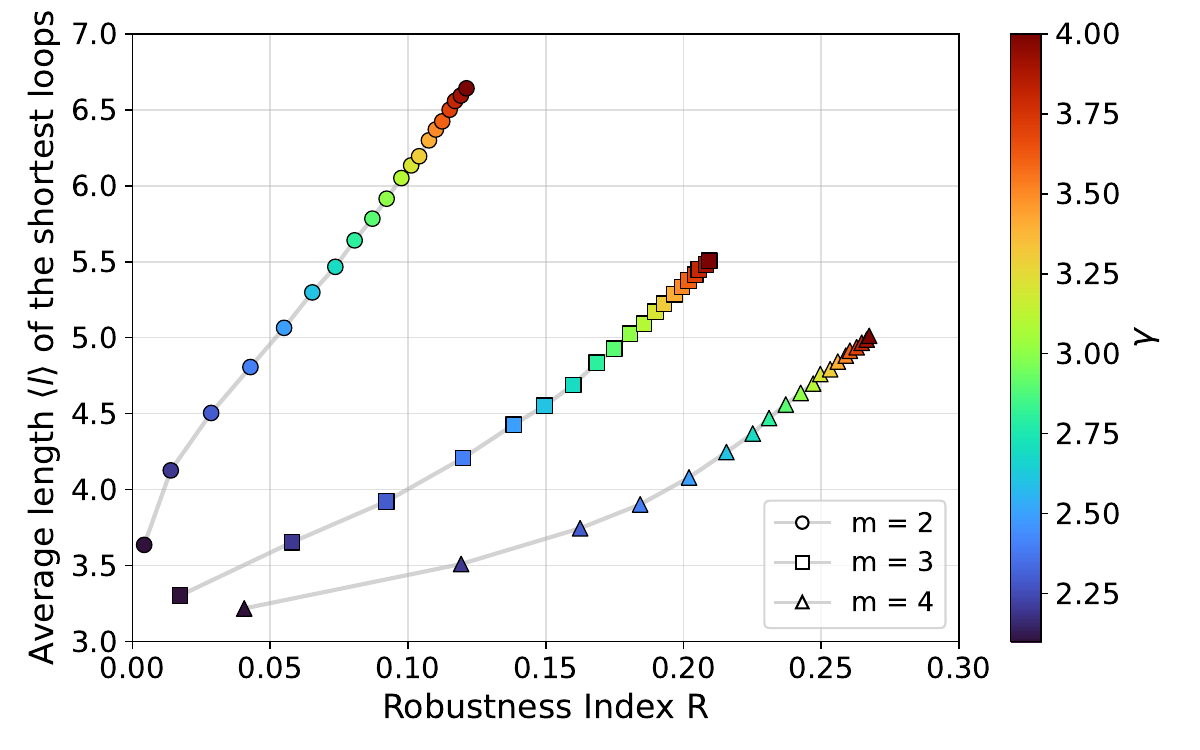}
    \caption{BP}
  \end{subfigure}  
  \caption{Robustness index R against (a)(d) degrees, (b)(e) betweenness, (c)(f) BP attacks versus the average length ⟨l⟩ of the shortest loops in randomized scale-free networks with $N=10^3$ (a-c), $N=10^4$ (b-d), and $m=2,3,4$. The horizontal axis represents the robustness index $R$, while the vertical axis shows the average length $\langle l \rangle$ of the shortest loops. Each colored point corresponds to a network generated with a power-law exponent in the range $2.1<\gamma<4.0$ shown by the color bar. Table \ref{tab:R_L_correlation} shows the positive correlation between $R$ and $\langle l \rangle$ as $\gamma$ increases (smaller variance of $P(k)$).}
\label{fig:R_vs_l}
\end{figure}

In this subsection, we show that the average length $\langle l \rangle$ of the shortest loops become larger as the variance $\sigma^2$ of $P(k)$ decreases. This means the emergence of large holes in the network. At the same time, the robustness index $R$ also increases as shown in Figure~\ref{fig:R_vs_l} (a)-(f). All figures present scatter plots of the robustness index $R$ versus the average length $\langle l \rangle$ of the shortest loops against degrees, betweenness centralities, and belief-propagation attacks, where colored points represent a power-law exponents $2.1<\gamma<4.0$ from dark purple to red. For all attacks and network sizes, a clear positive correlation between $R$ and $\langle l \rangle$ is observed as $\gamma$ increases, as summarized in Table \ref{tab:R_L_correlation}. This means larger $\gamma$ leads to both stronger robustness of connectivity and larger hole in networks as similar to the previous results in a wide class of randomized networks, which include not only scale-free networks (with $\gamma\approx3$), but also ER random graphs and regular networks \cite{kawato2025larger}. Moreover, by comparing (a)-(c) or (d)-(f), we find that for a fixed network size $N$, as $m$ increases, the slope of the scatter plots decreases, indicating that a small increases in the average length $\langle l \rangle$ of the shortest loops leads to a much larger improvement on the robustness index $R$. In addition, by comparing (a)(c), (b)(e), (c)(f), we find that for a fixed $m$, as $N$ becomes larger, the slope of the scatter plots increases, indicating that the influence of $\langle l \rangle$ on $R$ becomes weaker as the network size increases.

\begin{table}[!ht]
\centering
\caption{Pearson correlation coefficients between the robustness index $R$ and the average length $\langle l \rangle$ of the shortest loops against recalculated degree, recalculated betweenness, and recalculated BP attacks respectively in Figure \ref{fig:R_vs_l}.}
\begin{tabular}{c|c|c|c|c}
\hline
\textbf{$N$} & \textbf{$m$} & \textbf{$R^{\mathrm{Degrees}}$} & \textbf{$R^{\mathrm{Betweenness}}$} & \textbf{$R^{\mathrm{BP}}$} \\
\hline
1000  & 2 & 0.9953 & 0.9940 & 0.9952 \\
1000  & 3 & 0.9937 & 0.9969 & 0.9950 \\
1000  & 4 & 0.9733 & 0.9808 & 0.9870 \\
10000 & 2 & 0.9981 & 0.9979 & 0.9980 \\
10000 & 3 & 0.9865 & 0.9913 & 0.9894 \\
10000 & 4 & 0.9560 & 0.9633 & 0.9613 \\
\hline
\end{tabular}
\label{tab:R_L_correlation}
\end{table}

\section{Conclusion}
\label{sec:conclusion}
We have studied the robustness of connectivity in SF networks with tunable power-law exponents $\gamma$ in the realistic range $2.1 \leq \gamma \leq 4.0$ under the same condition of a fixed average degree $\langle k \rangle$. For investigating the pure effect of $P(k)$, the generated SF networks are randomized by the configuration model to eliminate the degree-degree or higher correlations. We have shown a relation that the robustness of connectivity becomes stronger as the degree distributions $P(k)$ are narrower by larger power-law exponent $\gamma$. Coincidentally, we have shown that the average length of $\langle l \rangle$ of the shortest loops becomes longer as $P(k)$ are narrower by larger exponent $\gamma$. These results are consistent with previous results obtained for synthetic randomized networks with continuously changing degree distributions which include SF networks with $\gamma=3$, ER random graphs, and regular networks \cite{Chujyo2023OptimalRobustness, kawato2025larger}. However, we have also find that the robustness index $R$ becomes bounded for $\gamma > 3$. This limitation is associated with the convergence of $k_{\max}$ and $\sigma^2$ (see Figure \ref{fig:kmax_variance_gamma}) to nonzero values. In other words, hub nodes still exist even for large $\gamma$. Consequently, SF networks remain vulnerable. Since many real-world networks have SF structures with $2 < \gamma < 3$, further enhancement of the robustness requires a drastic structural change from SF networks to regular networks. These implications are especially relevant to real networks such as the World Wide Web and the Internet. 

To examine whether our conclusions remain unchanged for denser or larger networks, we additionally analyzed denser SF networks with $N=10^3, m=3,4$, and larger networks with $N=10^4, m=2,3,4$ (see Supplementary Information for details). Among these networks, we observe the consistant results that narrower degree distributions lead to longer average length $\langle l \rangle$ of the shortest loops and larger robustness index $R$ as stronger robustness of connectivity, while the enhancement of $R$ for $\gamma>3$ is limited.

On the other hands, for a future work, the robustness may behave differently for geographical networks \cite{mou2025vulnerable} embedded on a space such as power grids or transportation systems. Extended analyses to these cases will also give important directions.

\clearpage
\nolinenumbers
\section*{Supplementary Information}
% 重置图和表格计数器
\setcounter{figure}{0}
\setcounter{table}{0}

% 保存原始的图表名称命令
\let\origfigurename\figurename
\let\origtablename\tablename

% 将图表名称设为空
\renewcommand{\figurename}{}
\renewcommand{\tablename}{}

% 重新定义图的编号格式
% \renewcommand{\thefigure}{S\arabic{figure} \origfigurename}
% \renewcommand{\thetable}{S\arabic{table} \origtablename}
\renewcommand{\thefigure}{S\arabic{figure}}
\renewcommand{\thetable}{S\arabic{table}}

\begin{figure}[H]
  \centering
  % 第一行
  \begin{subfigure}[b]{0.42\textwidth}
    \includegraphics[width=\textwidth]{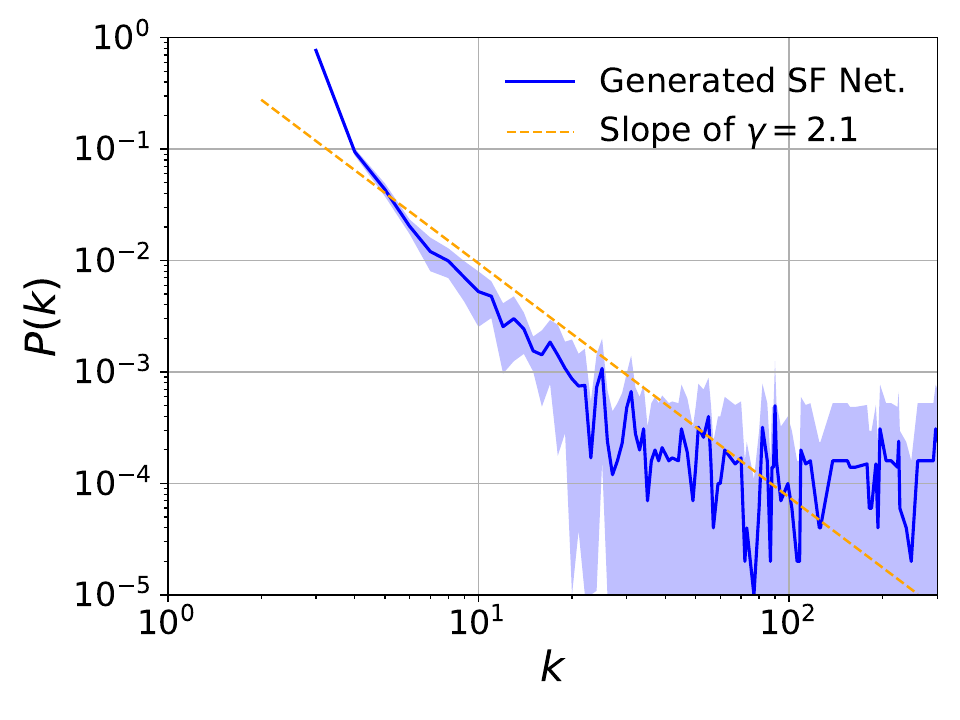}
    \caption{$\gamma=2.1$}
  \end{subfigure}
  \hfill
  \begin{subfigure}[b]{0.42\textwidth}
    \includegraphics[width=\textwidth]{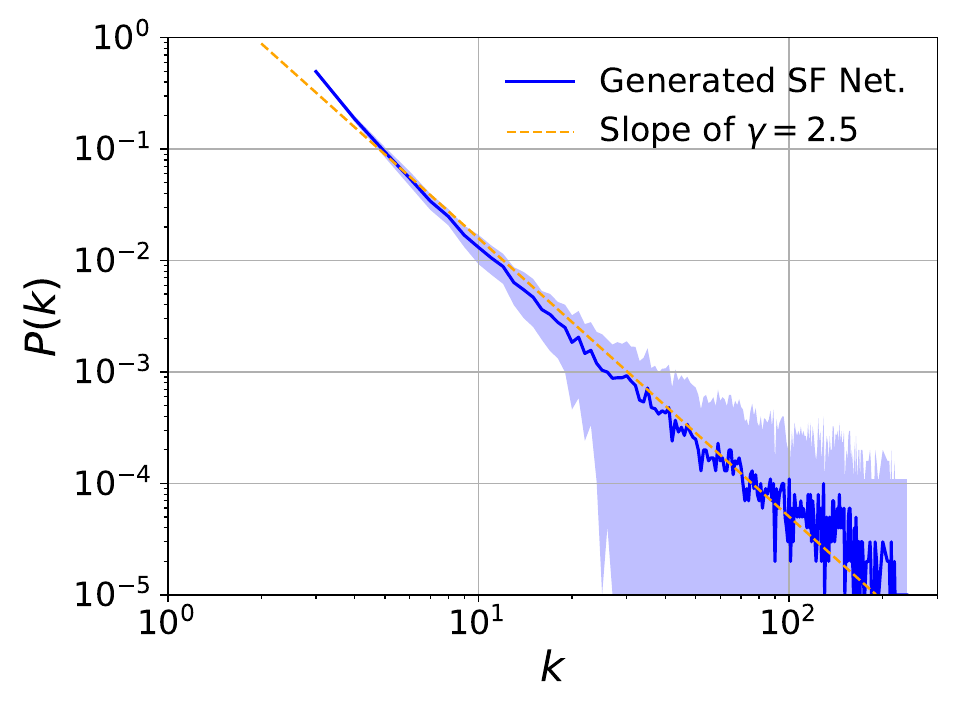}
    \caption{$\gamma=2.5$}
  \end{subfigure}

  \vspace{1em} % 两行之间的垂直间距
  
  \begin{subfigure}[b]{0.42\textwidth}
    \includegraphics[width=\textwidth]{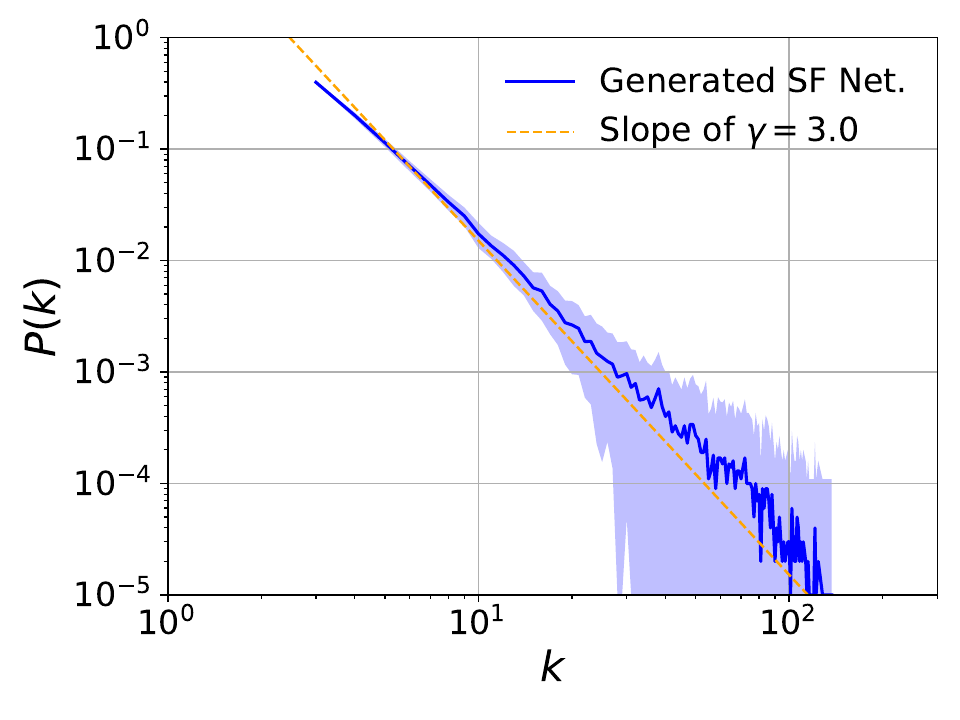}
    \caption{$\gamma=3.0$}
  \end{subfigure}
  \hfill
  \begin{subfigure}[b]{0.42\textwidth}
    \includegraphics[width=\textwidth]{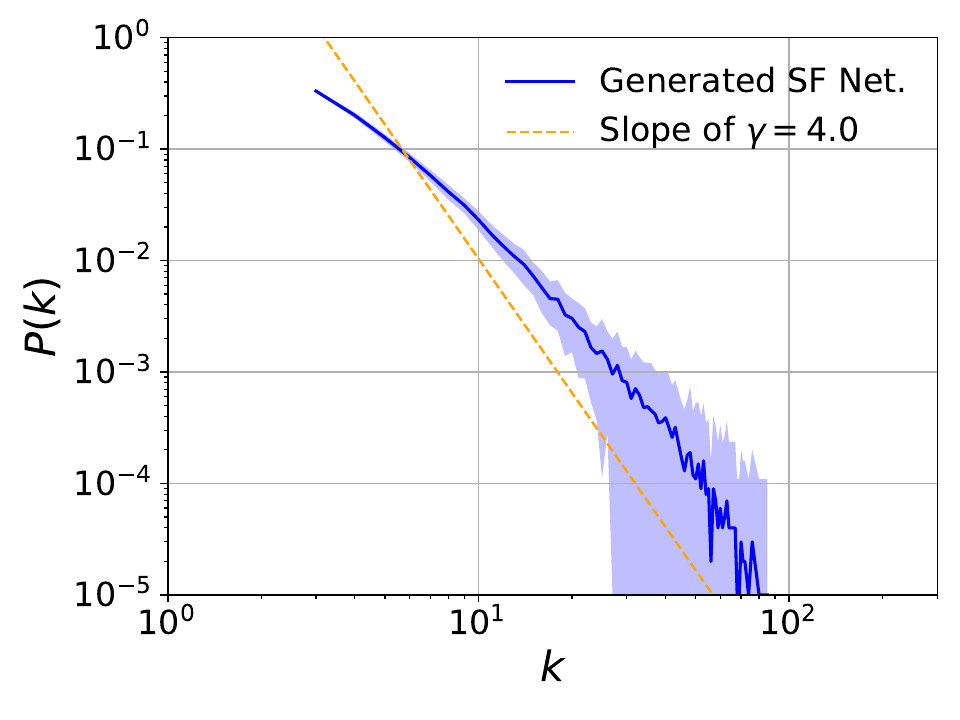}
    \caption{$\gamma=4.0$}
  \end{subfigure}
    \caption{Comparison with Figure \ref{fig:degree_distributions} in the case of $N=10^3$ and $m=2$. Degree distributions $P(k) \sim k^{-\gamma}$ in generated SF networks with power-law exponents (a) $\gamma=2.1$, (b) $\gamma=2.5$, (c) $\gamma=3.0$, and (d) $\gamma=4.0$ for \bm{$N=10^3$} and \bm{$m=3$}. Dashed lines guide the slope of power-law exponent $\gamma$ in the log-log plot. The shaded areas show the standard deviations in log-log scales.}
    \label{fig:degree_distributions_1000_m3}
\end{figure}

\begin{figure}[H]
  \centering
  % 第一行
  \begin{subfigure}[b]{0.42\textwidth}
    \includegraphics[width=\textwidth]{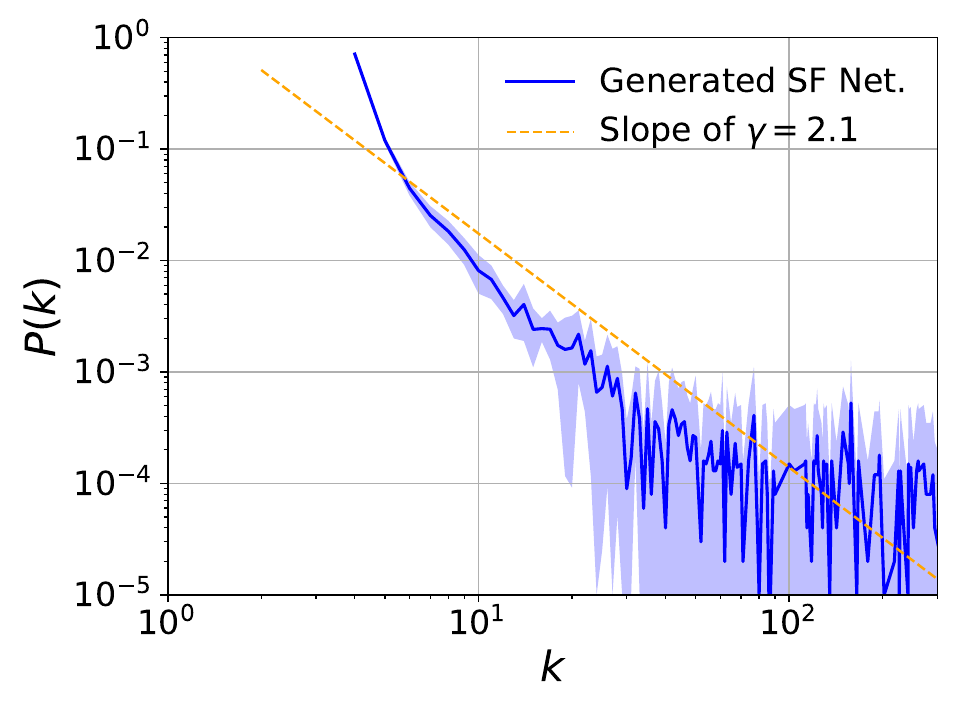}
    \caption{$\gamma=2.1$}
  \end{subfigure}
  \hfill
  \begin{subfigure}[b]{0.42\textwidth}
    \includegraphics[width=\textwidth]{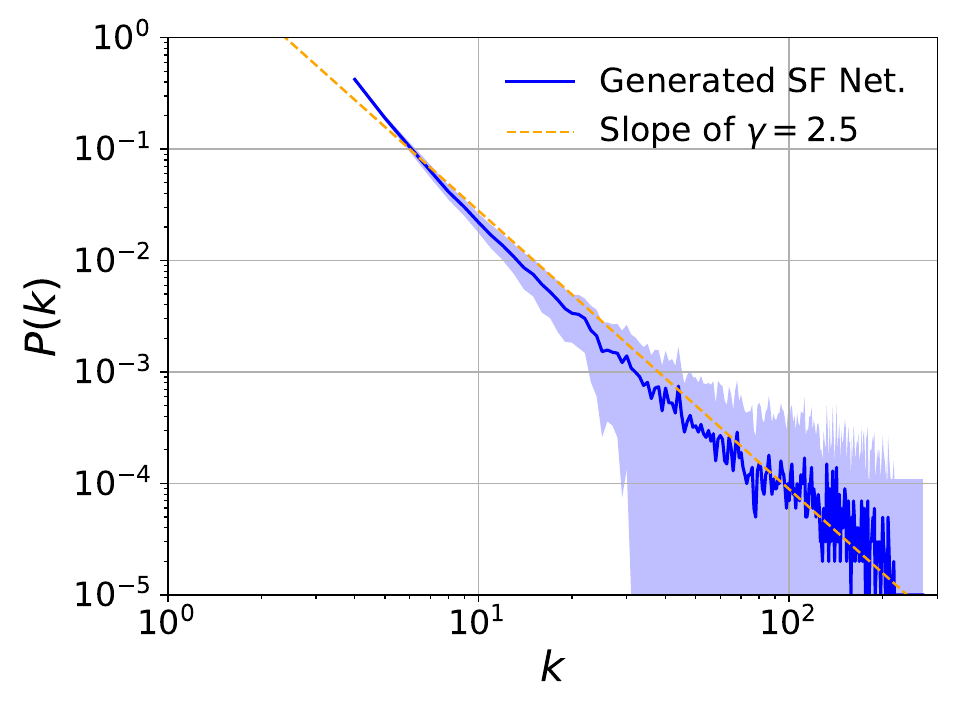}
    \caption{$\gamma=2.5$}
  \end{subfigure}

  \vspace{1em} % 两行之间的垂直间距
  
  \begin{subfigure}[b]{0.42\textwidth}
    \includegraphics[width=\textwidth]{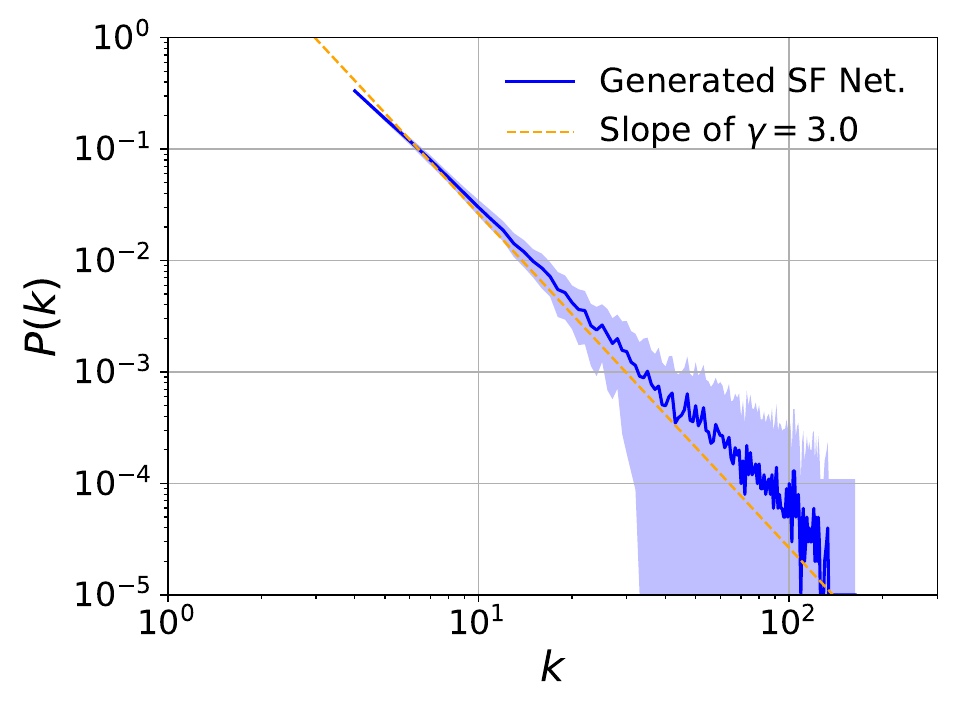}
    \caption{$\gamma=3.0$}
  \end{subfigure}
  \hfill
  \begin{subfigure}[b]{0.42\textwidth}
    \includegraphics[width=\textwidth]{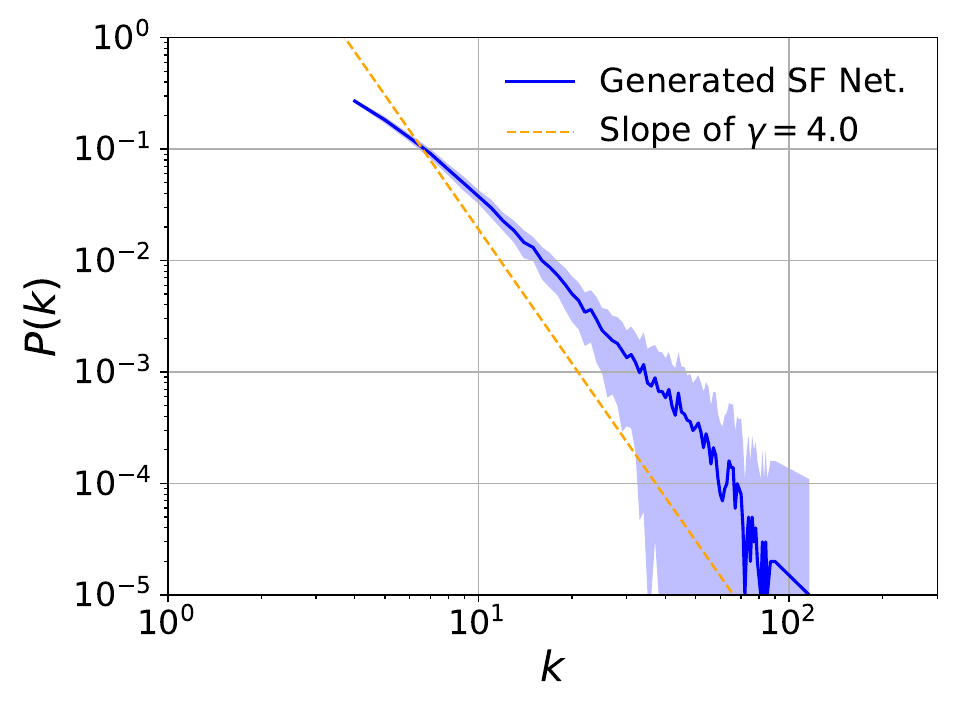}
    \caption{$\gamma=4.0$}
  \end{subfigure}
    \caption{Comparison with Figure \ref{fig:degree_distributions} in the case of $N=10^3$ and $m=2$. Degree distributions $P(k) \sim k^{-\gamma}$ in generated SF networks with power-law exponents (a) $\gamma=2.1$, (b) $\gamma=2.5$, (c) $\gamma=3.0$, and (d) $\gamma=4.0$ for \bm{$N=10^3$} and \bm{$m=4$}. Dashed lines guide the slope of power-law exponent $\gamma$ in the log-log plot. The shaded areas show the standard deviations in log-log scales.}
    \label{fig:degree_distributions_1000_m4}
\end{figure}

\begin{figure}[H]
  \centering
  % 第一行
  \begin{subfigure}[b]{0.42\textwidth}
    \includegraphics[width=\textwidth]{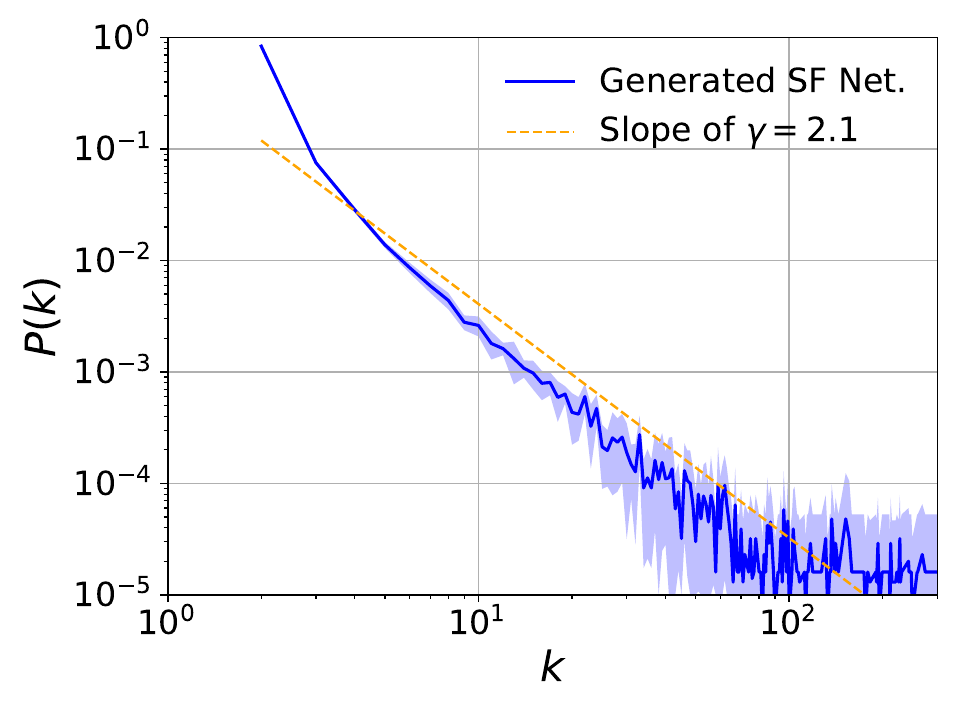}
    \caption{$\gamma=2.1$}
  \end{subfigure}
  \hfill
  \begin{subfigure}[b]{0.42\textwidth}
    \includegraphics[width=\textwidth]{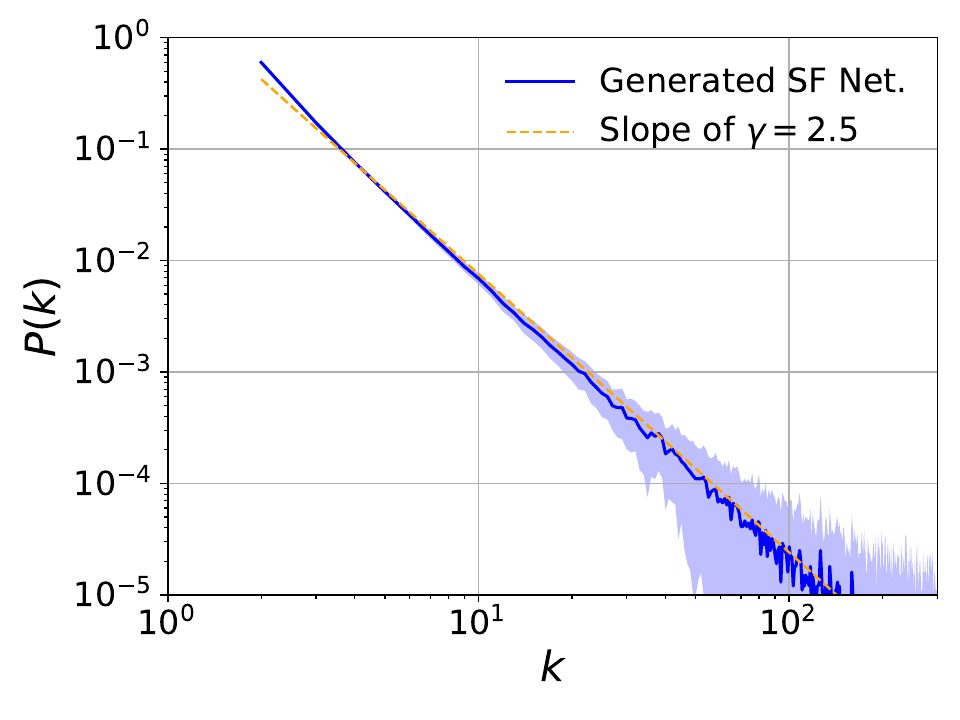}
    \caption{$\gamma=2.5$}
  \end{subfigure}

  \vspace{1em} % 两行之间的垂直间距
  
  \begin{subfigure}[b]{0.42\textwidth}
    \includegraphics[width=\textwidth]{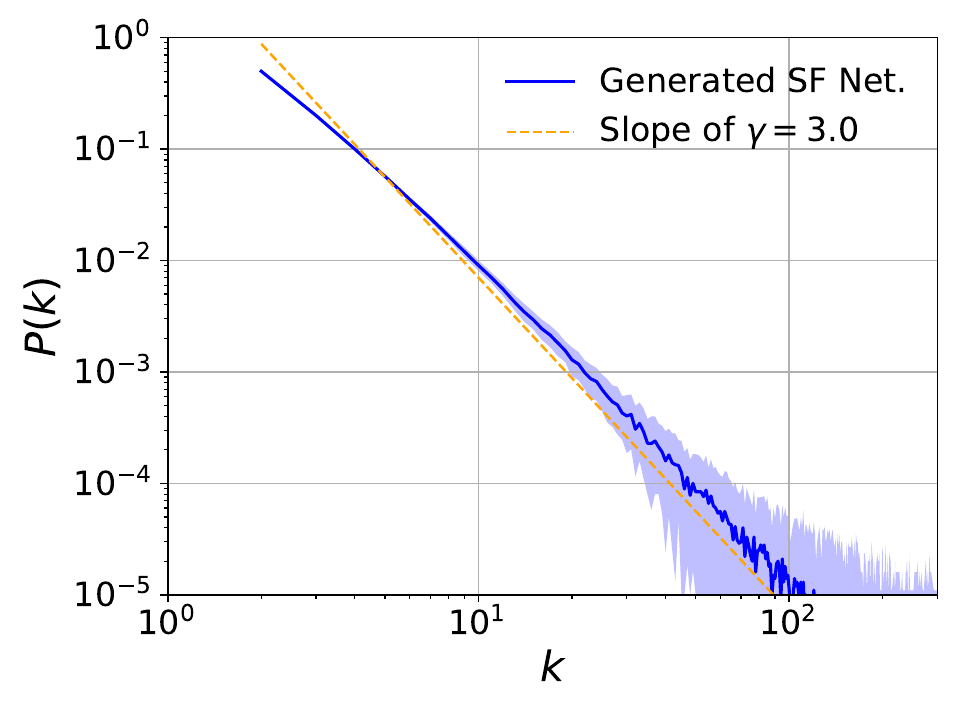}
    \caption{$\gamma=3.0$}
  \end{subfigure}
  \hfill
  \begin{subfigure}[b]{0.42\textwidth}
    \includegraphics[width=\textwidth]{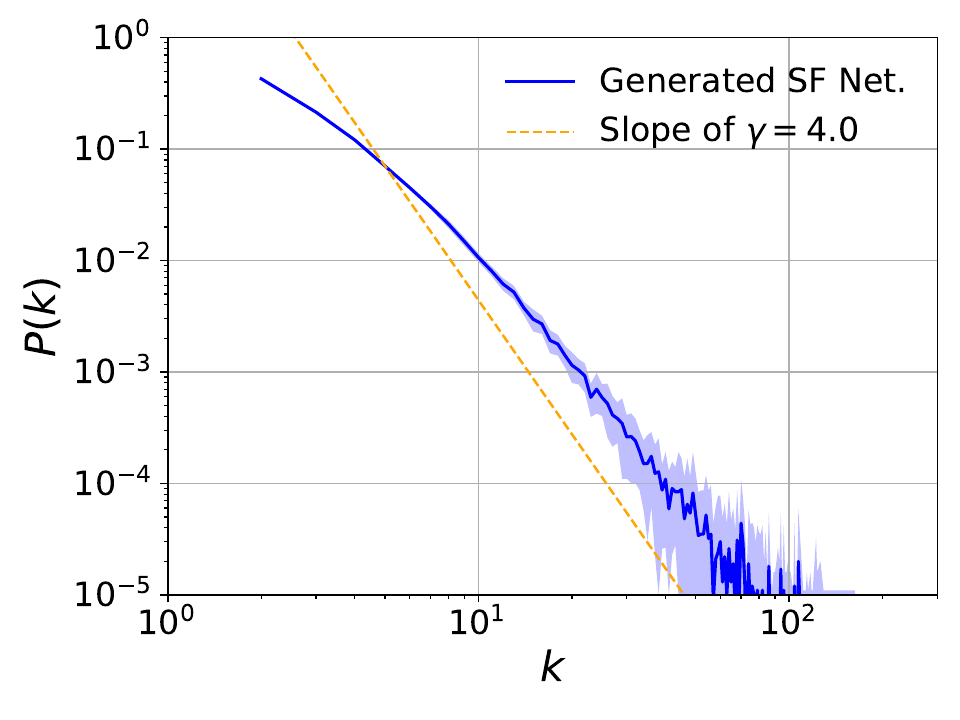}
    \caption{$\gamma=4.0$}
  \end{subfigure}
    \caption{Comparison with Figure \ref{fig:degree_distributions} in the case of $N=10^3$ and $m=2$. Degree distributions $P(k) \sim k^{-\gamma}$ in generated SF networks with power-law exponents (a) $\gamma=2.1$, (b) $\gamma=2.5$, (c) $\gamma=3.0$, and (d) $\gamma=4.0$ for \bm{$N=10^4$} and \bm{$m=2$}. Dashed lines guide the slope of power-law exponent $\gamma$ in the log-log plot. The shaded areas show the standard deviations in log-log scales.}
    \label{fig:degree_distributions_10000_m2}
\end{figure}

\begin{figure}[H]
  \centering
  % 第一行
  \begin{subfigure}[b]{0.42\textwidth}
    \includegraphics[width=\textwidth]{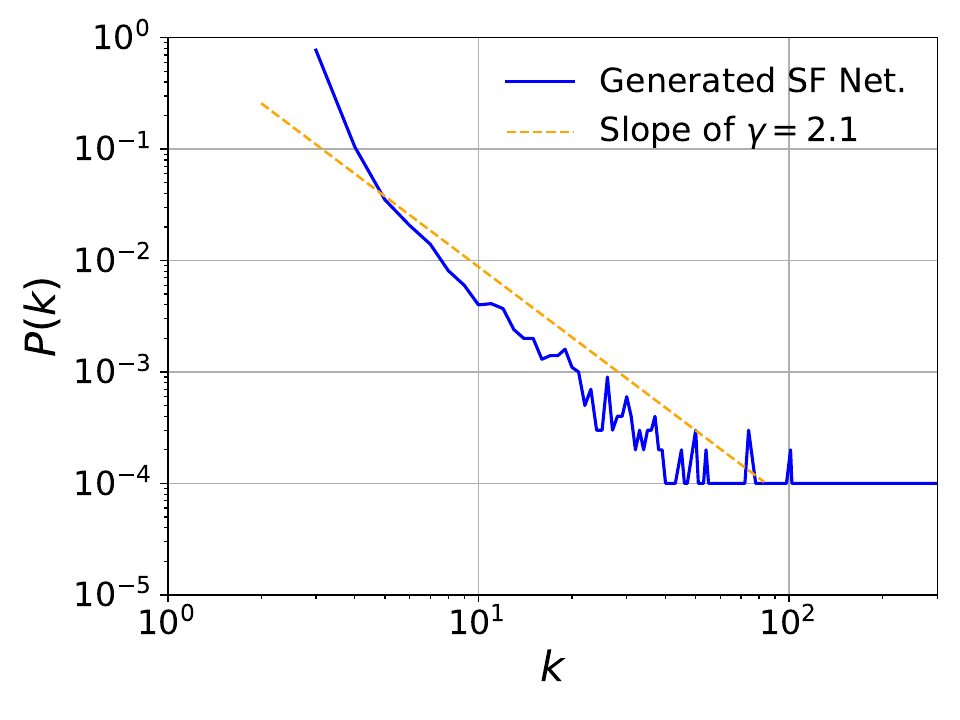}
    \caption{$\gamma=2.1$}
  \end{subfigure}
  \hfill
  \begin{subfigure}[b]{0.42\textwidth}
    \includegraphics[width=\textwidth]{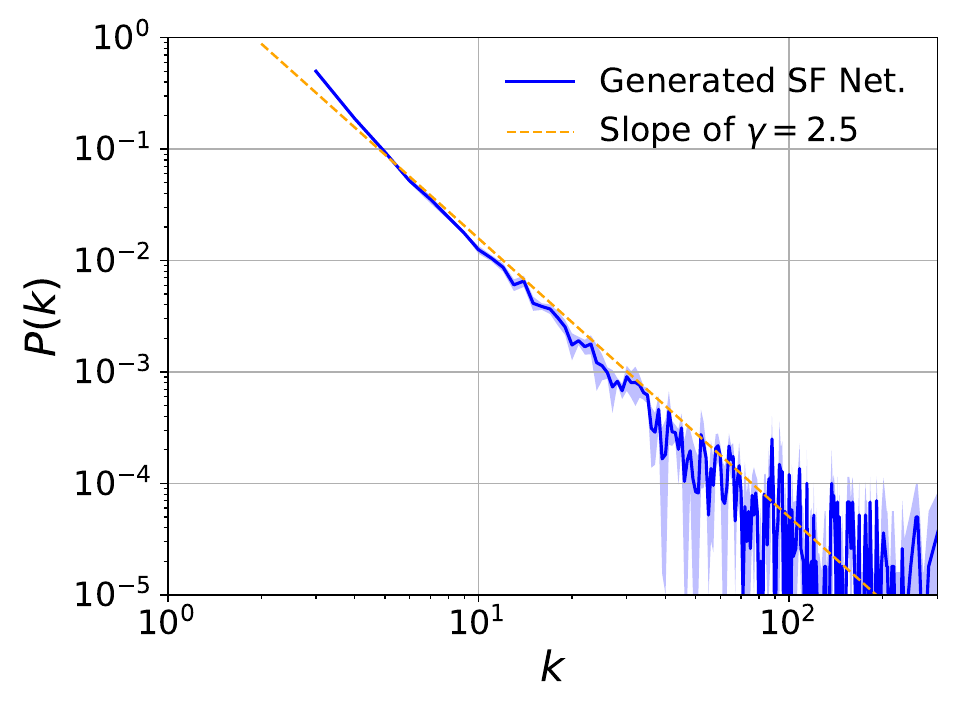}
    \caption{$\gamma=2.5$}
  \end{subfigure}

  \vspace{1em} % 两行之间的垂直间距
  
  \begin{subfigure}[b]{0.42\textwidth}
    \includegraphics[width=\textwidth]{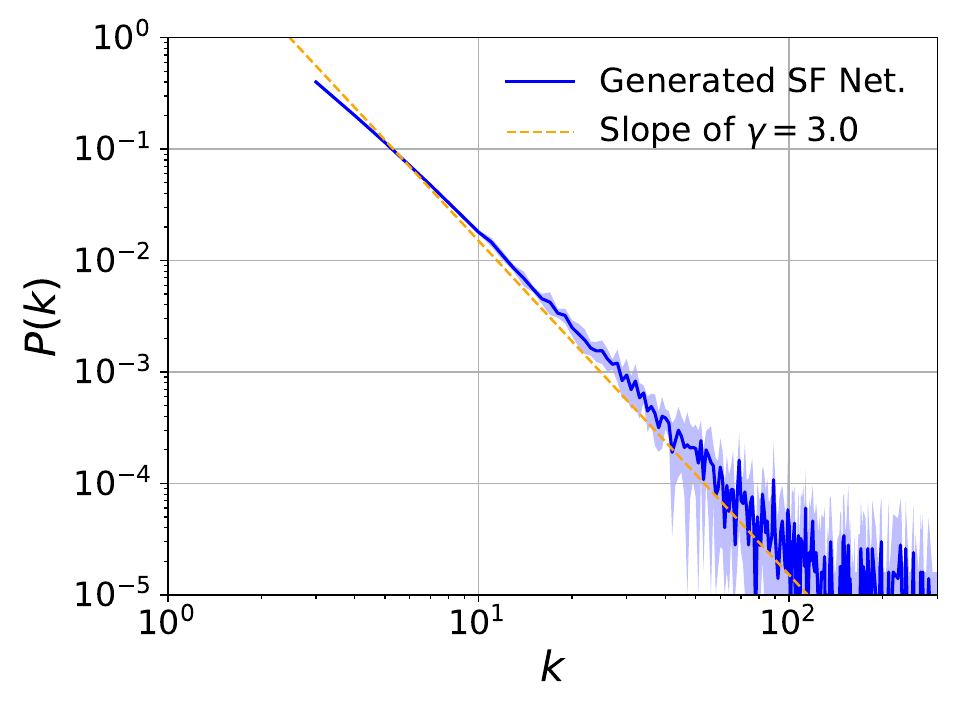}
    \caption{$\gamma=3.0$}
  \end{subfigure}
  \hfill
  \begin{subfigure}[b]{0.42\textwidth}
    \includegraphics[width=\textwidth]{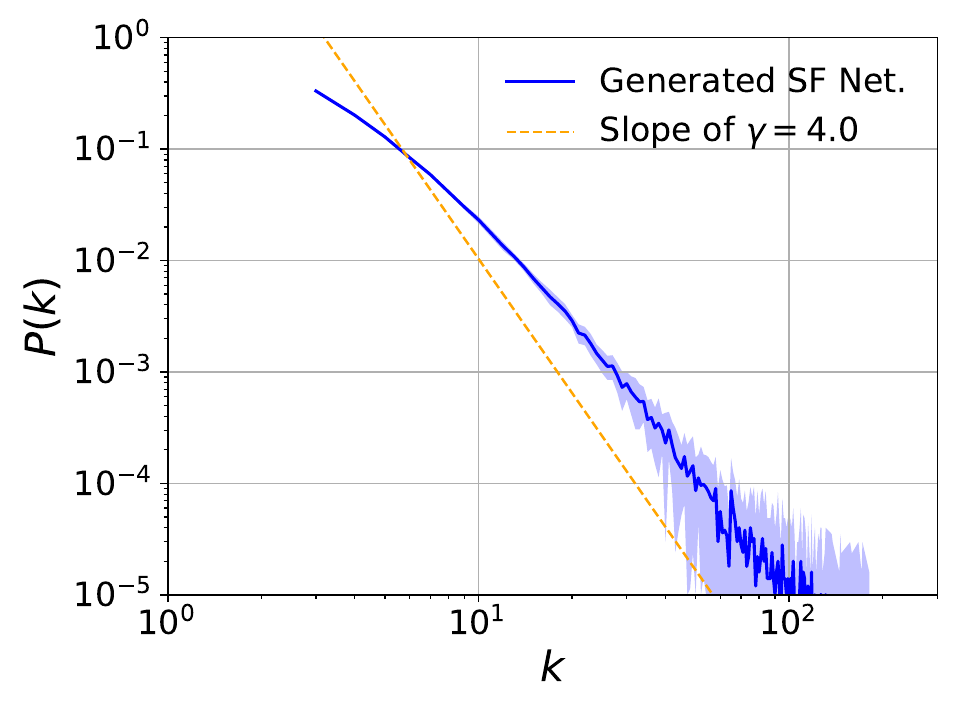}
    \caption{$\gamma=4.0$}
  \end{subfigure}
    \caption{Comparison with Figure \ref{fig:degree_distributions} in the case of $N=10^3$ and $m=2$. Degree distributions $P(k) \sim k^{-\gamma}$ in generated SF networks with power-law exponents (a) $\gamma=2.1$, (b) $\gamma=2.5$, (c) $\gamma=3.0$, and (d) $\gamma=4.0$ for \bm{$N=10^4$} and \bm{$m=3$}. Dashed lines guide the slope of power-law exponent $\gamma$ in the log-log plot. The shaded areas show the standard deviations in log-log scales.}
    \label{fig:degree_distributions_10000_m3}
\end{figure}

\begin{figure}[H]
  \centering
  % 第一行
  \begin{subfigure}[b]{0.42\textwidth}
    \includegraphics[width=\textwidth]{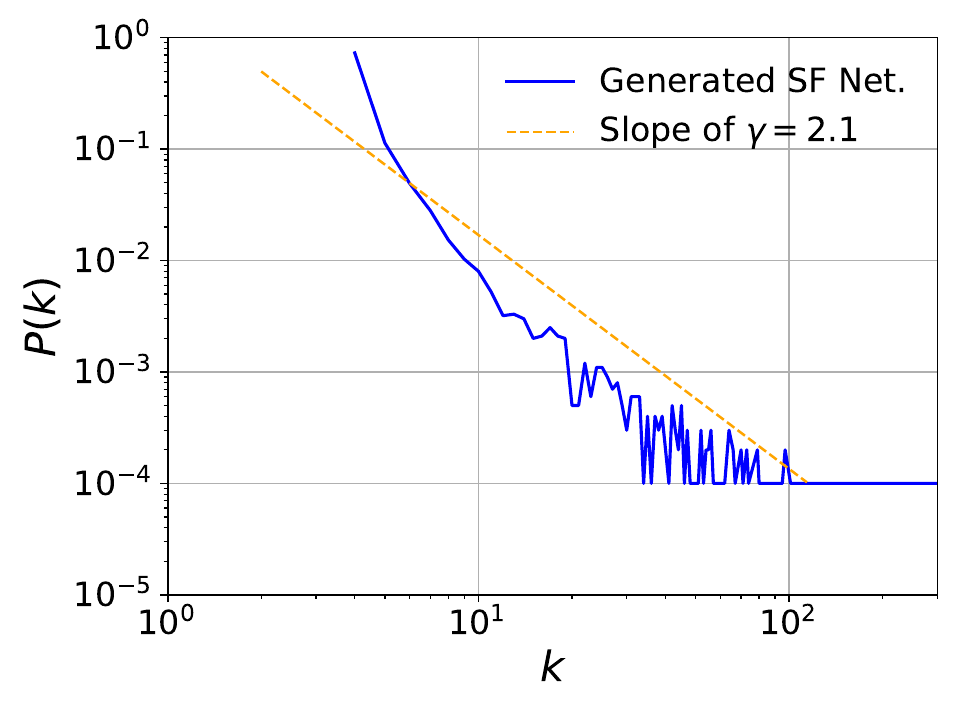}
    \caption{$\gamma=2.1$}
  \end{subfigure}
  \hfill
  \begin{subfigure}[b]{0.42\textwidth}
    \includegraphics[width=\textwidth]{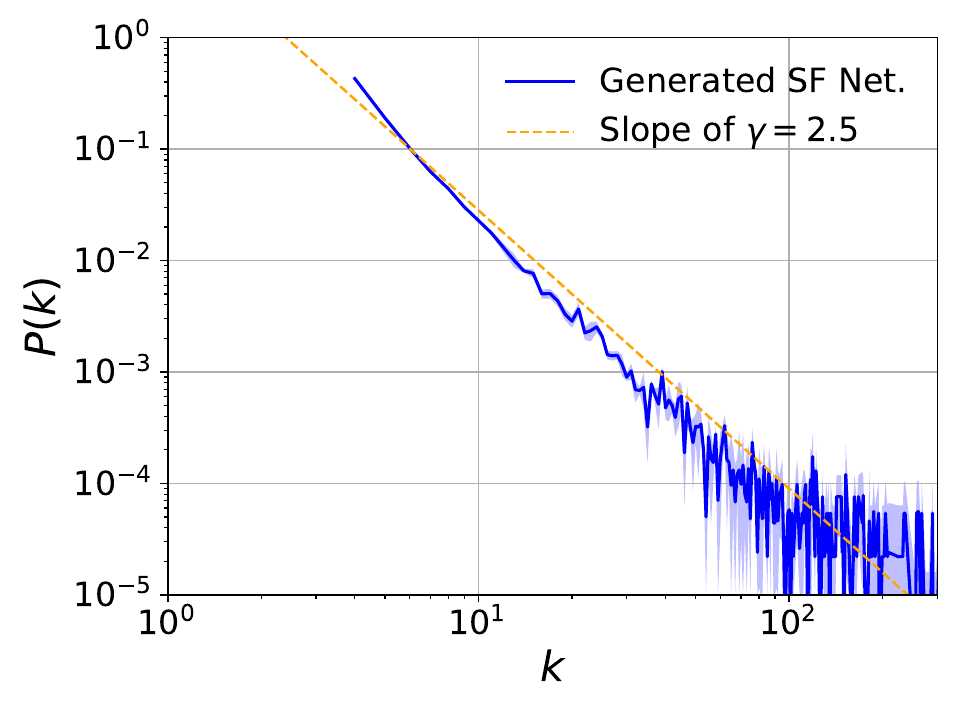}
    \caption{$\gamma=2.5$}
  \end{subfigure}

  \vspace{1em} % 两行之间的垂直间距
  
  \begin{subfigure}[b]{0.42\textwidth}
    \includegraphics[width=\textwidth]{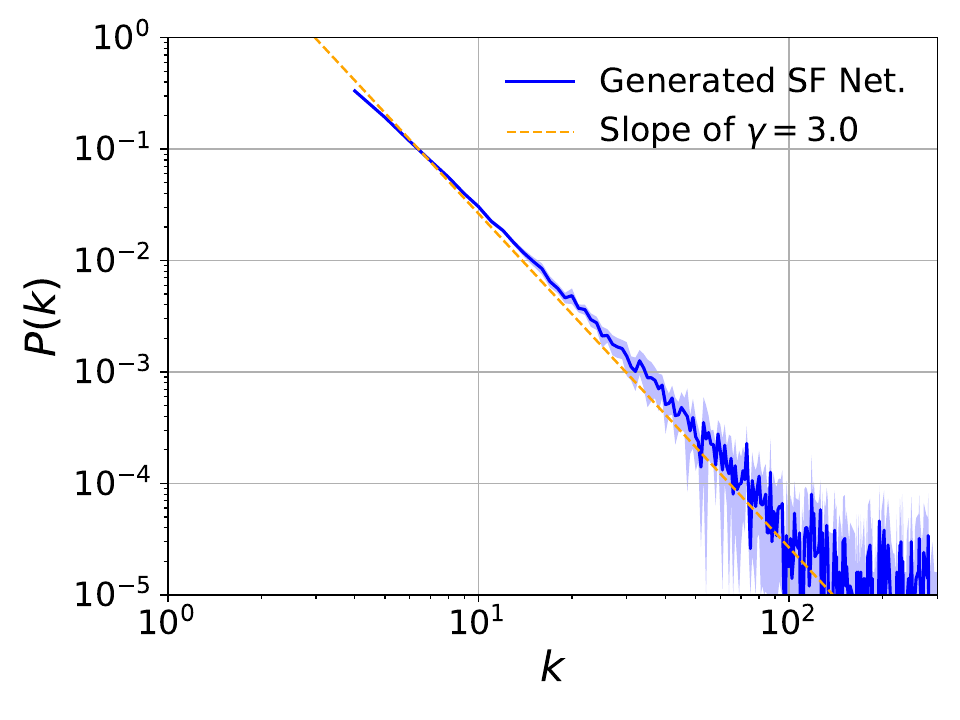}
    \caption{$\gamma=3.0$}
  \end{subfigure}
  \hfill
  \begin{subfigure}[b]{0.42\textwidth}
    \includegraphics[width=\textwidth]{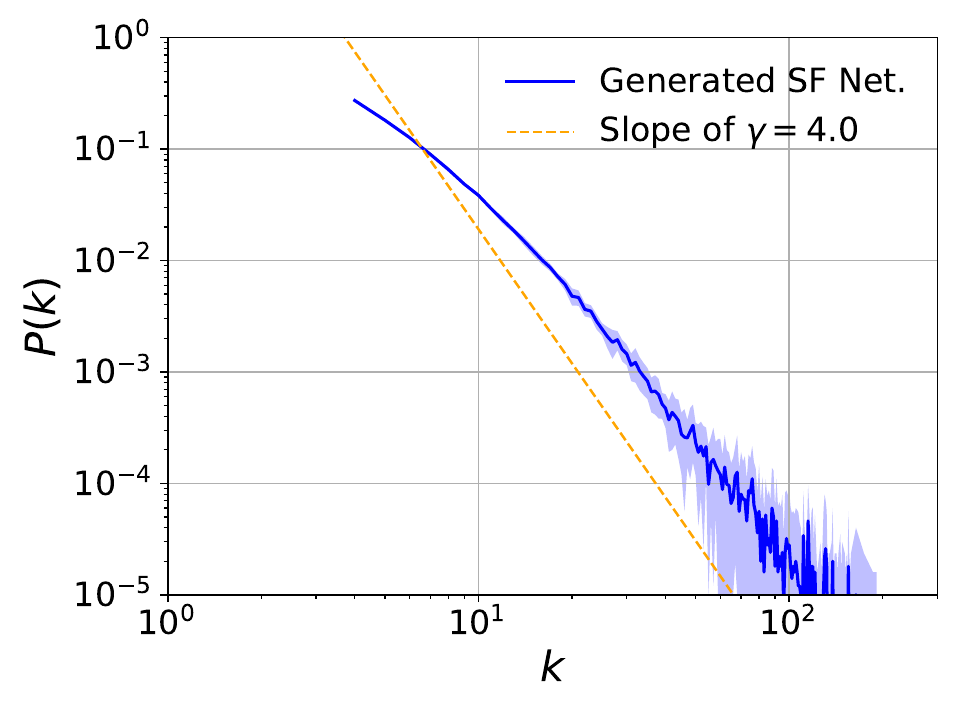}
    \caption{$\gamma=4.0$}
  \end{subfigure}
    \caption{Comparison with Figure \ref{fig:degree_distributions} in the case of $N=10^3$ and $m=2$. Degree distributions $P(k) \sim k^{-\gamma}$ in generated SF networks with power-law exponents (a) $\gamma=2.1$, (b) $\gamma=2.5$, (c) $\gamma=3.0$, and (d) $\gamma=4.0$ for \bm{$N=10^4$} and \bm{$m=4$}. Dashed lines guide the slope of power-law exponent $\gamma$ in the log-log plot. The shaded areas show the standard deviations in log-log scales.}
    \label{fig:degree_distributions_10000_m4}
\end{figure}

\begin{figure}[H]
  \centering
  % 第一行
  \begin{subfigure}[b]{0.42\textwidth}
    \includegraphics[width=\textwidth]{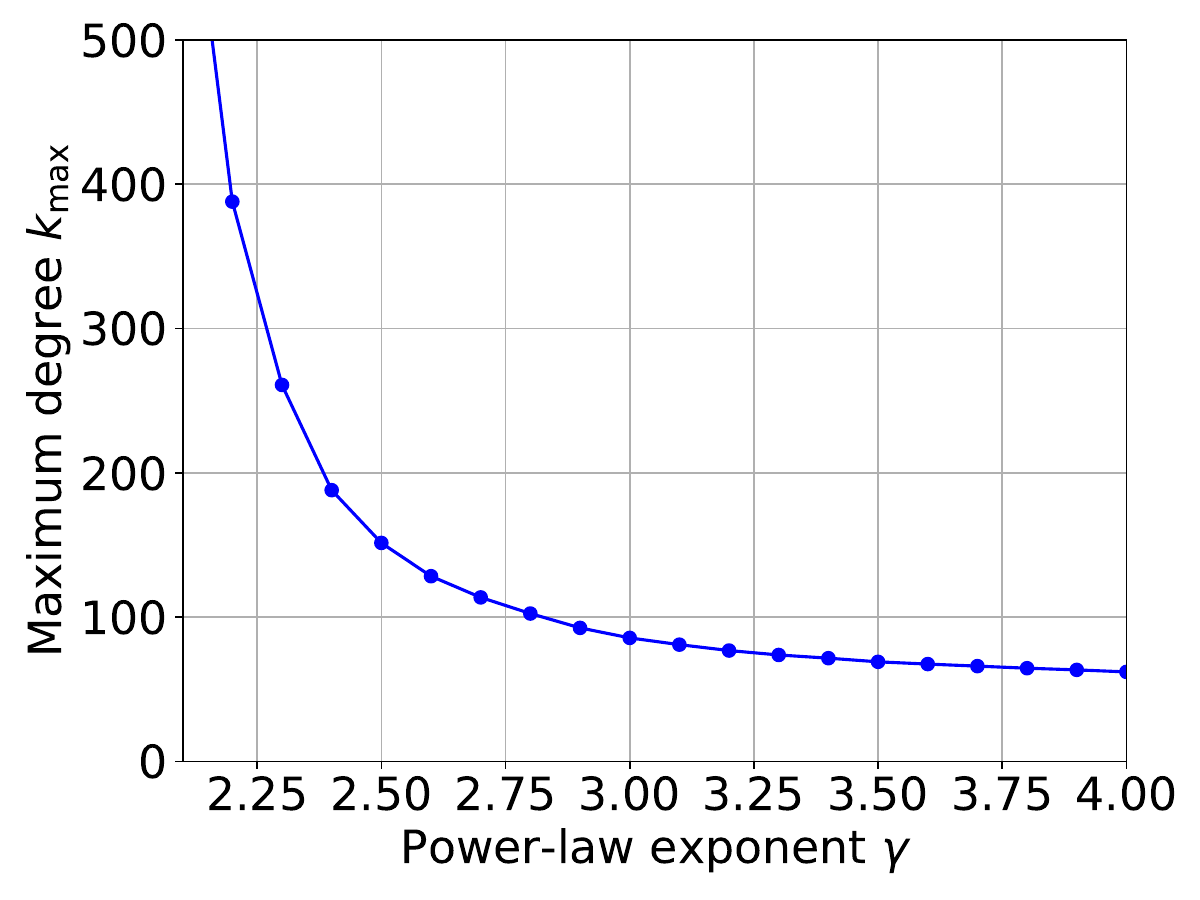}
    \caption{}
  \end{subfigure}
  \hfill
  \begin{subfigure}[b]{0.42\textwidth}
    \includegraphics[width=\textwidth]{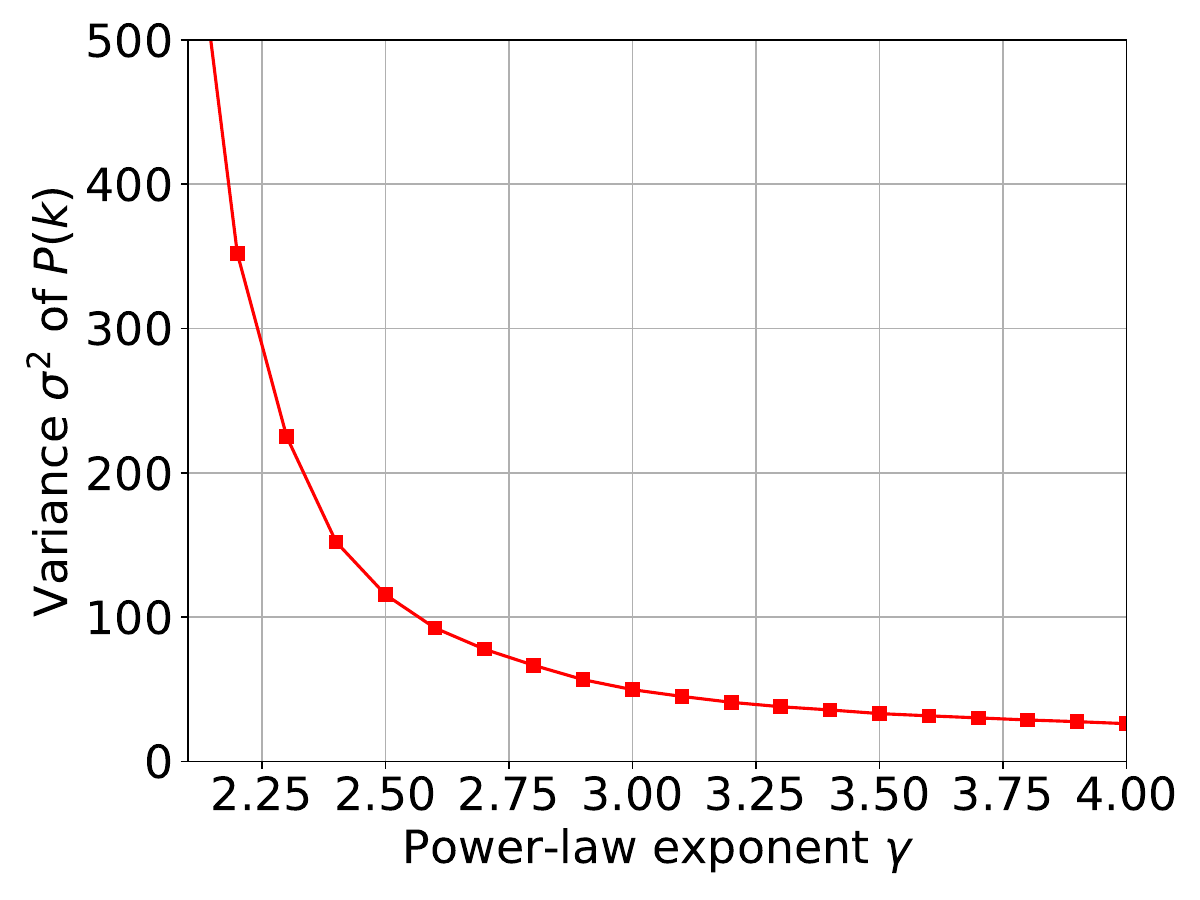}
    \caption{}
  \end{subfigure}
    \caption{Comparison with Figure \ref{fig:kmax_variance_gamma} in the case of $N=10^3$ and $m=2$. Monotone decreasing of (a) the maximum degree $k_{max}$ and (b) the variance $\sigma^2$ of degree distribution $P(k)$ with the power-law exponent $\gamma$ for \bm{$N=10^3$} and \bm{$m=3$}.}
    \label{fig:kmax_variance_gamma_1000_m3}
\end{figure}

\begin{figure}[H]
  \centering
  % 第一行
  \begin{subfigure}[b]{0.42\textwidth}
    \includegraphics[width=\textwidth]{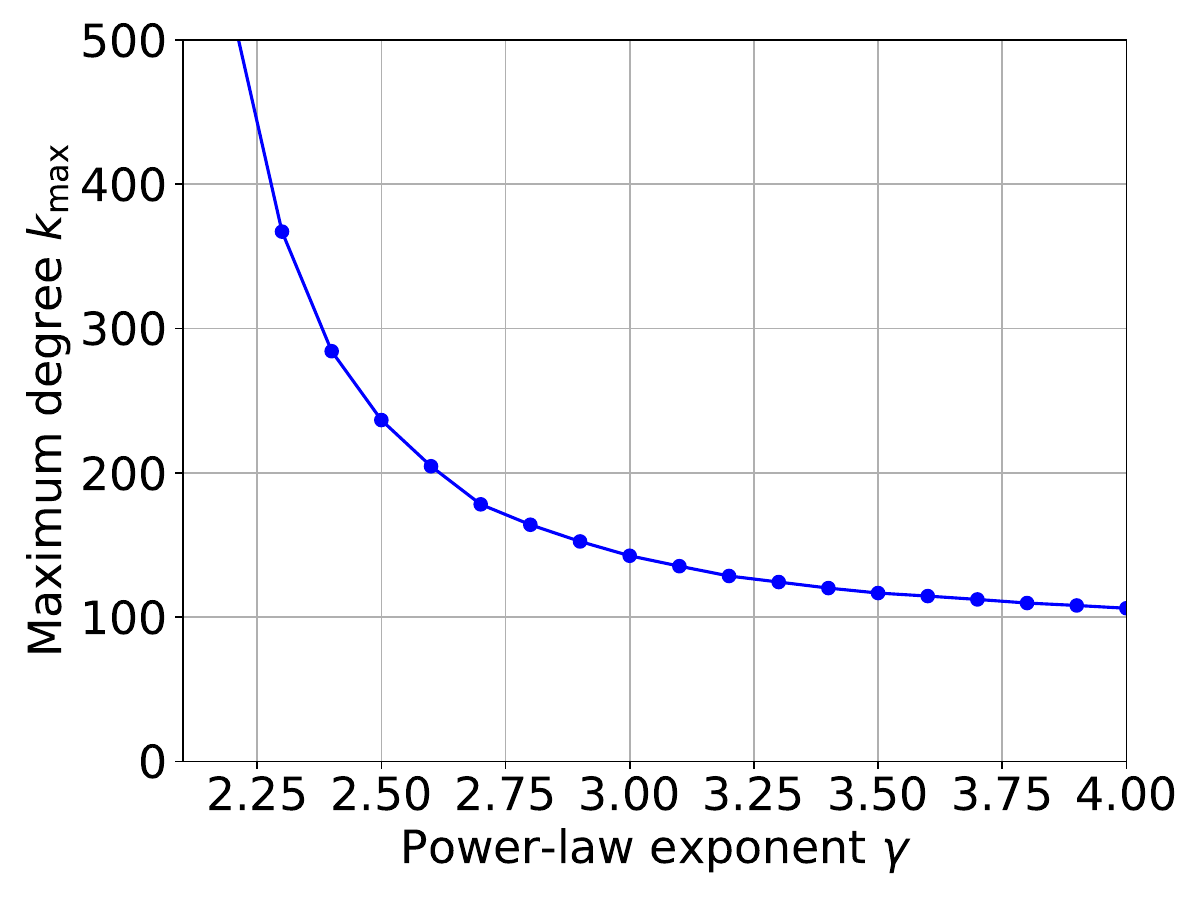}
    \caption{}
  \end{subfigure}
  \hfill
  \begin{subfigure}[b]{0.42\textwidth}
    \includegraphics[width=\textwidth]{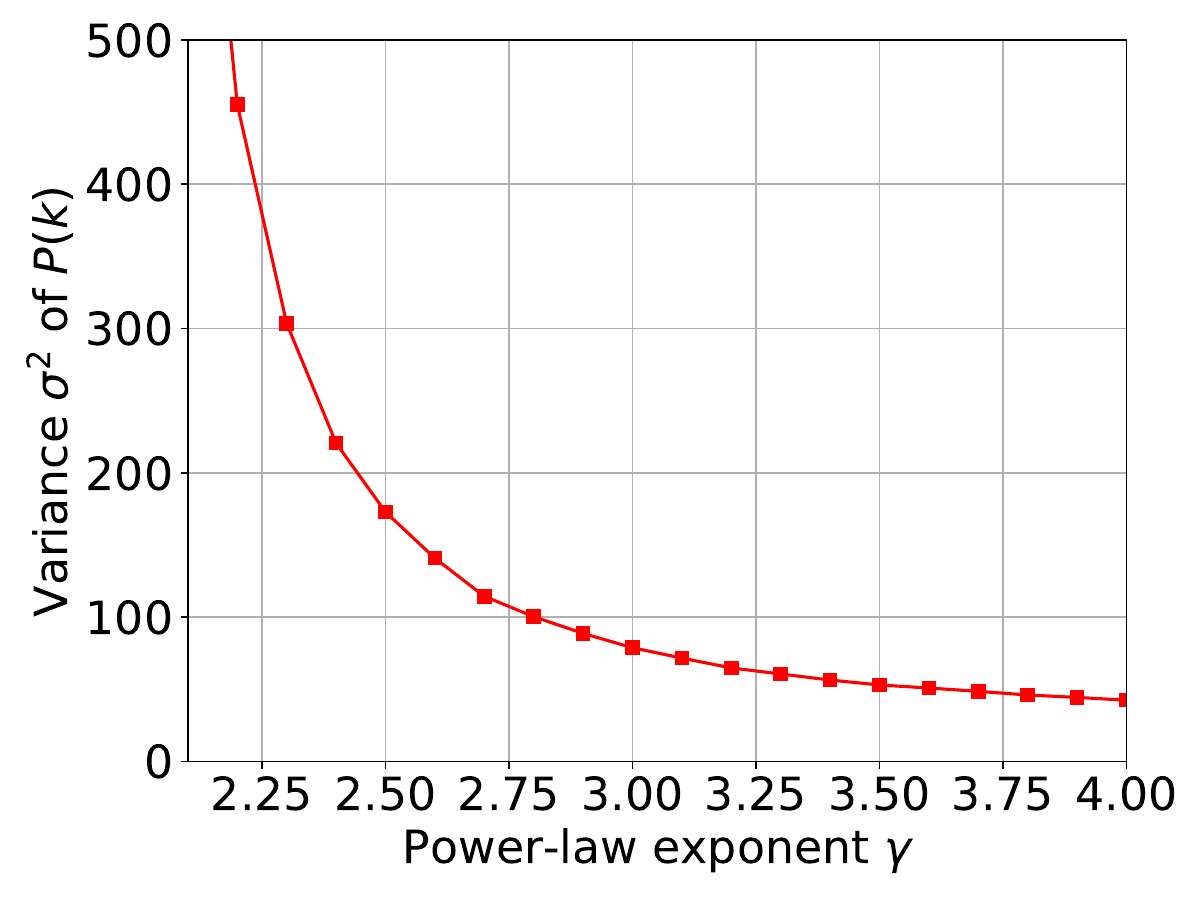}
    \caption{}
  \end{subfigure}
    \caption{Comparison with Figure \ref{fig:kmax_variance_gamma} in the case of $N=10^3$ and $m=2$. Monotone decreasing of (a) the maximum degree $k_{max}$ and (b) the variance $\sigma^2$ of degree distribution $P(k)$ with the power-law exponent $\gamma$ for \bm{$N=10^3$} and \bm{$m=4$}.}
    \label{fig:kmax_variance_gamma_1000_m4}
\end{figure}

\begin{figure}[H]
  \centering
  % 第一行
  \begin{subfigure}[b]{0.42\textwidth}
    \includegraphics[width=\textwidth]{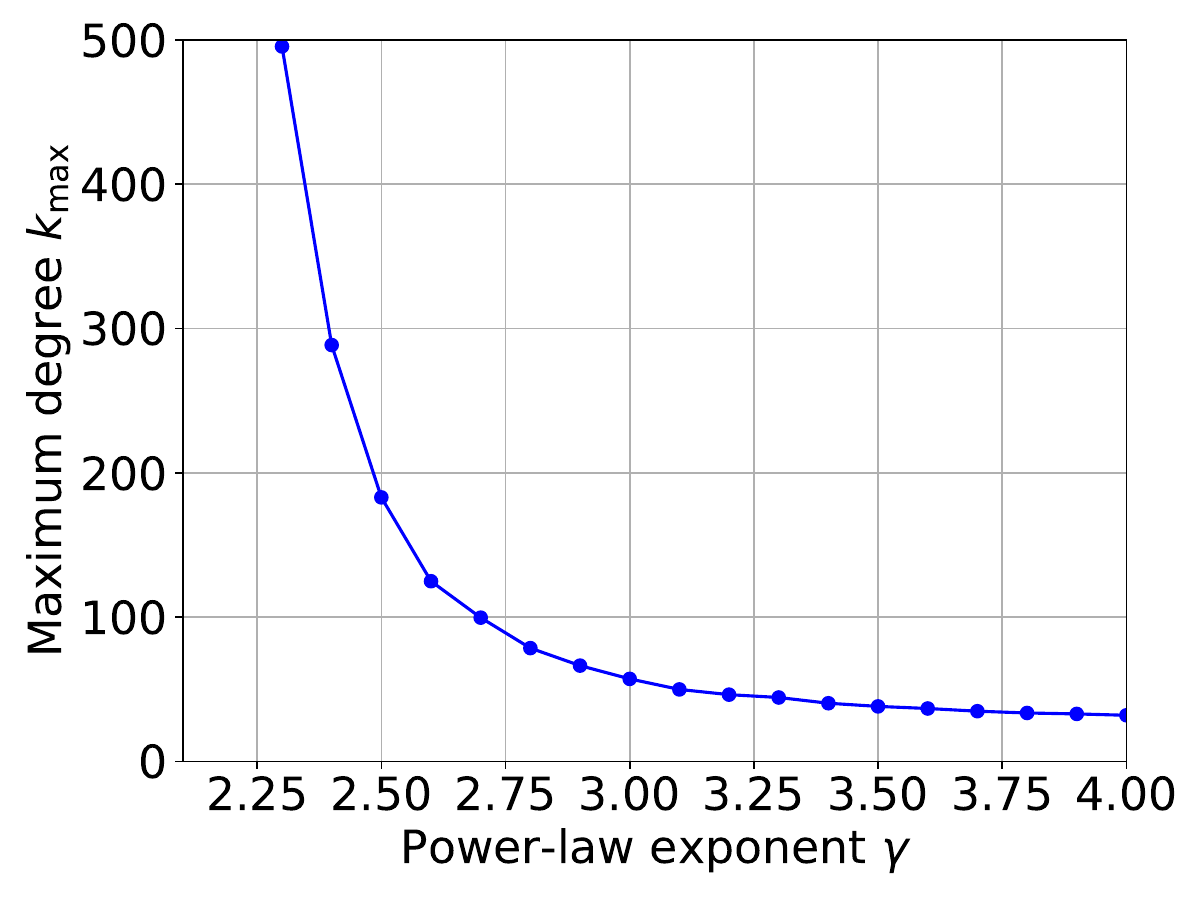}
    \caption{}
  \end{subfigure}
  \hfill
  \begin{subfigure}[b]{0.42\textwidth}
    \includegraphics[width=\textwidth]{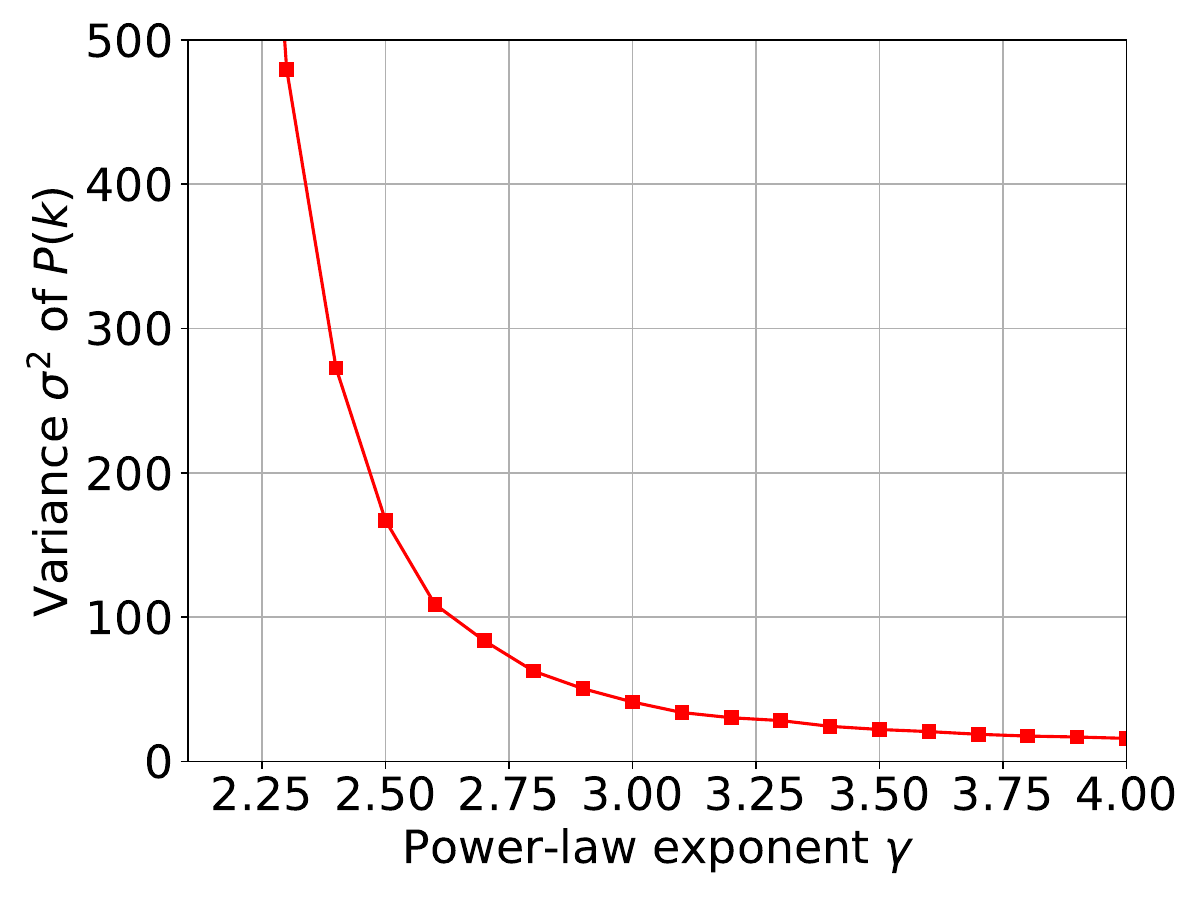}
    \caption{}
  \end{subfigure}
    \caption{Comparison with Figure \ref{fig:kmax_variance_gamma} in the case of $N=10^3$ and $m=2$. Monotone decreasing of (a) the maximum degree $k_{max}$ and (b) the variance $\sigma^2$ of degree distribution $P(k)$ with the power-law exponent $\gamma$ for \bm{$N=10^4$} and \bm{$m=2$}.}
    \label{fig:kmax_variance_gamma_10000_m2}
\end{figure}

\begin{figure}[H]
  \centering
  % 第一行
  \begin{subfigure}[b]{0.42\textwidth}
    \includegraphics[width=\textwidth]{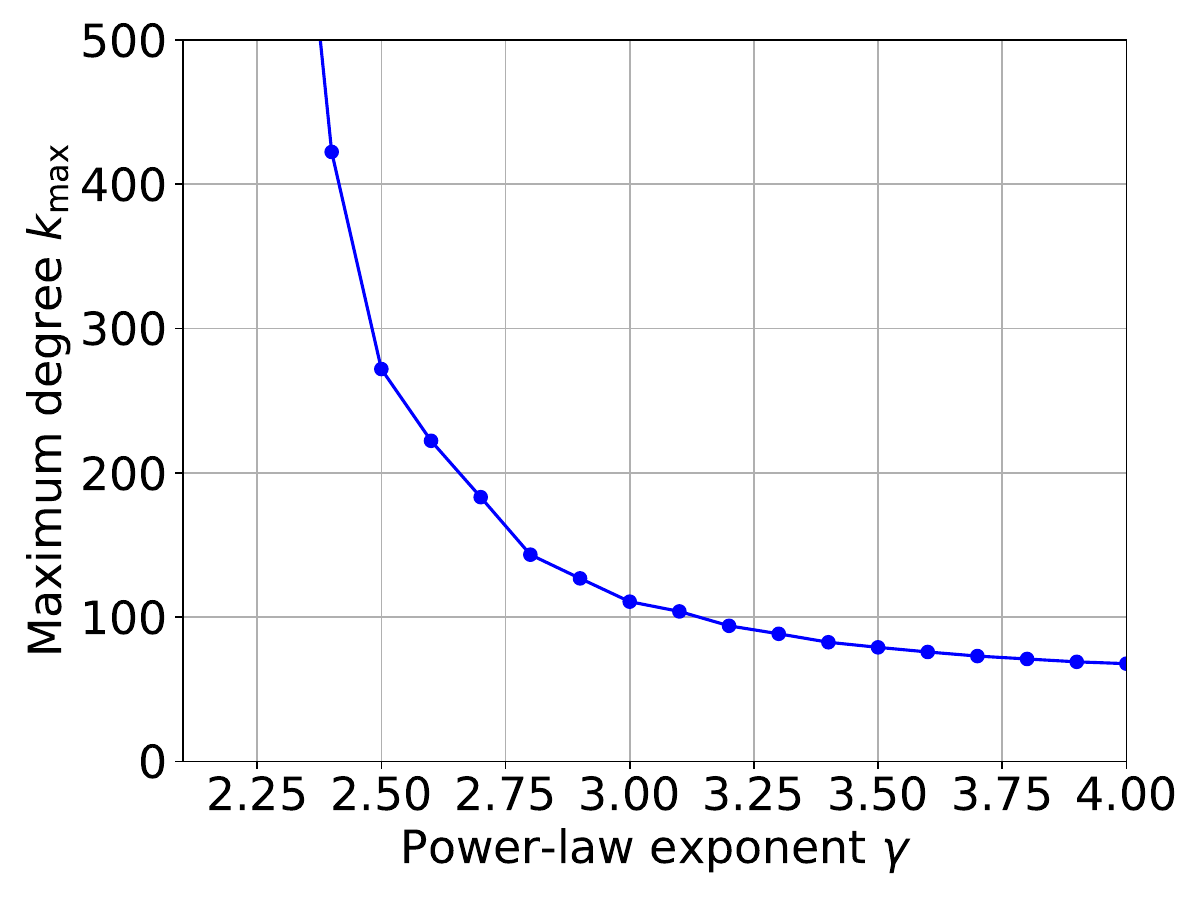}
    \caption{}
  \end{subfigure}
  \hfill
  \begin{subfigure}[b]{0.42\textwidth}
    \includegraphics[width=\textwidth]{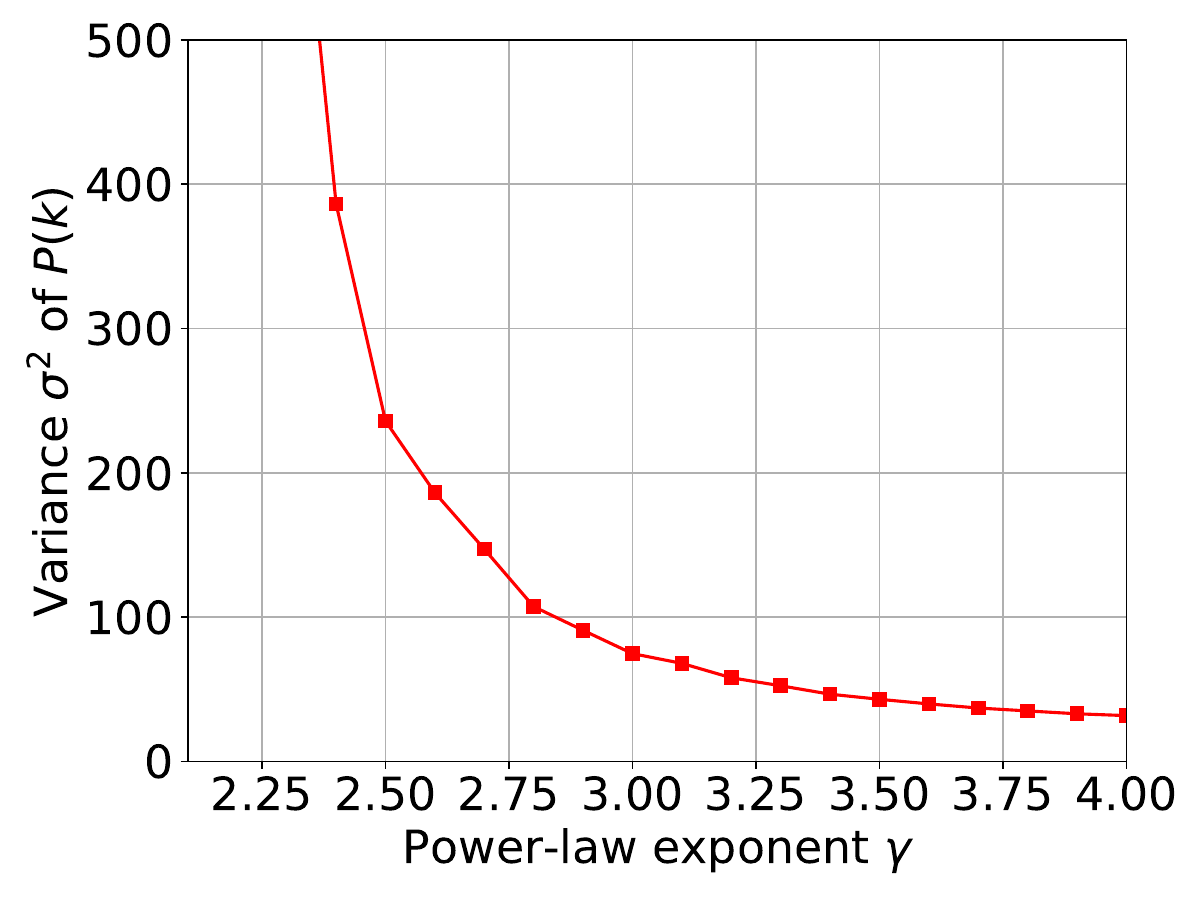}
    \caption{}
  \end{subfigure}
    \caption{Comparison with Figure \ref{fig:kmax_variance_gamma} in the case of $N=10^3$ and $m=2$. Monotone decreasing of (a) the maximum degree $k_{max}$ and (b) the variance $\sigma^2$ of degree distribution $P(k)$ with the power-law exponent $\gamma$ for \bm{$N=10^4$} and \bm{$m=3$}.}
    \label{fig:kmax_variance_gamma_10000_m3}
\end{figure}

\begin{figure}[H]
  \centering
  % 第一行
  \begin{subfigure}[b]{0.42\textwidth}
    \includegraphics[width=\textwidth]{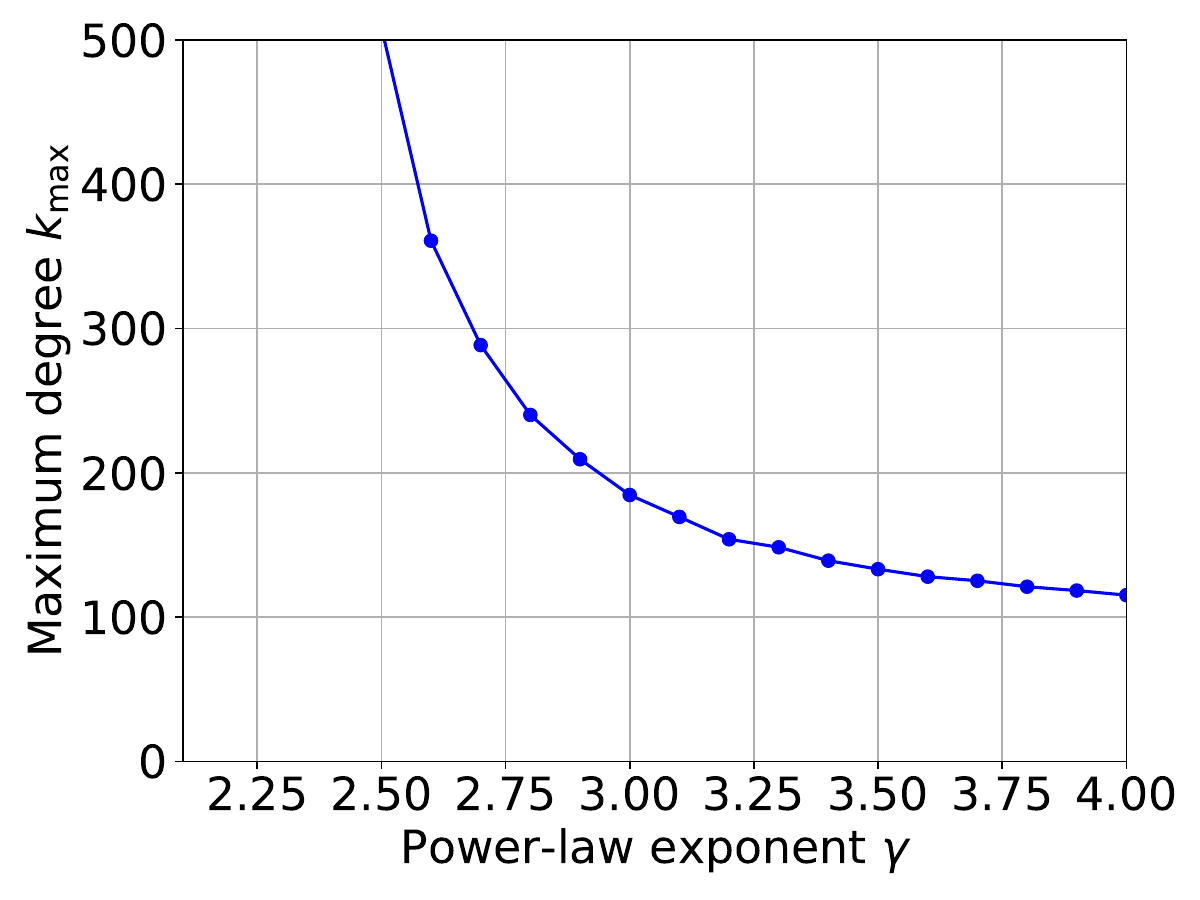}
    \caption{}
  \end{subfigure}
  \hfill
  \begin{subfigure}[b]{0.42\textwidth}
    \includegraphics[width=\textwidth]{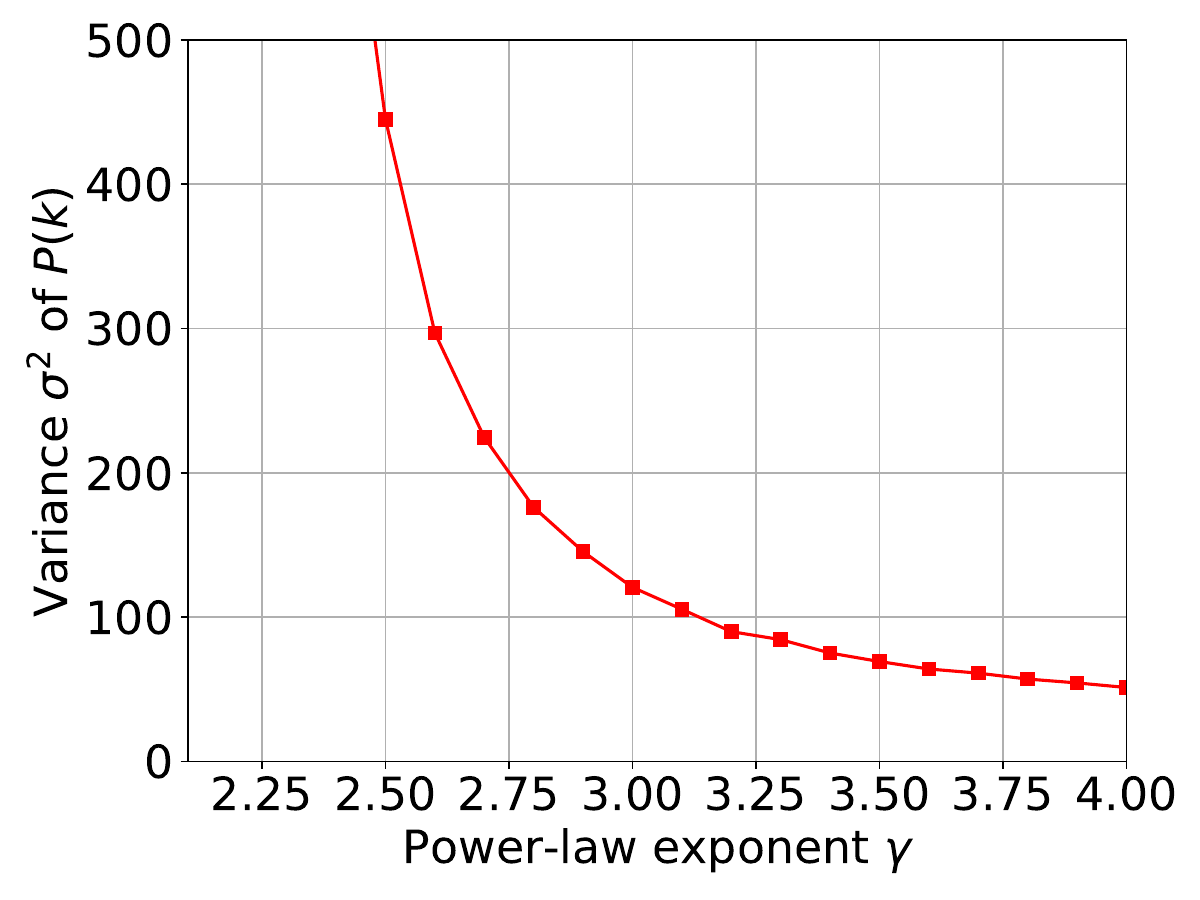}
    \caption{}
  \end{subfigure}
    \caption{Comparison with Figure \ref{fig:kmax_variance_gamma} in the case of $N=10^3$ and $m=2$. Monotone decreasing of (a) the maximum degree $k_{max}$ and (b) the variance $\sigma^2$ of degree distribution $P(k)$ with the power-law exponent $\gamma$ for \bm{$N=10^4$} and \bm{$m=4$}.}
    \label{fig:kmax_variance_gamma_10000_m4}
\end{figure}

\begin{figure}
\centering   
\hfill
\begin{subfigure}[b]{0.42\textwidth}
\includegraphics[width=\textwidth]{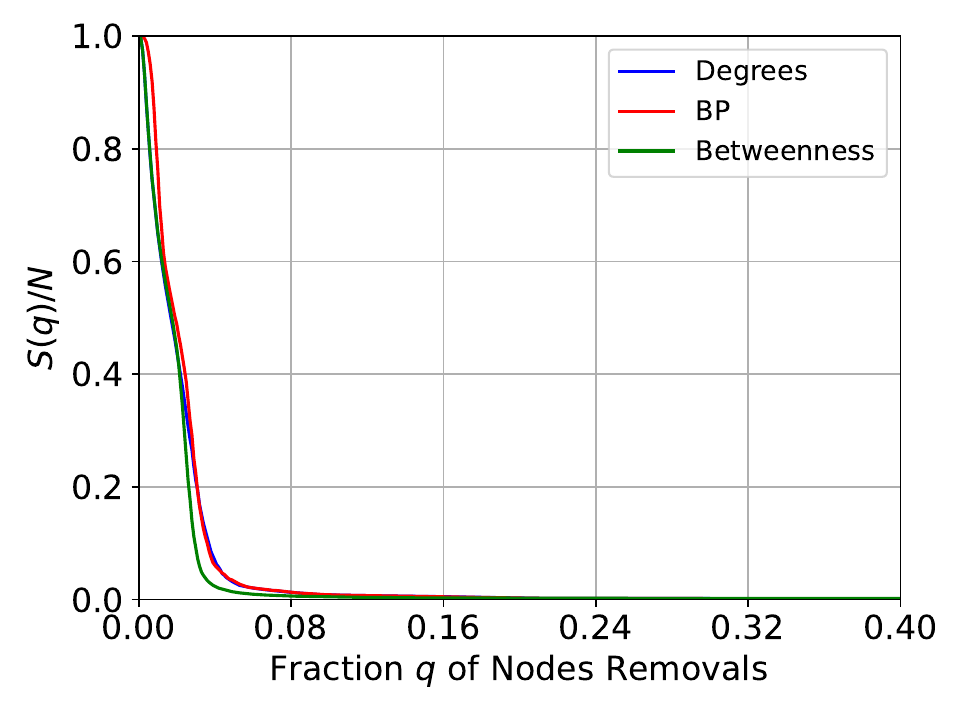}
\caption{$\gamma=2.1$}
\end{subfigure}
\hfill
\begin{subfigure}[b]{0.42\textwidth}
\includegraphics[width=\textwidth]{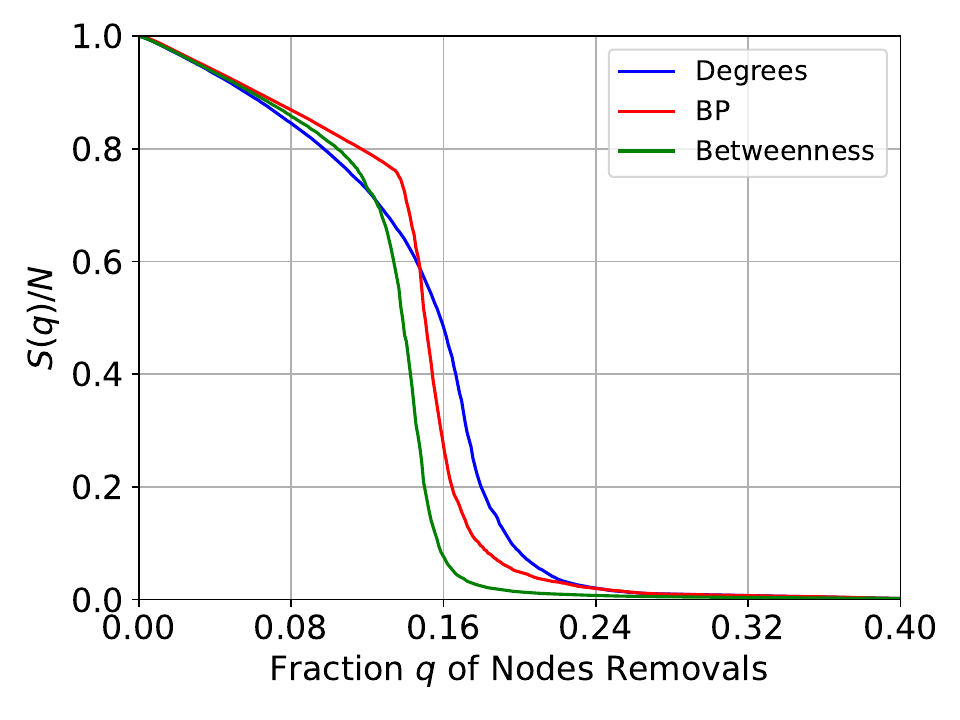}
\caption{$\gamma=2.5$}
\end{subfigure}

\vspace{1em}
\hfill
\begin{subfigure}[b]{0.42\textwidth}
\includegraphics[width=\textwidth]{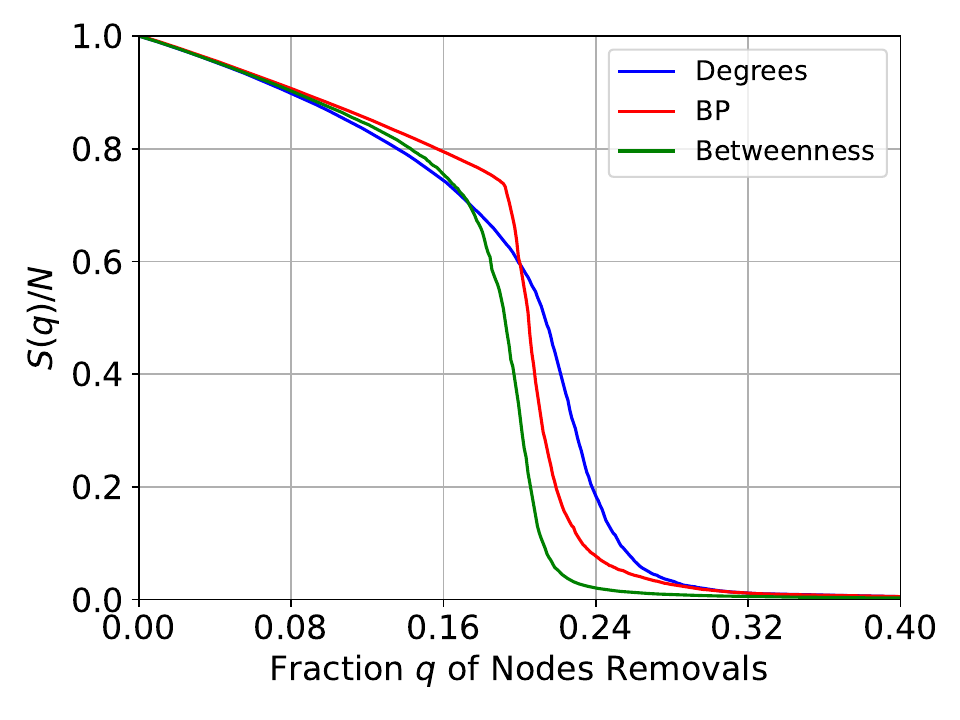}
\caption{$\gamma=3.0$}
\end{subfigure}
\hfill
\begin{subfigure}[b]{0.42\textwidth}
\includegraphics[width=\textwidth]{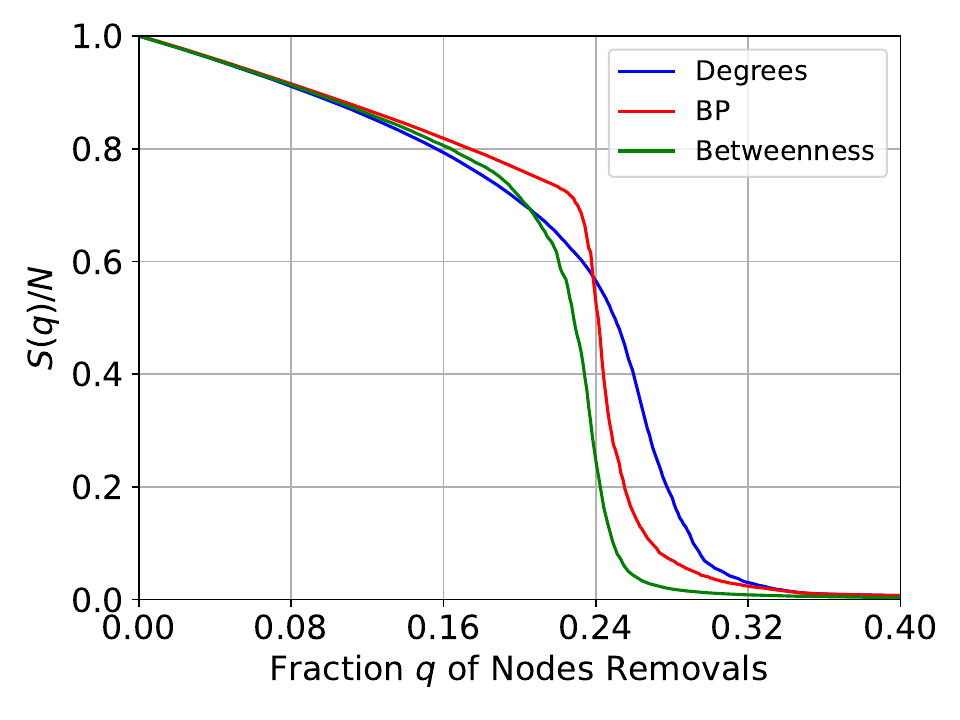}
\caption{$\gamma=4.0$}
\end{subfigure}
\caption{Comparison with Figure \ref{fig:r_by_attacks} in the case of $N=10^3$ and $m=2$. The relative size $S(q)/N$ of the largest connected component (LCC) against different attacks in randomized SF networks with the power-law exponents (a) $\gamma = 2.1$, (b) $\gamma = 2.5$, (c) $\gamma = 3.0$, and (d) $\gamma = 4.0$ for \bm{$N=10^3$} and \bm{$m=3$}. Blue, red, and green curves correspond to recalculated degrees, betweenness centralities, and BP attacks, respectively. In comparing the areas under curves, red (BP attacks) and green (betweenness centralities) curves show more destructive with smaller areas than blue curves (degrees attacks).}
\label{fig:r_by_attacks_1000_m3}
\end{figure}

\begin{figure}
\centering   
\hfill
\begin{subfigure}[b]{0.42\textwidth}
\includegraphics[width=\textwidth]{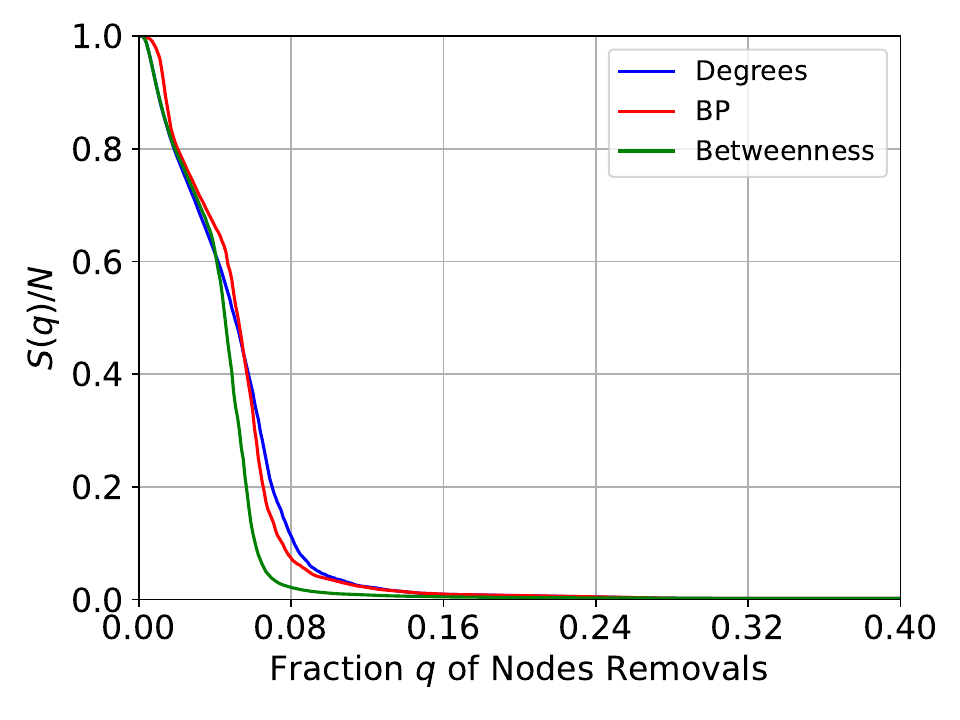}
\caption{$\gamma=2.1$}
\end{subfigure}
\hfill
\begin{subfigure}[b]{0.42\textwidth}
\includegraphics[width=\textwidth]{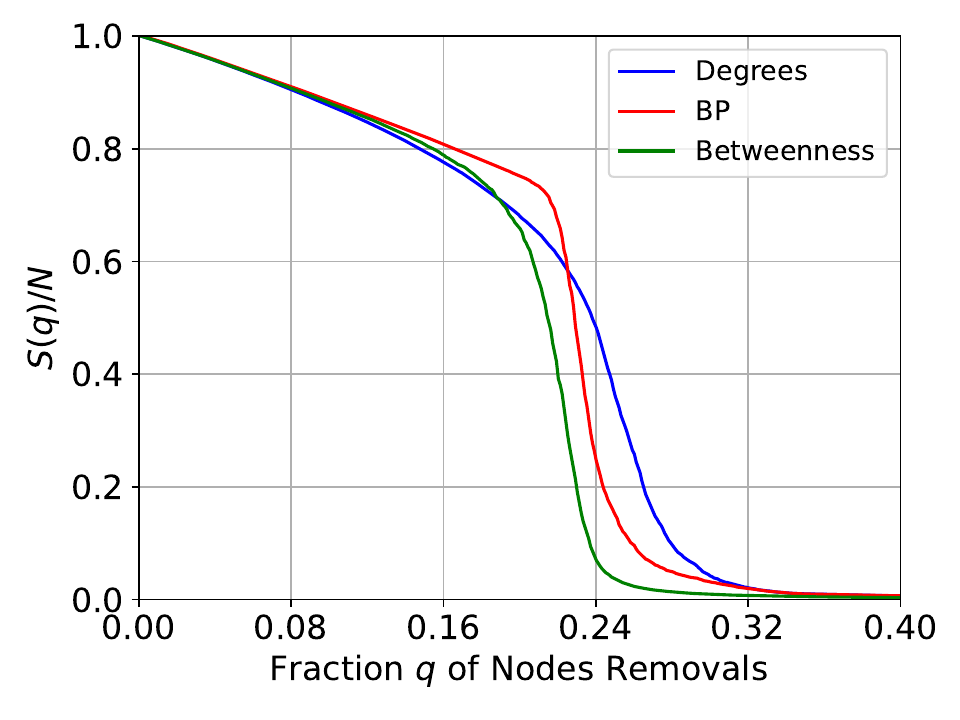}
\caption{$\gamma=2.5$}
\end{subfigure}

\vspace{1em}
\hfill
\begin{subfigure}[b]{0.42\textwidth}
\includegraphics[width=\textwidth]{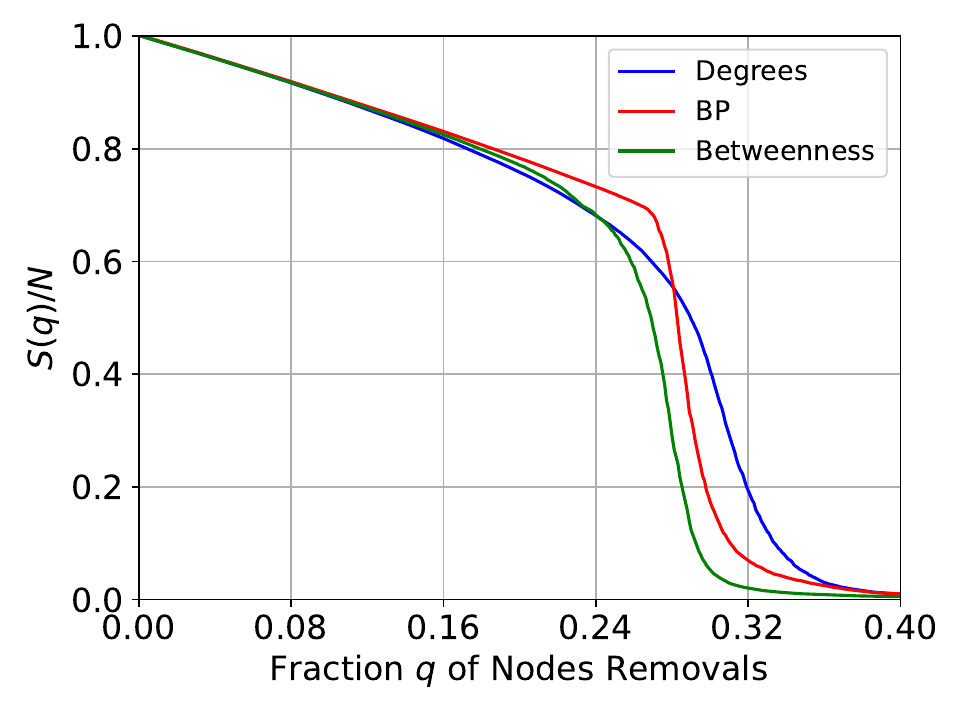}
\caption{$\gamma=3.0$}
\end{subfigure}
\hfill
\begin{subfigure}[b]{0.42\textwidth}
\includegraphics[width=\textwidth]{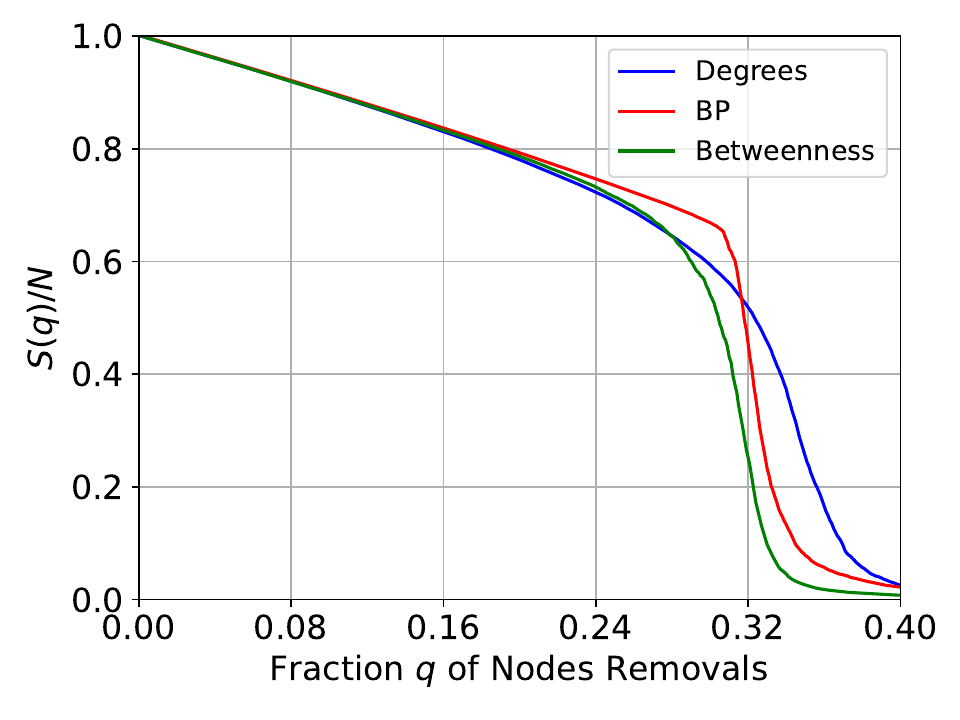}
\caption{$\gamma=4.0$}
\end{subfigure}
\caption{Comparison with Figure \ref{fig:r_by_attacks} in the case of $N=10^3$ and $m=2$. The relative size $S(q)/N$ of the largest connected component (LCC) against different attacks in randomized SF networks with the power-law exponents (a) $\gamma = 2.1$, (b) $\gamma = 2.5$, (c) $\gamma = 3.0$, and (d) $\gamma = 4.0$ for \bm{$N=10^3$} and \bm{$m=4$}. Blue, red, and green curves correspond to recalculated degrees, betweenness centralities, and BP attacks, respectively. In comparing the areas under curves, red (BP attacks) and green (betweenness centralities) curves show more destructive with smaller areas than blue curves (degrees attacks).}
\label{fig:r_by_attacks_1000_m4}
\end{figure}

\begin{figure}
\centering   
\hfill
\begin{subfigure}[b]{0.42\textwidth}
\includegraphics[width=\textwidth]{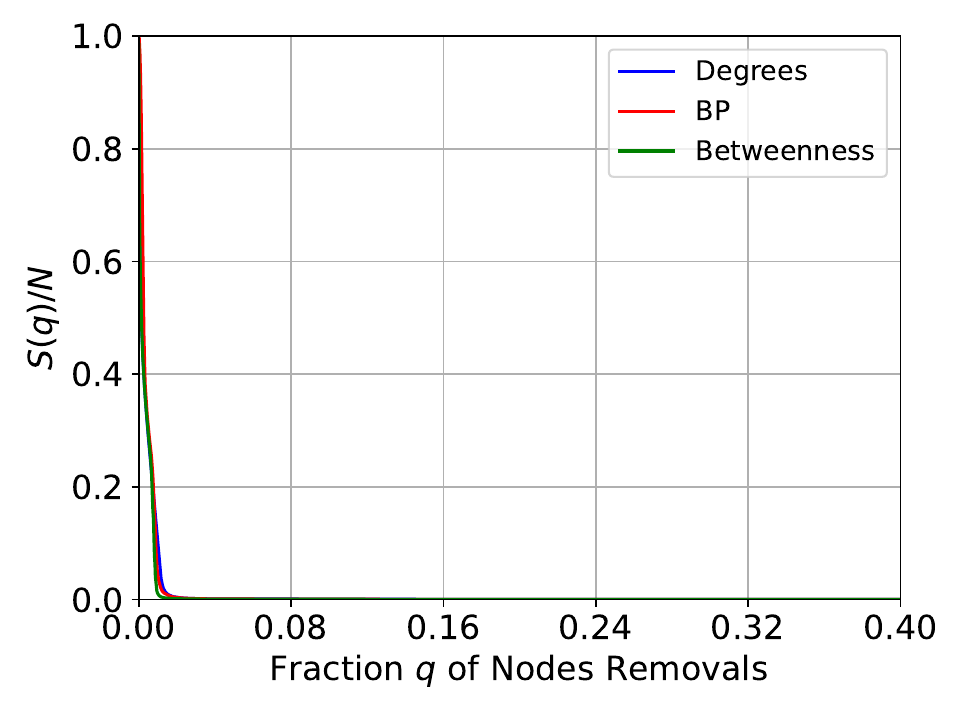}
\caption{$\gamma=2.1$}
\end{subfigure}
\hfill
\begin{subfigure}[b]{0.42\textwidth}
\includegraphics[width=\textwidth]{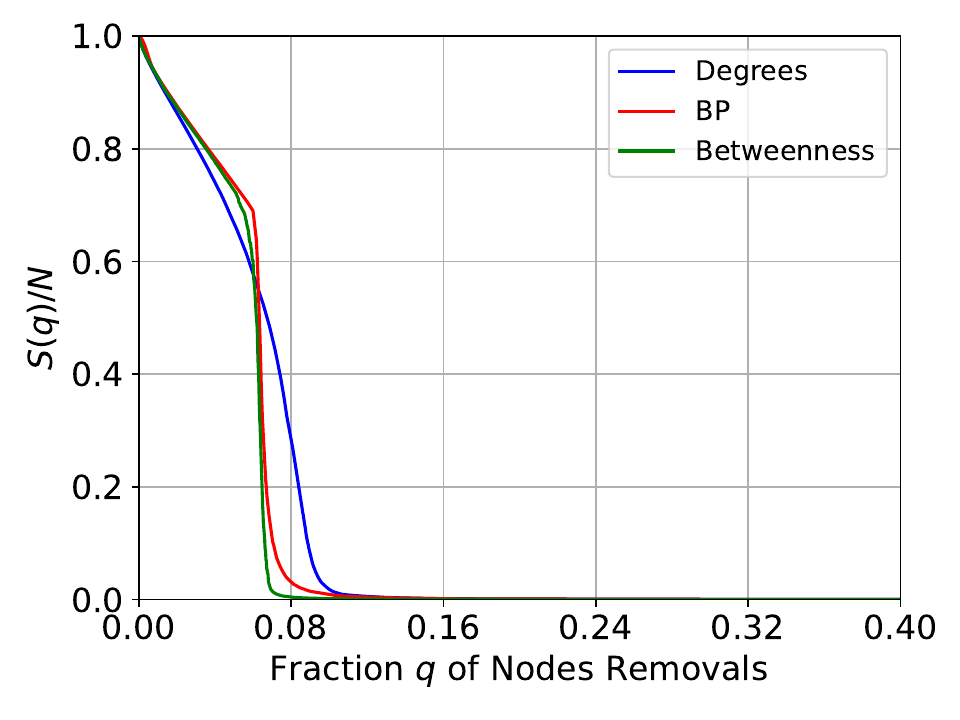}
\caption{$\gamma=2.5$}
\end{subfigure}

\vspace{1em}
\hfill
\begin{subfigure}[b]{0.42\textwidth}
\includegraphics[width=\textwidth]{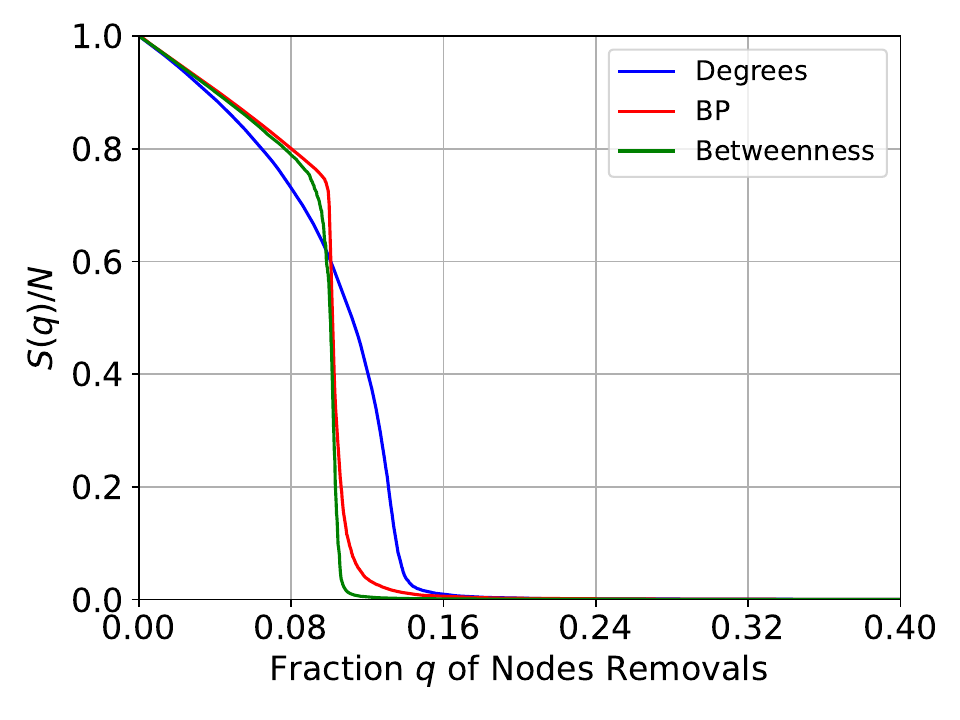}
\caption{$\gamma=3.0$}
\end{subfigure}
\hfill
\begin{subfigure}[b]{0.42\textwidth}
\includegraphics[width=\textwidth]{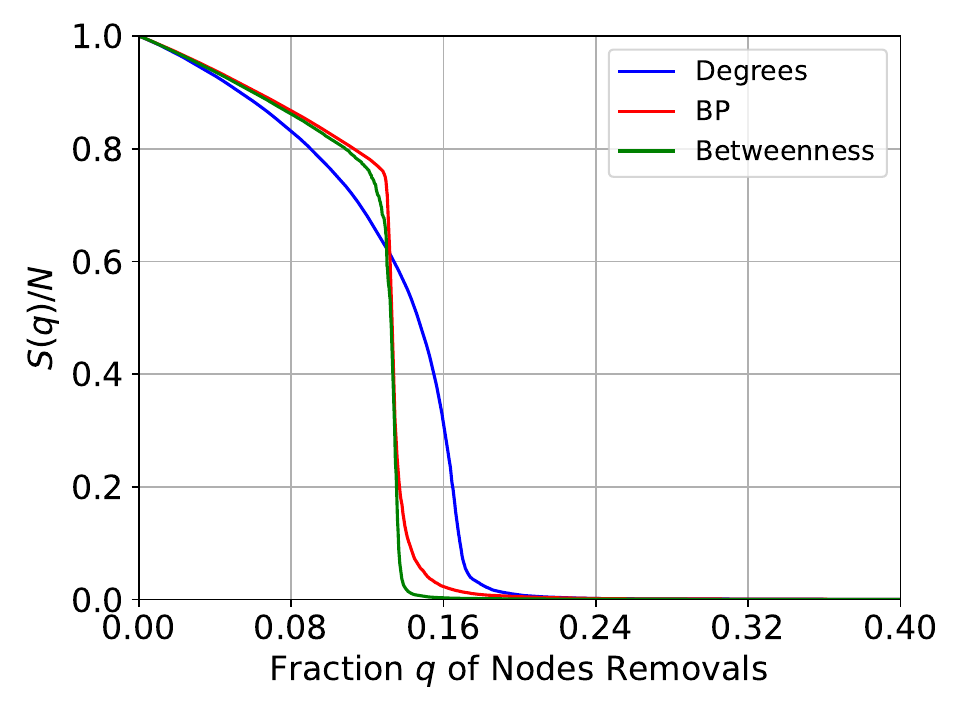}
\caption{$\gamma=4.0$}
\end{subfigure}
\caption{Comparison with Figure \ref{fig:r_by_attacks} in the case of $N=10^3$ and $m=2$. The relative size $S(q)/N$ of the largest connected component (LCC) against different attacks in randomized SF networks with the power-law exponents (a) $\gamma = 2.1$, (b) $\gamma = 2.5$, (c) $\gamma = 3.0$, and (d) $\gamma = 4.0$ for \bm{$N=10^4$} and \bm{$m=2$}. Blue, red, and green curves correspond to recalculated degrees, betweenness centralities, and BP attacks, respectively. In comparing the areas under curves, red (BP attacks) and green (betweenness centralities) curves show more destructive with smaller areas than blue curves (degrees attacks).}
\label{fig:r_by_attacks_10000_m2}
\end{figure}

\begin{figure}
\centering   
\hfill
\begin{subfigure}[b]{0.42\textwidth}
\includegraphics[width=\textwidth]{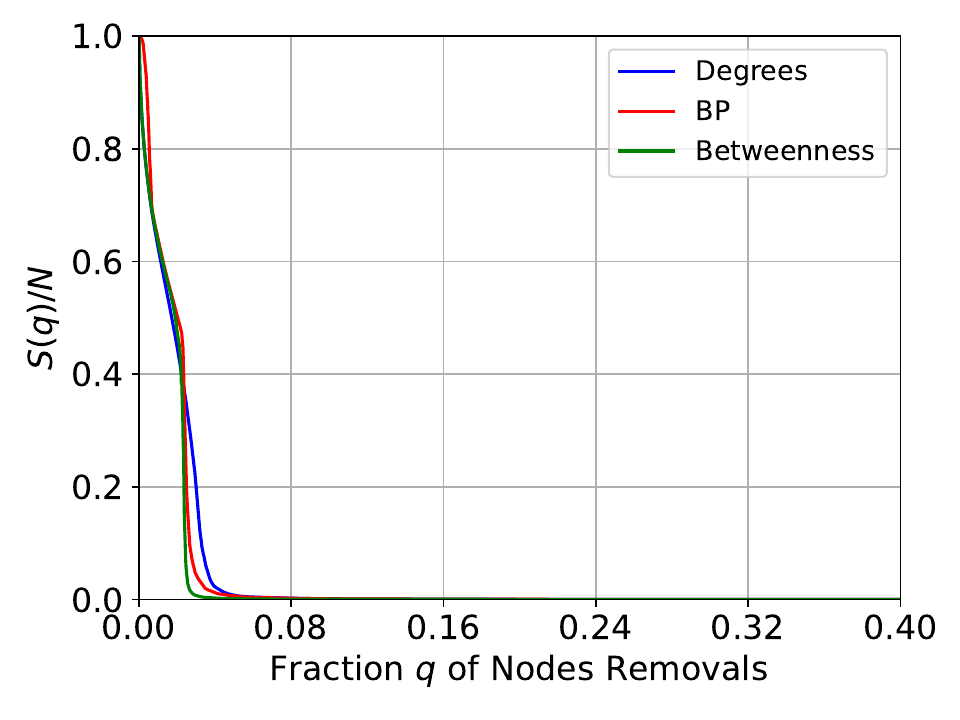}
\caption{$\gamma=2.1$}
\end{subfigure}
\hfill
\begin{subfigure}[b]{0.42\textwidth}
\includegraphics[width=\textwidth]{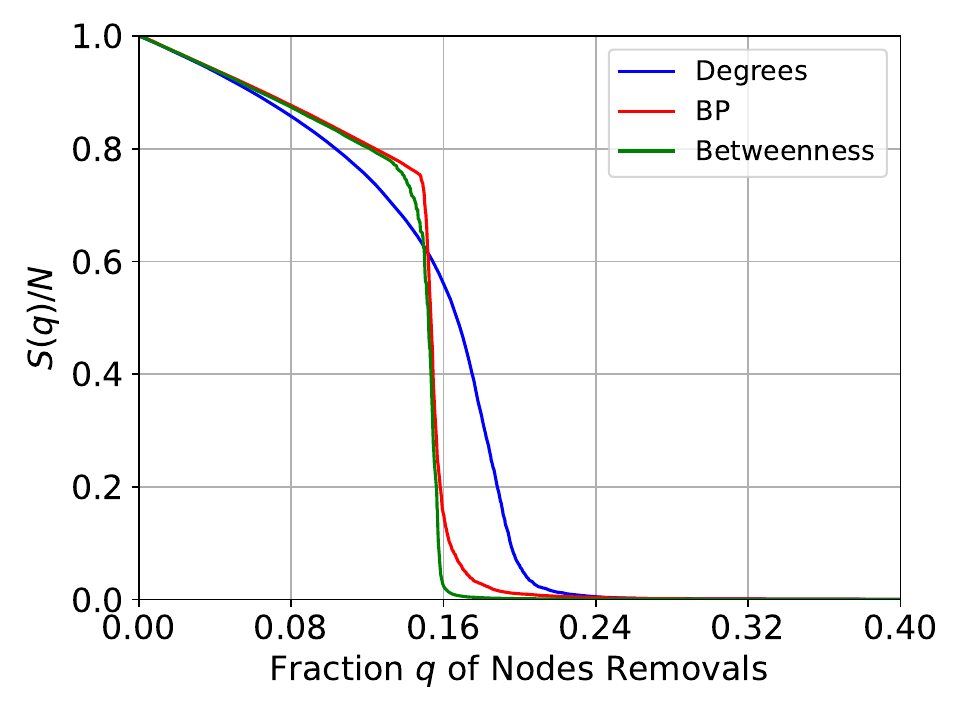}
\caption{$\gamma=2.5$}
\end{subfigure}

\vspace{1em}
\hfill
\begin{subfigure}[b]{0.42\textwidth}
\includegraphics[width=\textwidth]{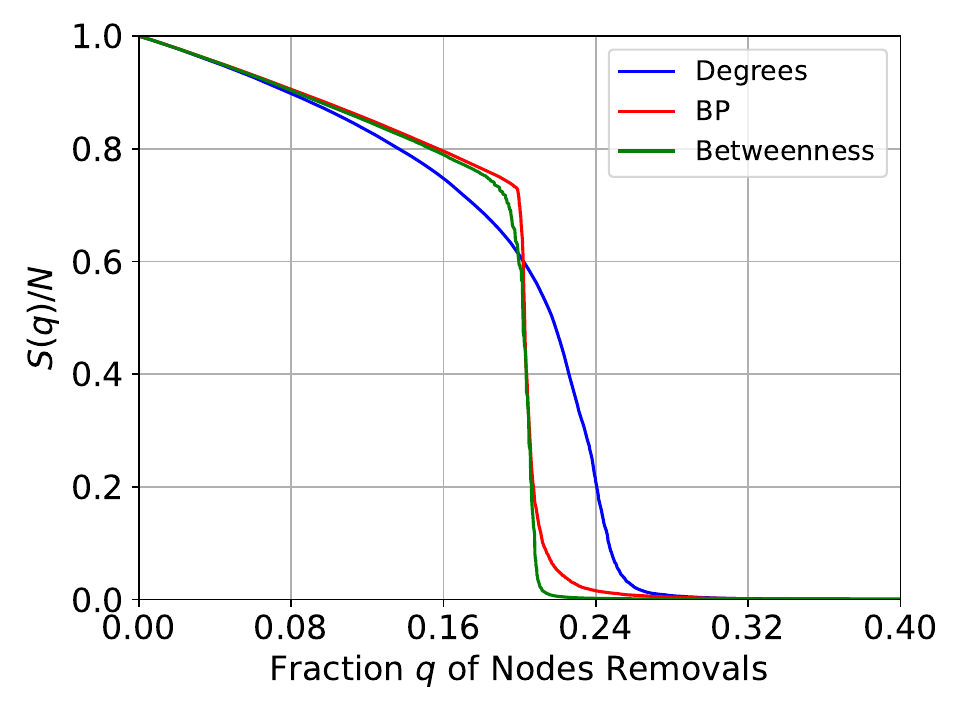}
\caption{$\gamma=3.0$}
\end{subfigure}
\hfill
\begin{subfigure}[b]{0.42\textwidth}
\includegraphics[width=\textwidth]{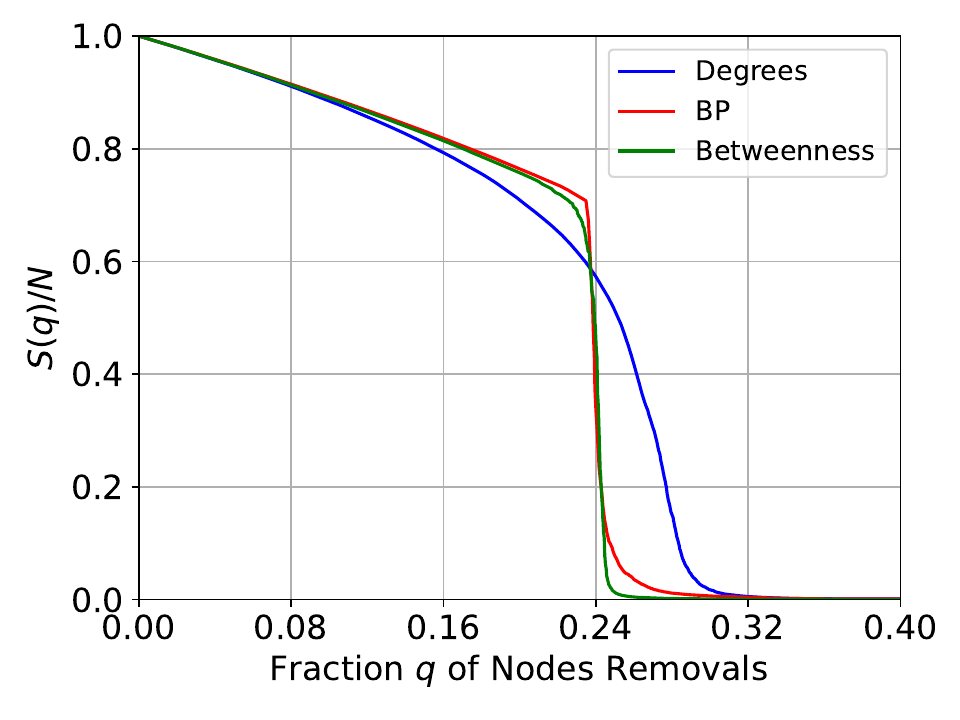}
\caption{$\gamma=4.0$}
\end{subfigure}
\caption{Comparison with Figure \ref{fig:r_by_attacks} in the case of $N=10^3$ and $m=2$. The relative size $S(q)/N$ of the largest connected component (LCC) against different attacks in randomized SF networks with the power-law exponents (a) $\gamma = 2.1$, (b) $\gamma = 2.5$, (c) $\gamma = 3.0$, and (d) $\gamma = 4.0$ for \bm{$N=10^4$} and \bm{$m=3$}. Blue, red, and green curves correspond to recalculated degrees, betweenness centralities, and BP attacks, respectively. In comparing the areas under curves, red (BP attacks) and green (betweenness centralities) curves show more destructive with smaller areas than blue curves (degrees attacks).}
\label{fig:r_by_attacks_10000_m3}
\end{figure}

\begin{figure}
\centering   
\hfill
\begin{subfigure}[b]{0.42\textwidth}
\includegraphics[width=\textwidth]{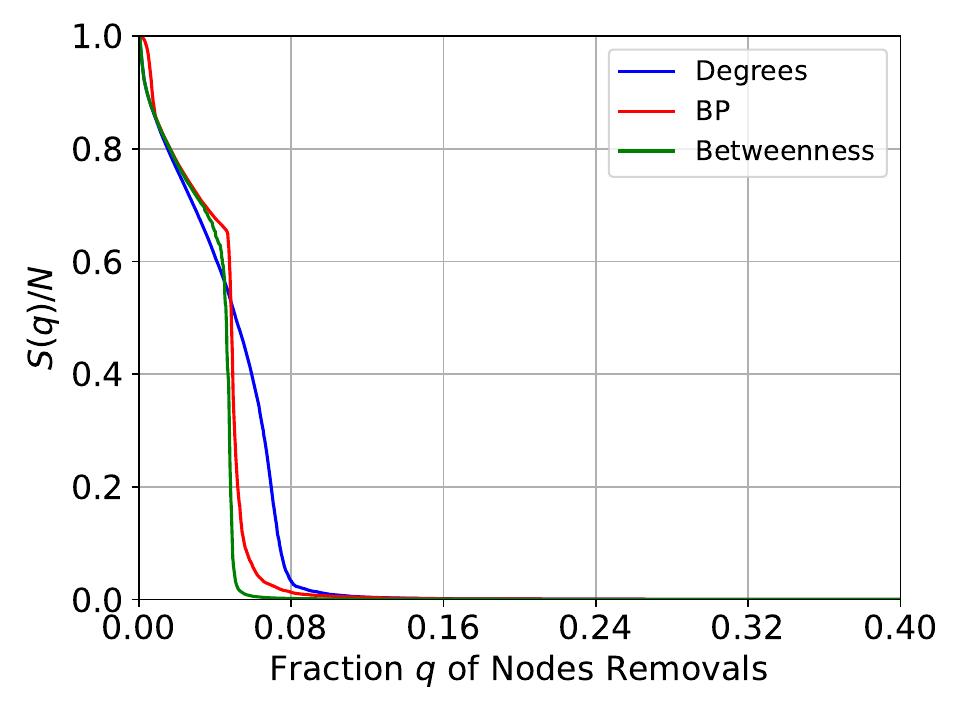}
\caption{$\gamma=2.1$}
\end{subfigure}
\hfill
\begin{subfigure}[b]{0.42\textwidth}
\includegraphics[width=\textwidth]{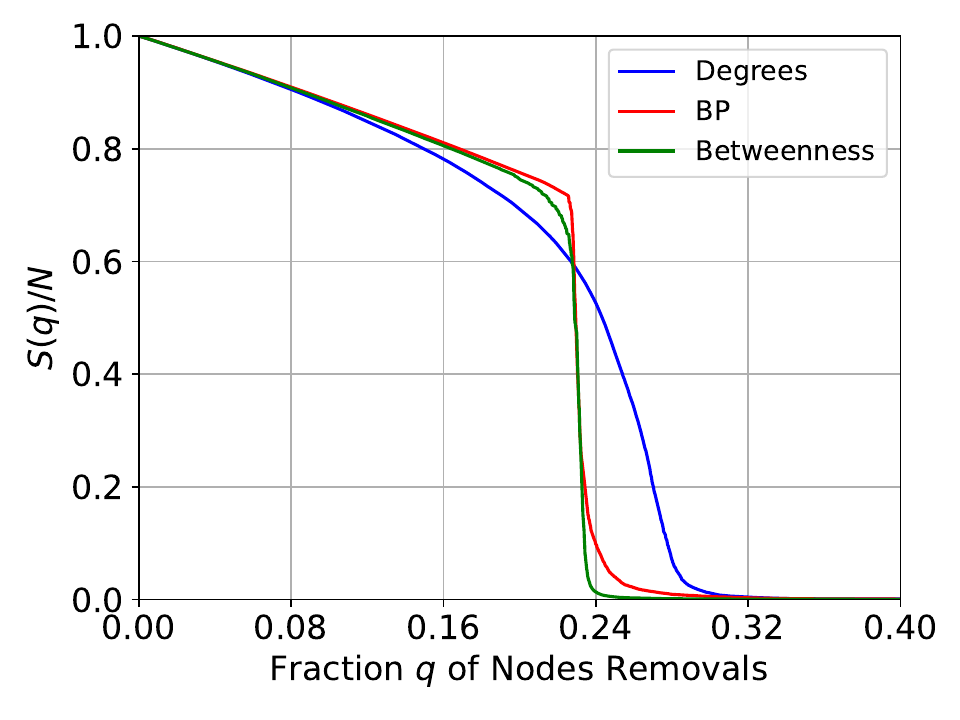}
\caption{$\gamma=2.5$}
\end{subfigure}

\vspace{1em}
\hfill
\begin{subfigure}[b]{0.42\textwidth}
\includegraphics[width=\textwidth]{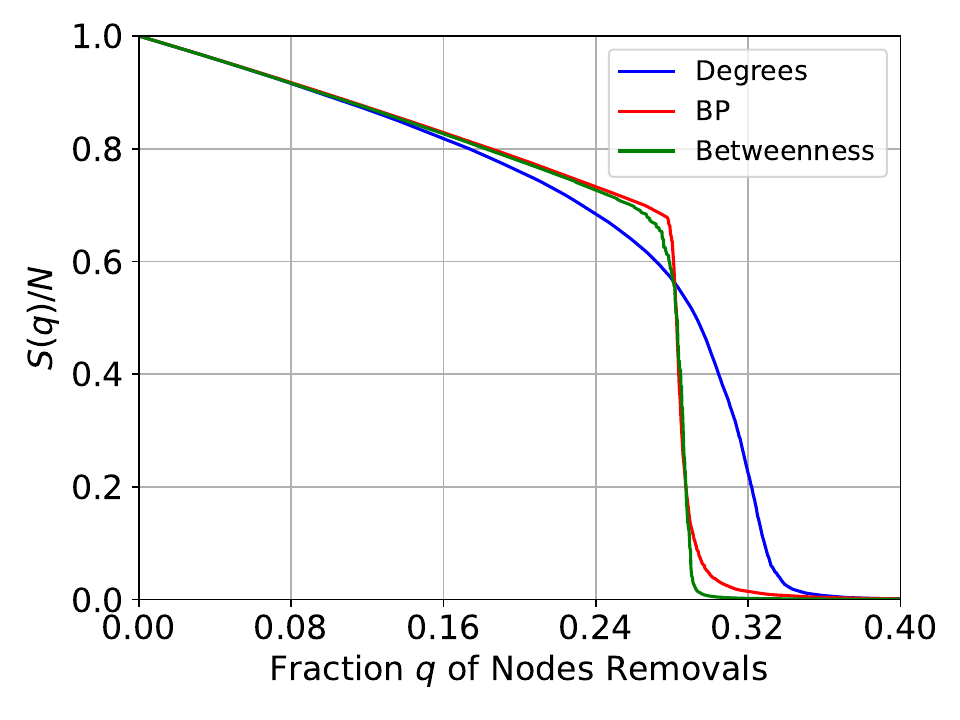}
\caption{$\gamma=3.0$}
\end{subfigure}
\hfill
\begin{subfigure}[b]{0.42\textwidth}
\includegraphics[width=\textwidth]{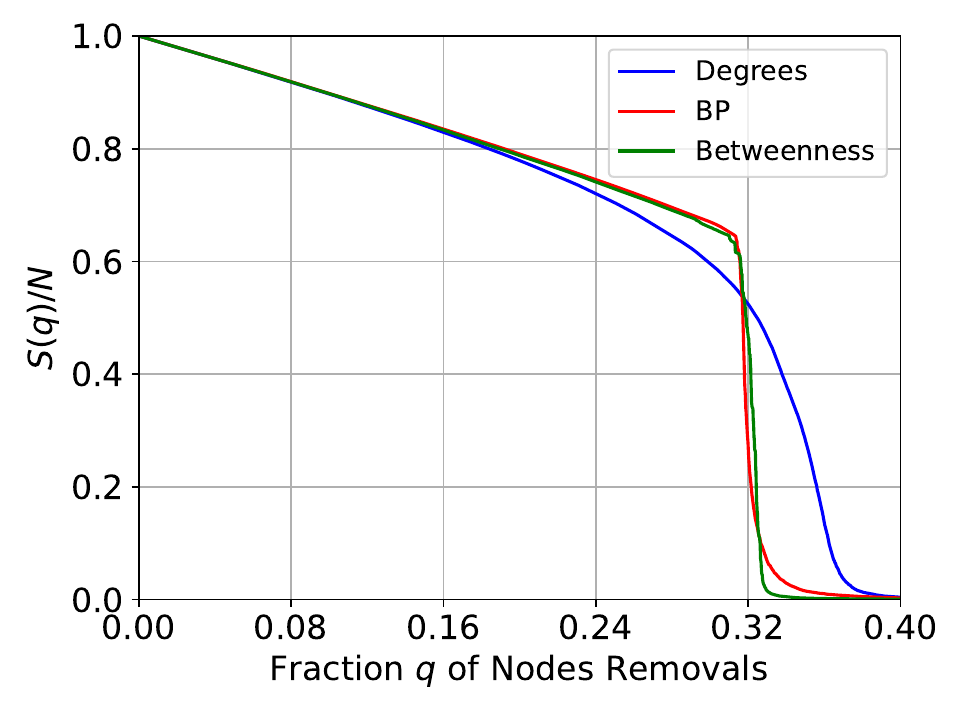}
\caption{$\gamma=4.0$}
\end{subfigure}
\caption{Comparison with Figure \ref{fig:r_by_attacks} in the case of $N=10^3$ and $m=2$. The relative size $S(q)/N$ of the largest connected component (LCC) against different attacks in randomized SF networks with the power-law exponents (a) $\gamma = 2.1$, (b) $\gamma = 2.5$, (c) $\gamma = 3.0$, and (d) $\gamma = 4.0$ for \bm{$N=10^4$} and \bm{$m=4$}. Blue, red, and green curves correspond to recalculated degrees, betweenness centralities, and BP attacks, respectively. In comparing the areas under curves, red (BP attacks) and green (betweenness centralities) curves show more destructive with smaller areas than blue curves (degrees attacks).}
\label{fig:r_by_attacks_10000_m4}
\end{figure}

\begin{figure}[H]
\centering
% 第一行
\hfill
\begin{subfigure}[b]{0.42\textwidth}
\includegraphics[width=\textwidth]{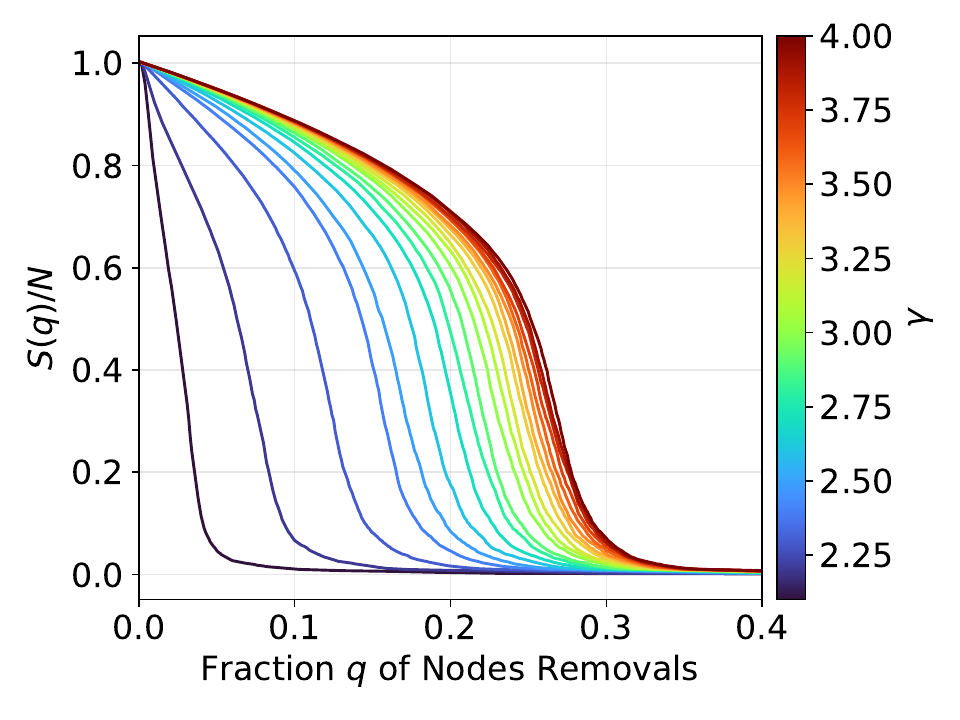}
\caption{Degrees attacks}
\end{subfigure}
\hfill
  \begin{subfigure}[b]{0.42\textwidth}
    \includegraphics[width=\textwidth]{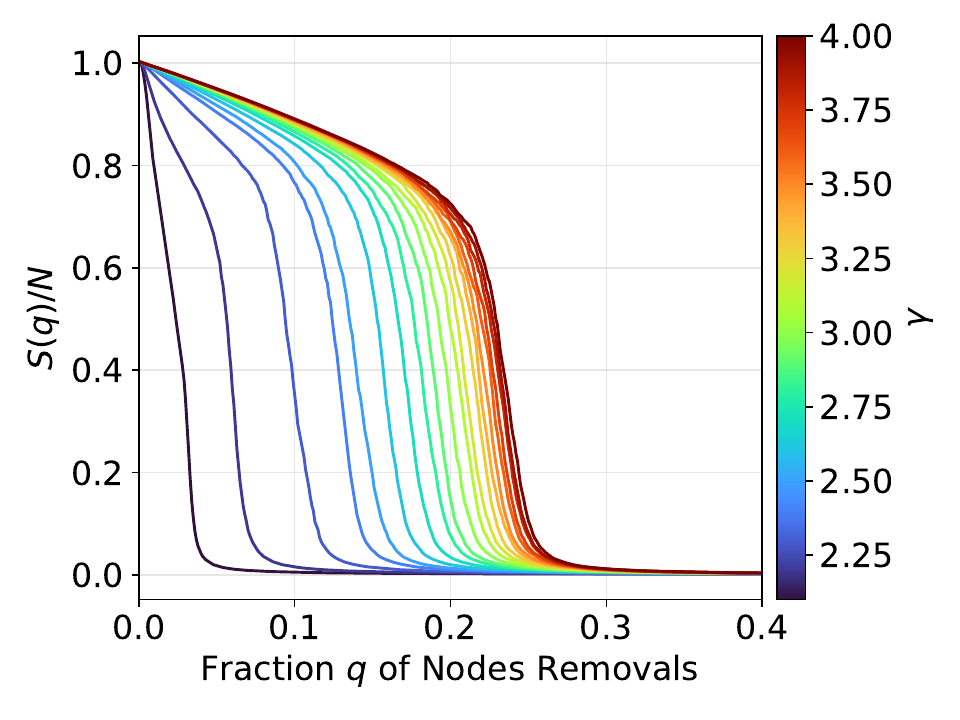}
    \caption{Betweenness centralities attacks}
  \end{subfigure}

\vspace{1em}
\hfill
\begin{subfigure}[b]{0.42\textwidth}
\includegraphics[width=\textwidth]{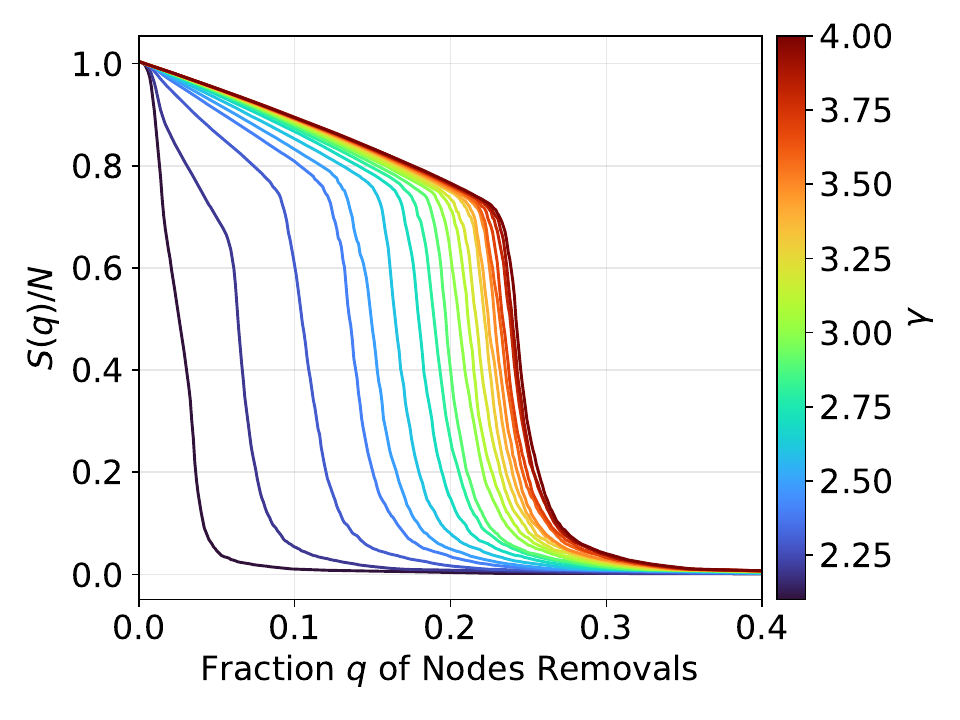}
\caption{Belief propagation attacks}
\end{subfigure}
\caption{Comparison with Figure \ref{fig:r_by_gamma} in the case of $N=10^3$ and $m=2$. More detailed results for the robustness against recalculated (a) degrees, (b) betweenness centralities, and (c) belief propagation (BP) attacks for \bm{$N=10^3$} and \bm{$m=3$}. The areas under colored curves represent the robustness index $R$ in SF networks with power-law exponents from $\gamma = 2.1$ (dark purple) to $\gamma = 4.0$ (red). As $\gamma$ increases, the areas under curves become larger from dark purple to red lines.}
\label{fig:r_gamma_1000_m3}
\end{figure}

\begin{figure}[H]
\centering
% 第一行
\hfill
\begin{subfigure}[b]{0.42\textwidth}
\includegraphics[width=\textwidth]{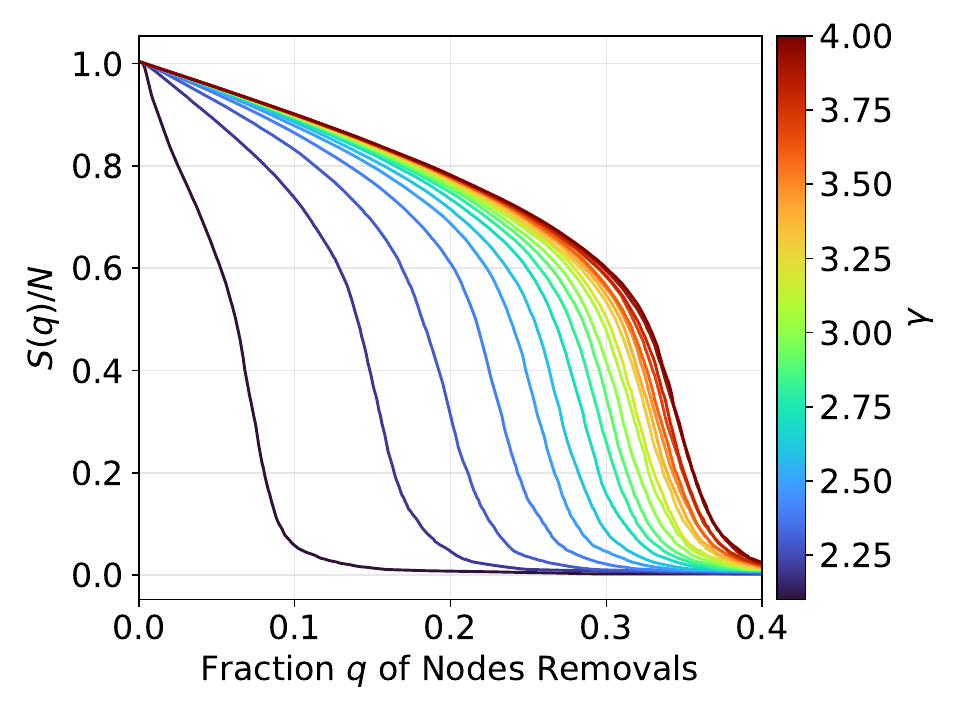}
\caption{Degrees attacks}
\end{subfigure}
\hfill
  \begin{subfigure}[b]{0.42\textwidth}
    \includegraphics[width=\textwidth]{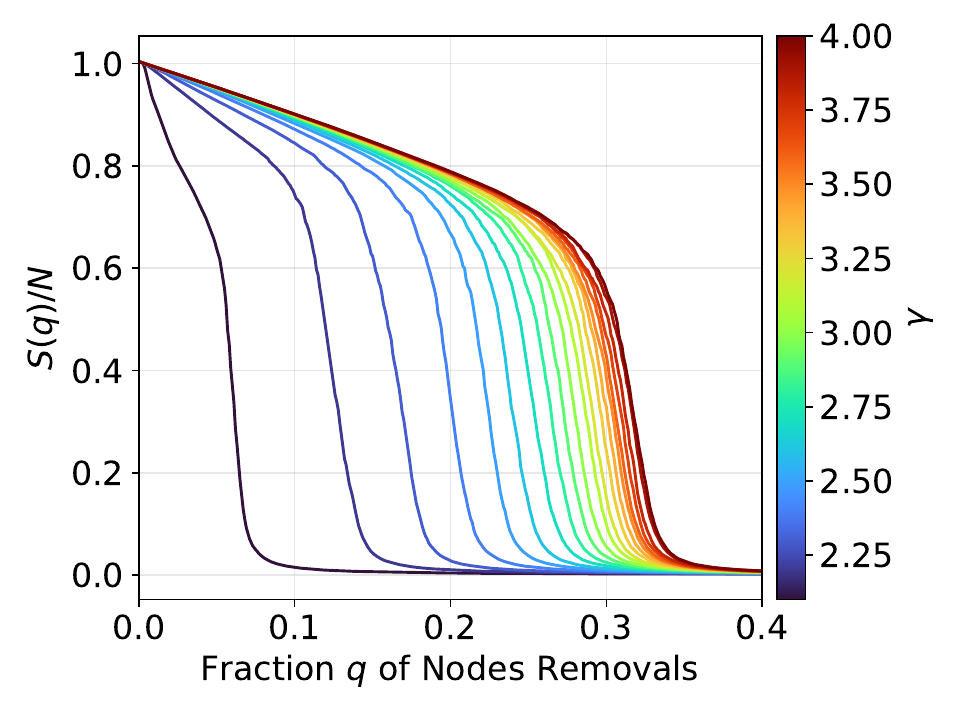}
    \caption{Betweenness centralities attacks}
  \end{subfigure}

\vspace{1em}
\hfill
\begin{subfigure}[b]{0.42\textwidth}
\includegraphics[width=\textwidth]{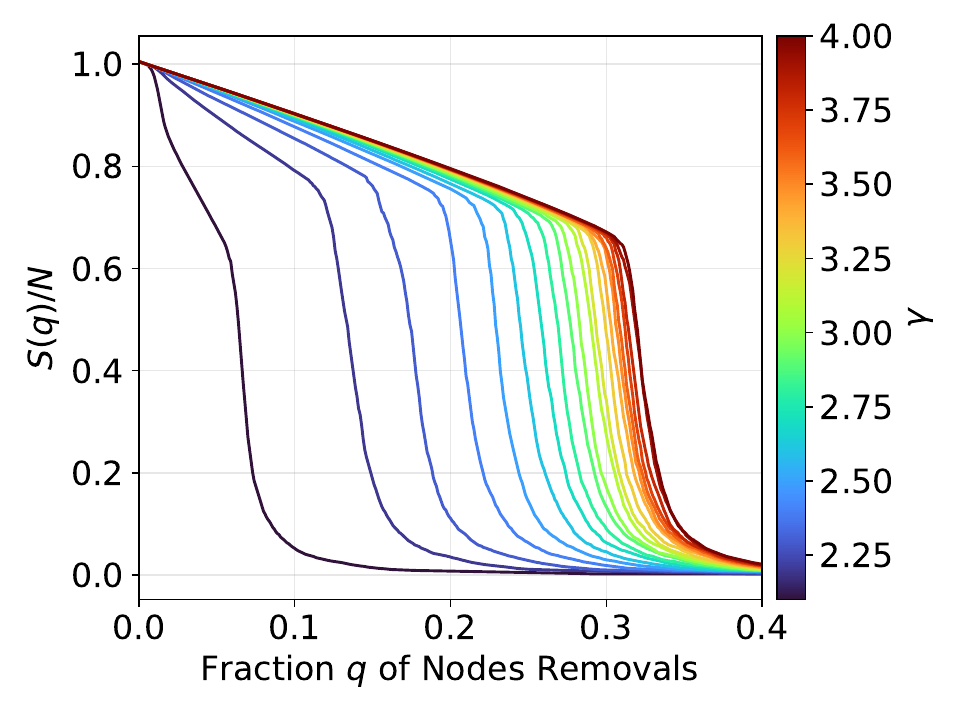}
\caption{Belief propagation attacks}
\end{subfigure}
\caption{Comparison with Figure \ref{fig:r_by_gamma} in the case of $N=10^3$ and $m=2$. More detailed results for the robustness against recalculated (a) degrees, (b) betweenness centralities, and (c) belief propagation (BP) attacks for \bm{$N=10^3$} and \bm{$m=4$}. The areas under colored curves represent the robustness index $R$ in SF networks with power-law exponents from $\gamma = 2.1$ (dark purple) to $\gamma = 4.0$ (red). As $\gamma$ increases, the areas under curves become larger from dark purple to red lines.}
\label{fig:r_gamma_1000_m4}
\end{figure}

\begin{figure}[H]
\centering
% 第一行
\hfill
\begin{subfigure}[b]{0.42\textwidth}
\includegraphics[width=\textwidth]{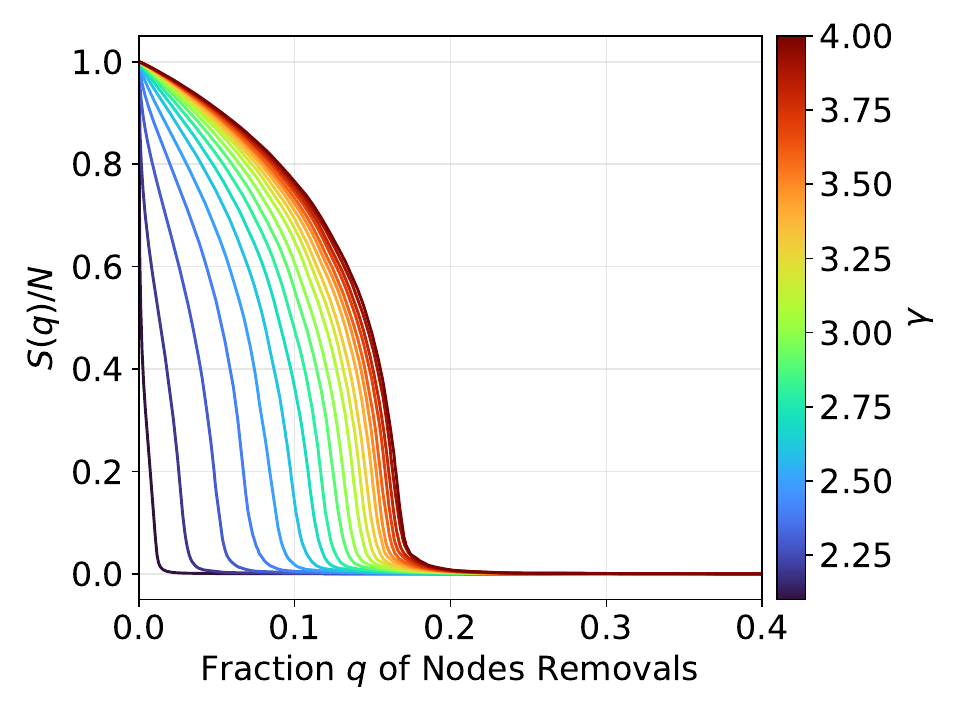}
\caption{Degrees attacks}
\end{subfigure}
\hfill
  \begin{subfigure}[b]{0.42\textwidth}
    \includegraphics[width=\textwidth]{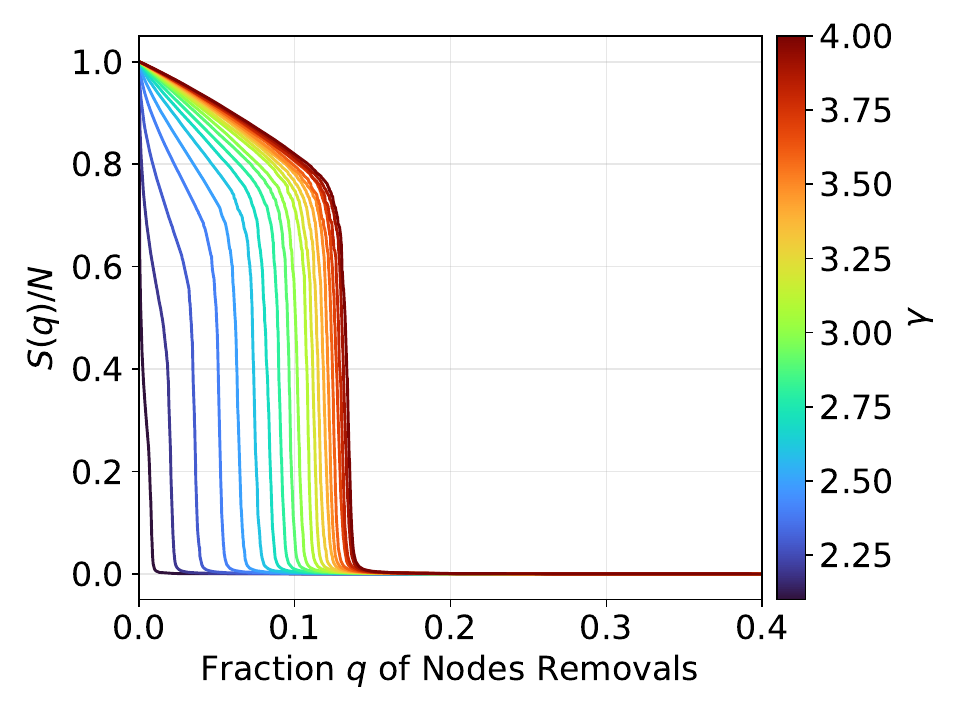}
    \caption{Betweenness centralities attacks}
  \end{subfigure}

\vspace{1em}
\hfill
\begin{subfigure}[b]{0.42\textwidth}
\includegraphics[width=\textwidth]{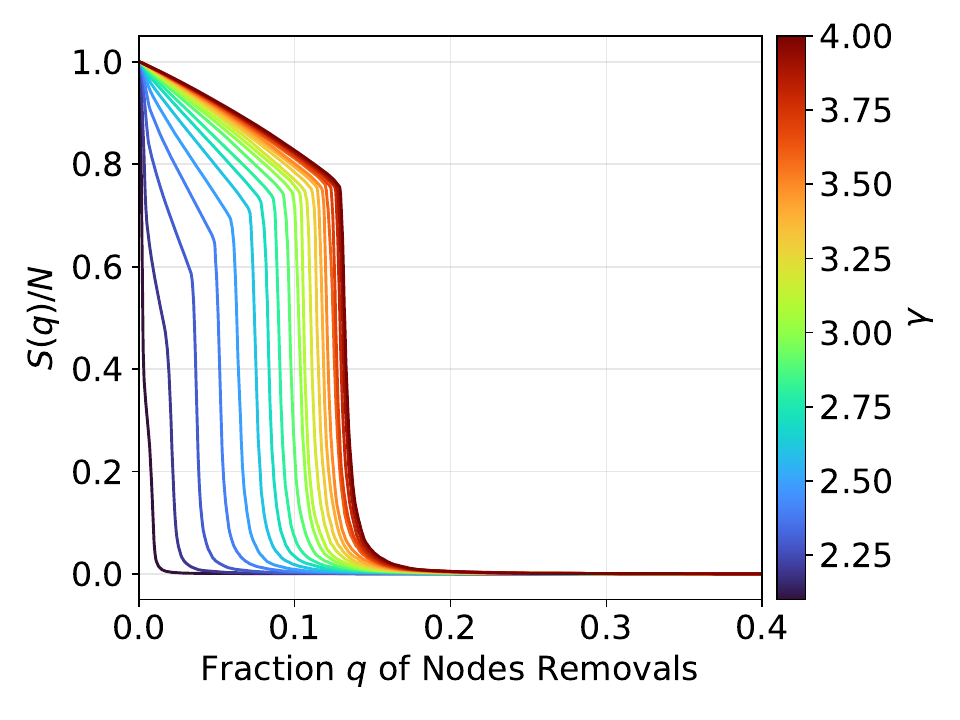}
\caption{Belief propagation attacks}
\end{subfigure}
\caption{Comparison with Figure \ref{fig:r_by_gamma} in the case of $N=10^3$ and $m=2$. More detailed results for the robustness against recalculated (a) degrees, (b) betweenness centralities, and (c) belief propagation (BP) attacks for \bm{$N=10^4$} and \bm{$m=2$}. The areas under colored curves represent the robustness index $R$ in SF networks with power-law exponents from $\gamma = 2.1$ (dark purple) to $\gamma = 4.0$ (red). As $\gamma$ increases, the areas under curves become larger from dark purple to red lines.}
\label{fig:r_gamma_10000_m2}
\end{figure}

\begin{figure}[H]
\centering
% 第一行
\hfill
\begin{subfigure}[b]{0.42\textwidth}
\includegraphics[width=\textwidth]{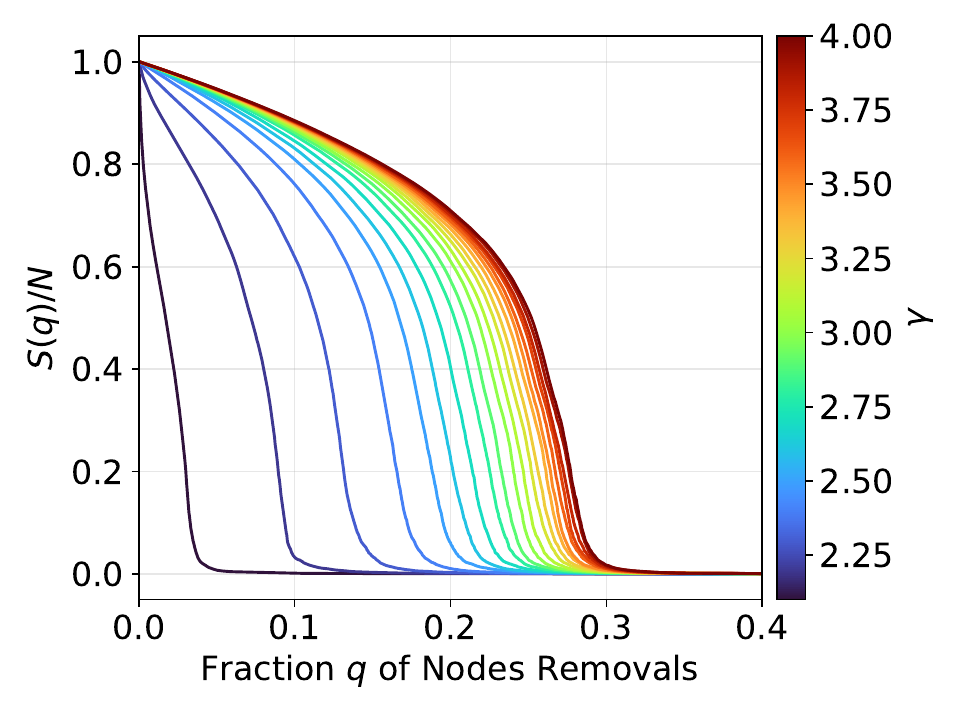}
\caption{Degrees attacks}
\end{subfigure}
\hfill
  \begin{subfigure}[b]{0.42\textwidth}
    \includegraphics[width=\textwidth]{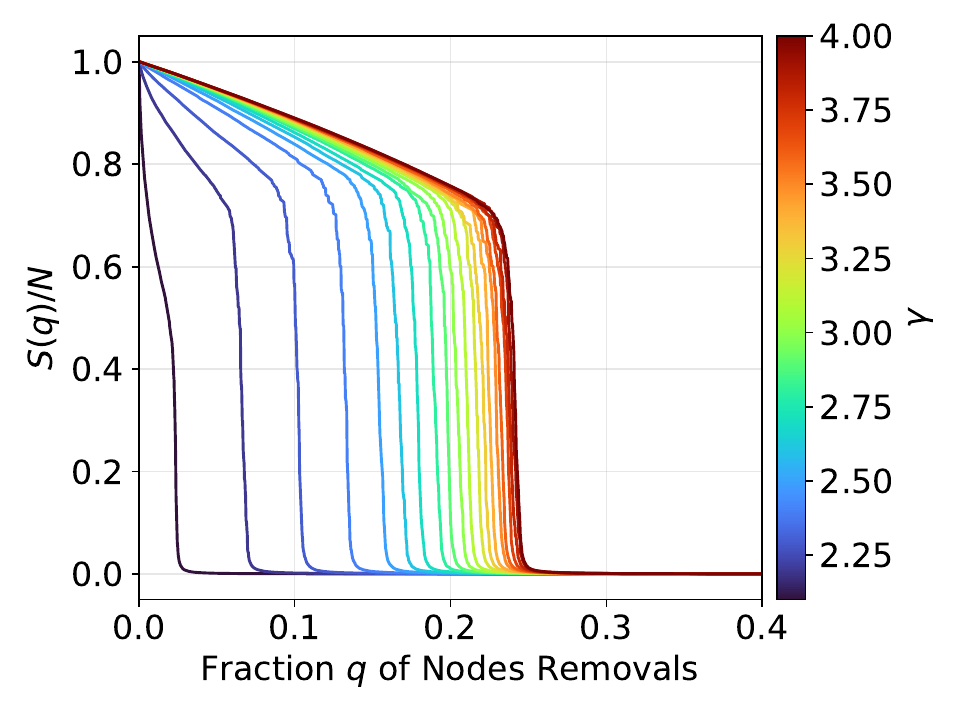}
    \caption{Betweenness centralities attacks}
  \end{subfigure}

\vspace{1em}
\hfill
\begin{subfigure}[b]{0.42\textwidth}
\includegraphics[width=\textwidth]{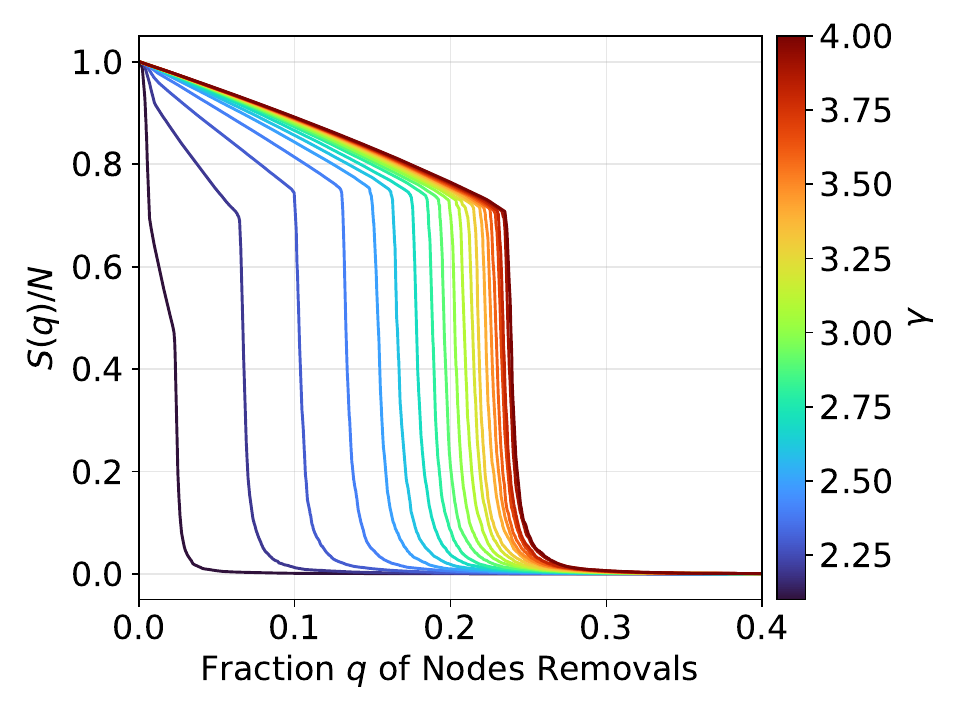}
\caption{Belief propagation attacks}
\end{subfigure}
\caption{Comparison with Figure \ref{fig:r_by_gamma} in the case of $N=10^3$ and $m=2$. More detailed results for the robustness against recalculated (a) degrees, (b) betweenness centralities, and (c) belief propagation (BP) attacks for \bm{$N=10^4$} and \bm{$m=3$}. The areas under colored curves represent the robustness index $R$ in SF networks with power-law exponents from $\gamma = 2.1$ (dark purple) to $\gamma = 4.0$ (red). As $\gamma$ increases, the areas under curves become larger from dark purple to red lines.}
\label{fig:r_gamma_10000_m3}
\end{figure}

\begin{figure}[H]
\centering
% 第一行
\hfill
\begin{subfigure}[b]{0.42\textwidth}
\includegraphics[width=\textwidth]{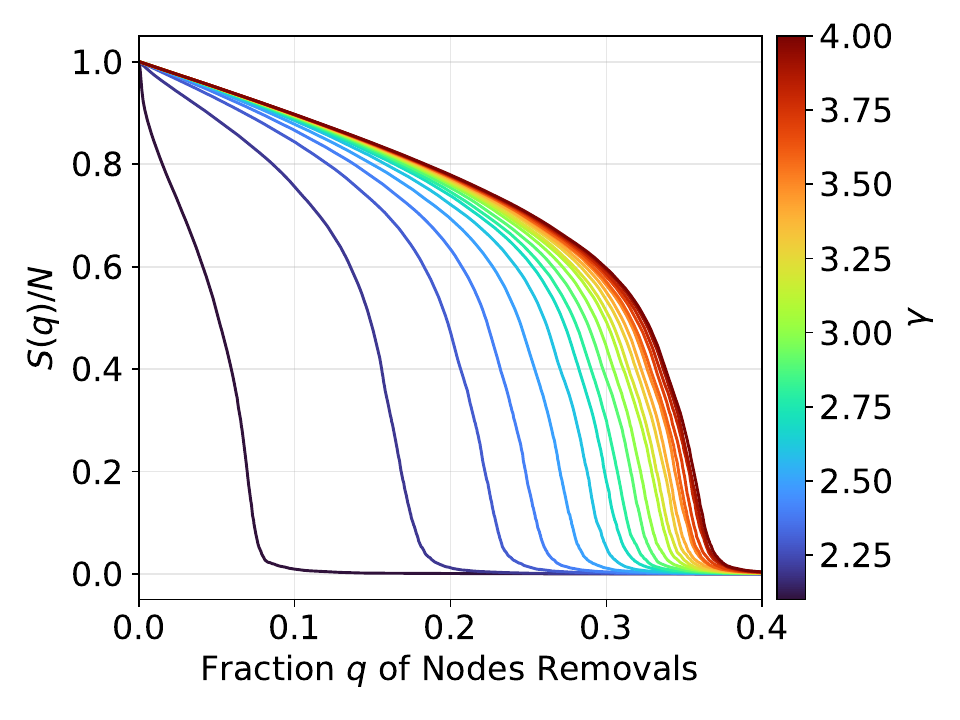}
\caption{Degrees attacks}
\end{subfigure}
\hfill
  \begin{subfigure}[b]{0.42\textwidth}
    \includegraphics[width=\textwidth]{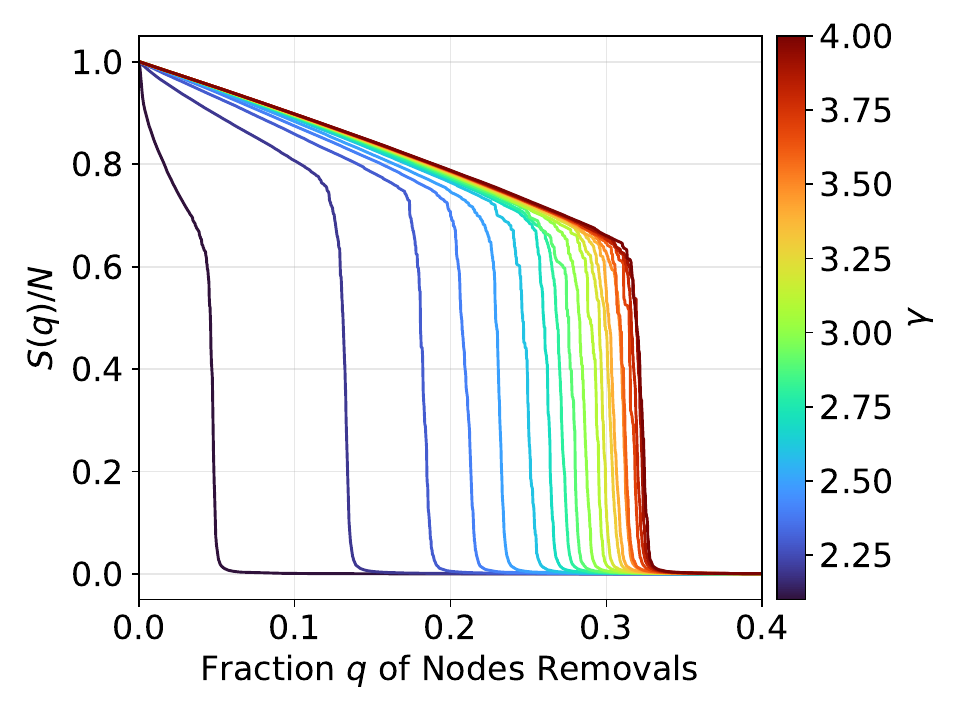}
    \caption{Betweenness centralities attacks}
  \end{subfigure}

\vspace{1em}
\hfill
\begin{subfigure}[b]{0.42\textwidth}
\includegraphics[width=\textwidth]{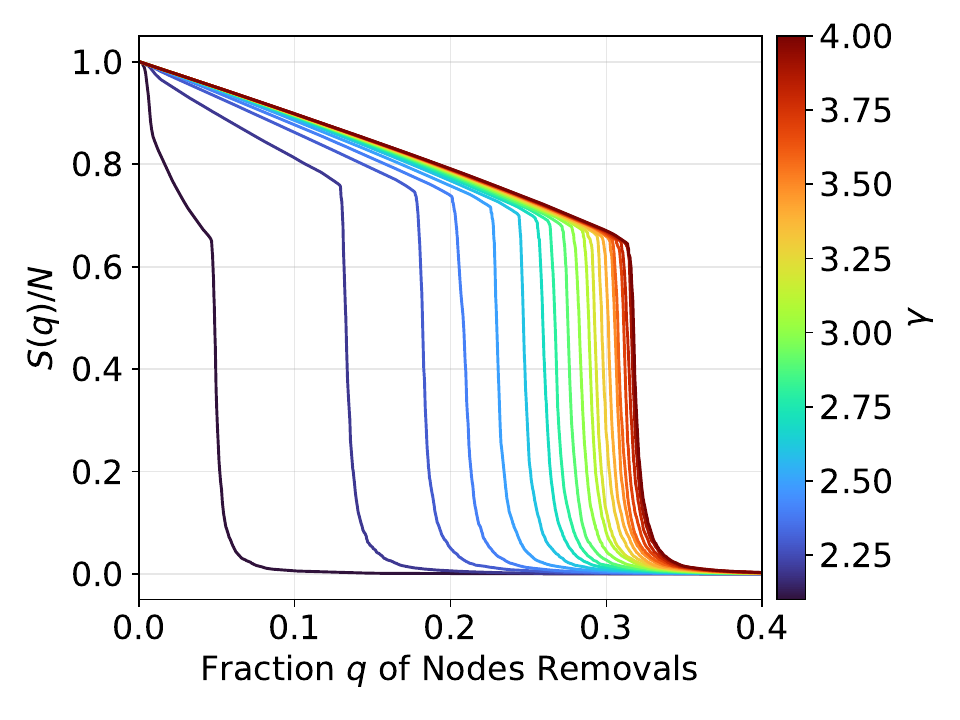}
\caption{Belief propagation attacks}
\end{subfigure}
\caption{Comparison with Figure \ref{fig:r_by_gamma} in the case of $N=10^3$ and $m=2$. More detailed results for the robustness against recalculated (a) degrees, (b) betweenness centralities, and (c) belief propagation (BP) attacks for \bm{$N=10^4$} and \bm{$m=4$}. The areas under colored curves represent the robustness index $R$ in SF networks with power-law exponents from $\gamma = 2.1$ (dark purple) to $\gamma = 4.0$ (red). As $\gamma$ increases, the areas under curves become larger from dark purple to red lines.}
\label{fig:r_gamma_10000_m4}
\end{figure}

\begin{figure}
\centering   
\hfill
\begin{subfigure}[b]{0.42\textwidth}
\includegraphics[width=\textwidth]{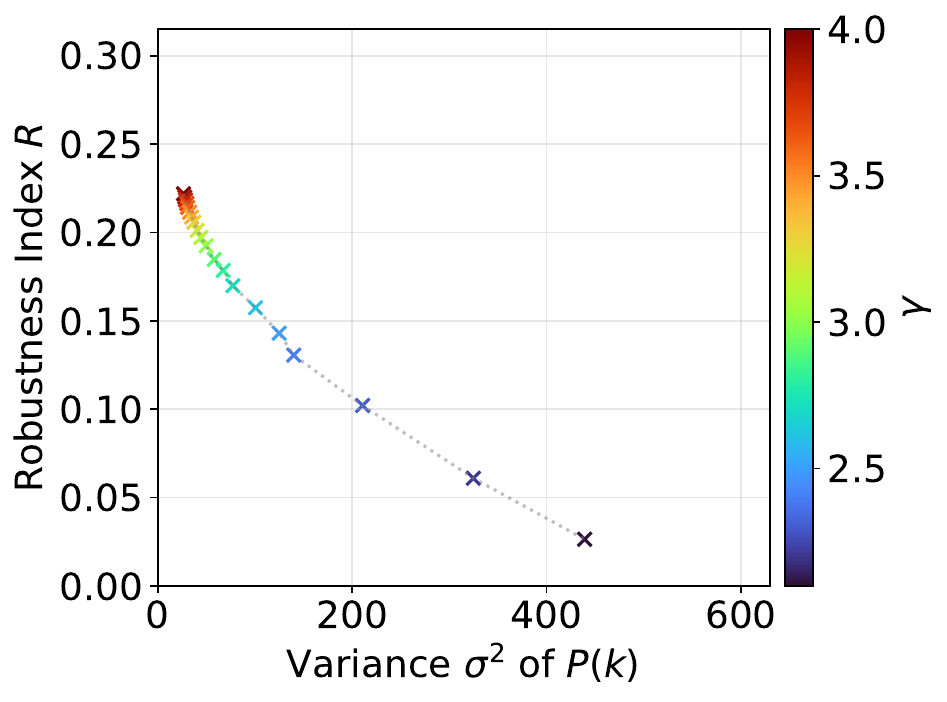}
\caption{Degrees attacks}
\end{subfigure}
\hfill
\begin{subfigure}[b]{0.42\textwidth}
\includegraphics[width=\textwidth]{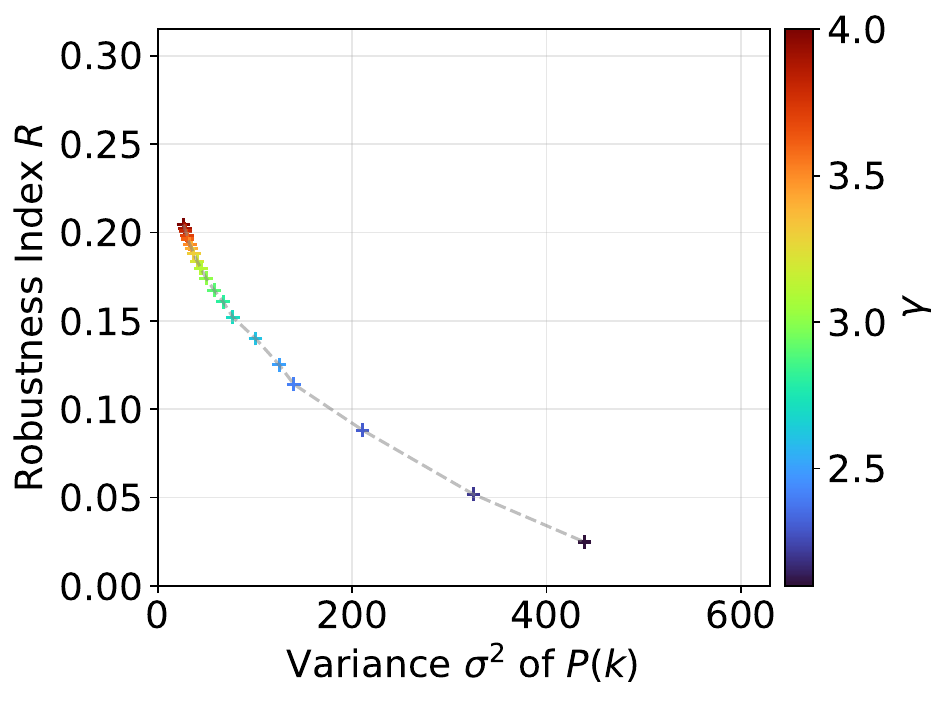}
\caption{Betweenness centralities attacks}
\end{subfigure}

\vspace{1em}
\hfill
\begin{subfigure}[b]{0.42\textwidth}
\includegraphics[width=\textwidth]{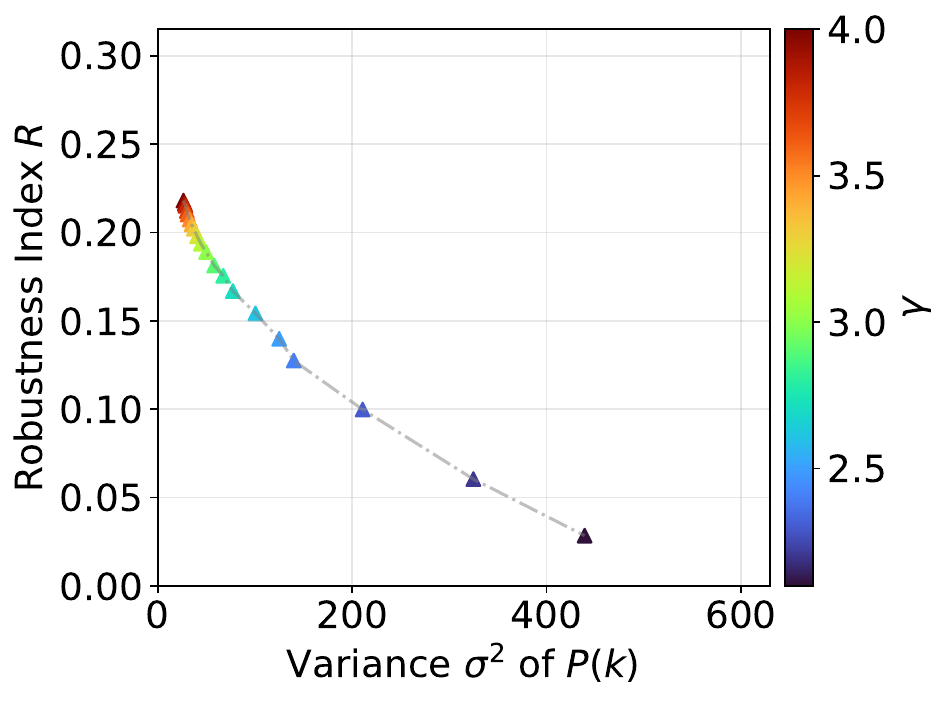}
\caption{Belief propagation attacks}
\end{subfigure}
\hfill
\begin{subfigure}[b]{0.42\textwidth}
\includegraphics[width=\textwidth]{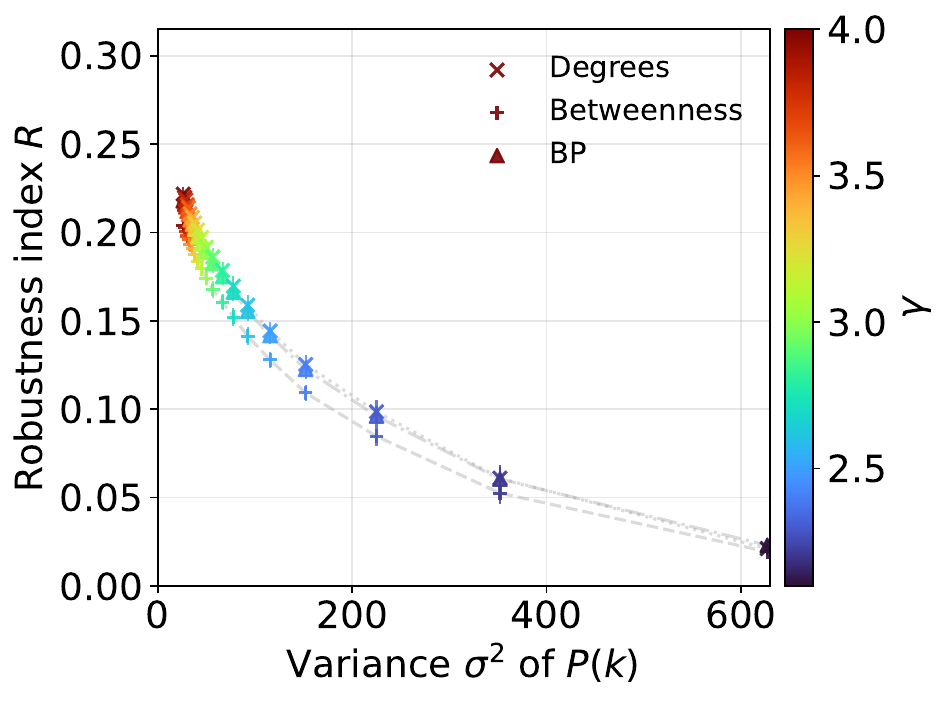}
\caption{Comparison of three attacks}
\end{subfigure}
\caption{Comparison with Figure \ref{fig:r_vs_var} in the case of $N=10^3$ and $m=2$. Robustness index $R$ versus the variance $\sigma^2$ of degree distribution $P(k)$ in randomized SF networks against recalculated (a) degrees, (b) betweenness centralities, (c) belief propagation (BP) attacks, and (d) the comparison of robustness against these attacks for \bm{$N=10^3$} and \bm{$m=3$}. Colored points represent the results for networks with power-law exponents $\gamma$ ranging from $\gamma = 2.1$ (dark purple points) to $\gamma = 4.0$ (red points). It is common that $R$ becomes larger as $\gamma$ increases. However, for $\gamma > 3$ (from green to red points), the improvement of $R$ is bounded.}
\label{fig:r_vs_var_1000_m3}
\end{figure}

\begin{figure}
\centering   
\hfill
\begin{subfigure}[b]{0.42\textwidth}
\includegraphics[width=\textwidth]{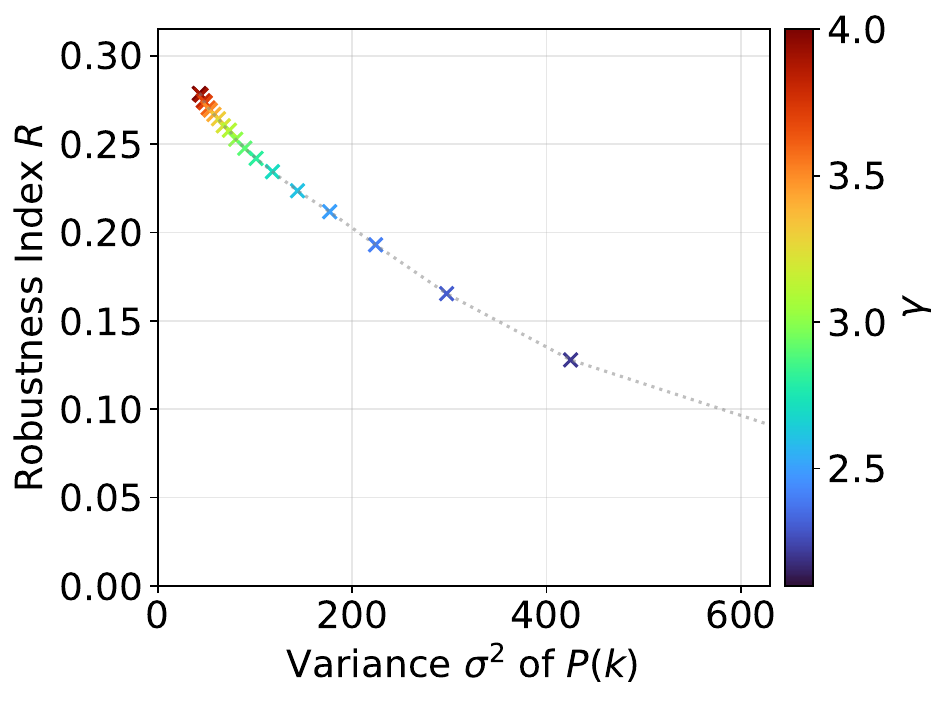}
\caption{Degrees attacks}
\end{subfigure}
\hfill
\begin{subfigure}[b]{0.42\textwidth}
\includegraphics[width=\textwidth]{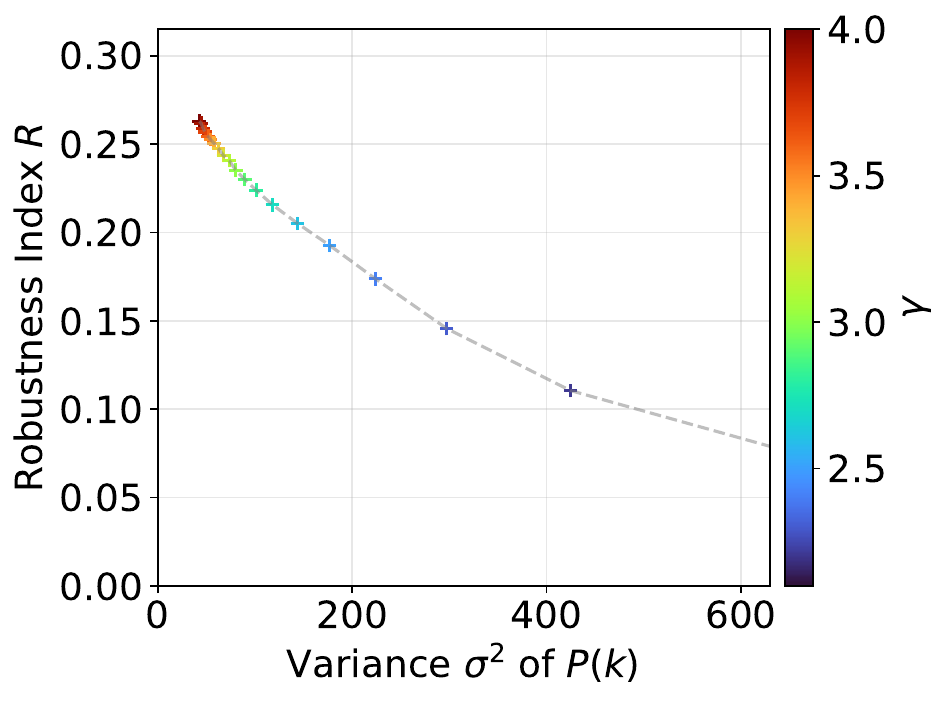}
\caption{Betweenness centralities attacks}
\end{subfigure}

\vspace{1em}
\hfill
\begin{subfigure}[b]{0.42\textwidth}
\includegraphics[width=\textwidth]{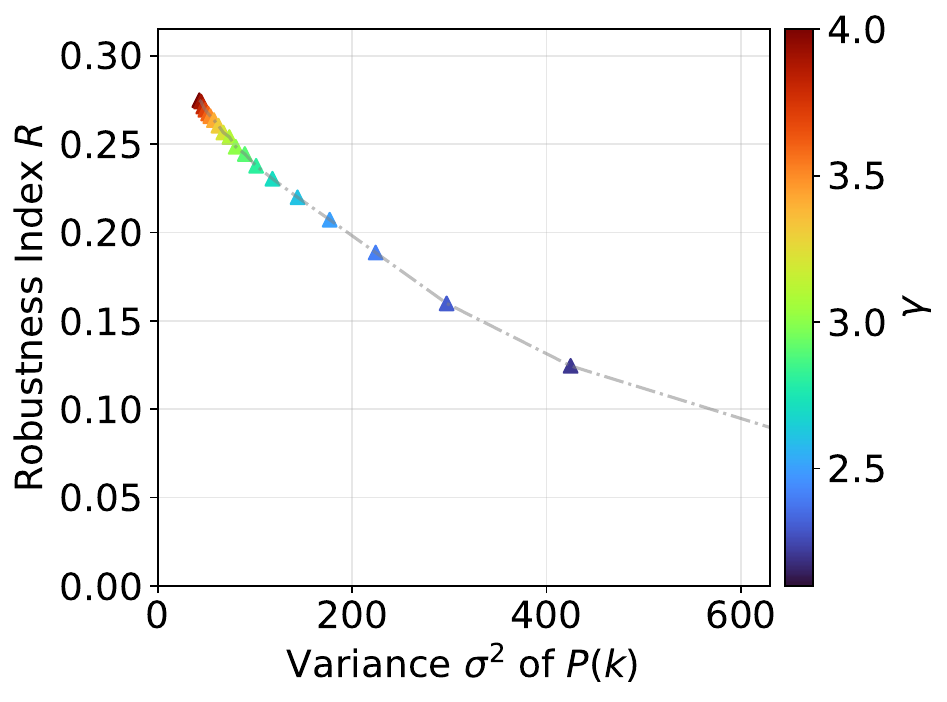}
\caption{Belief propagation attacks}
\end{subfigure}
\hfill
\begin{subfigure}[b]{0.42\textwidth}
\includegraphics[width=\textwidth]{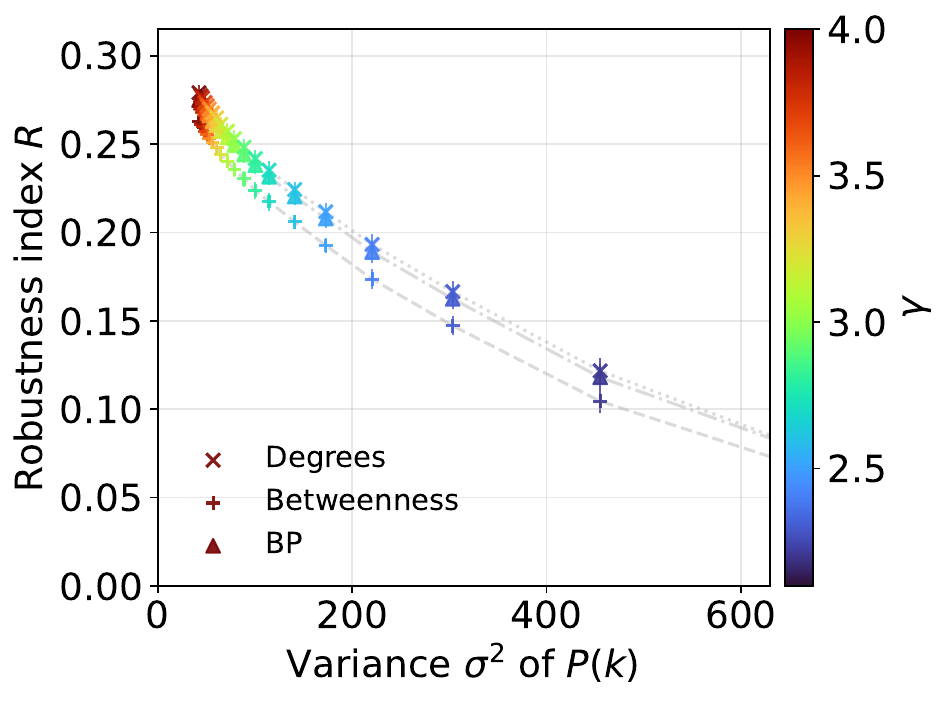}
\caption{Comparison of three attacks}
\end{subfigure}
\caption{Comparison with Figure \ref{fig:r_vs_var} in the case of $N=10^3$ and $m=2$. Robustness index $R$ versus the variance $\sigma^2$ of degree distribution $P(k)$ in randomized SF networks against recalculated (a) degrees, (b) betweenness centralities, (c) belief propagation (BP) attacks, and (d) the comparison of robustness against these attacks for \bm{$N=10^3$} and \bm{$m=4$}. Colored points represent the results for networks with power-law exponents $\gamma$ ranging from $\gamma = 2.1$ (dark purple points) to $\gamma = 4.0$ (red points). It is common that $R$ becomes larger as $\gamma$ increases. However, for $\gamma > 3$ (from green to red points), the improvement of $R$ is bounded.}
\label{fig:r_vs_var_1000_m4}
\end{figure}

\begin{figure}
\centering   
\hfill
\begin{subfigure}[b]{0.42\textwidth}
\includegraphics[width=\textwidth]{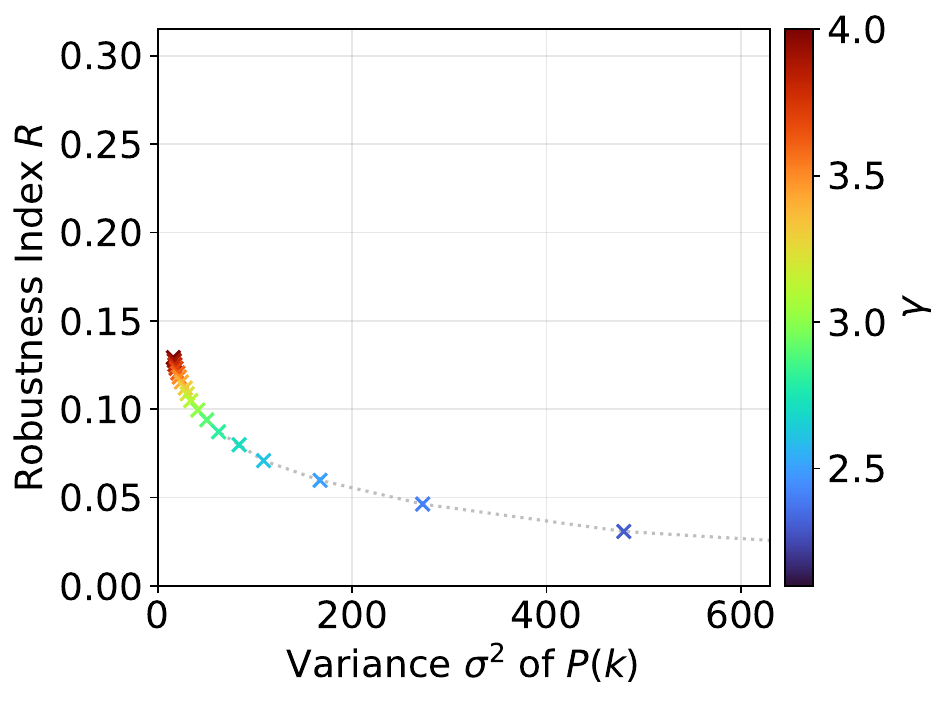}
\caption{Degrees attacks}
\end{subfigure}
\hfill
\begin{subfigure}[b]{0.42\textwidth}
\includegraphics[width=\textwidth]{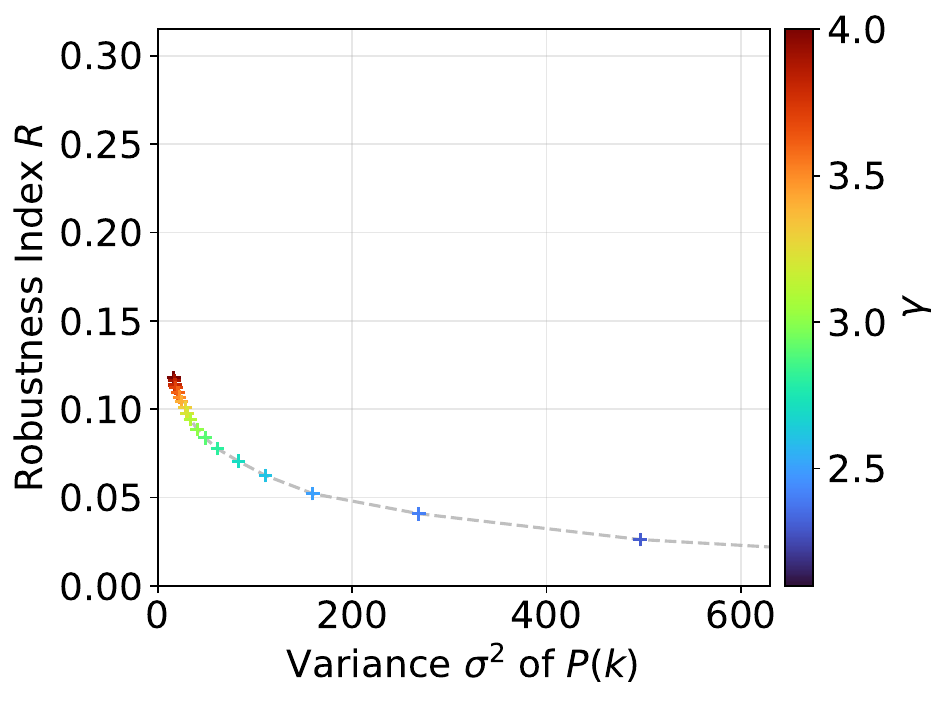}
\caption{Betweenness centralities attacks}
\end{subfigure}

\vspace{1em}
\hfill
\begin{subfigure}[b]{0.42\textwidth}
\includegraphics[width=\textwidth]{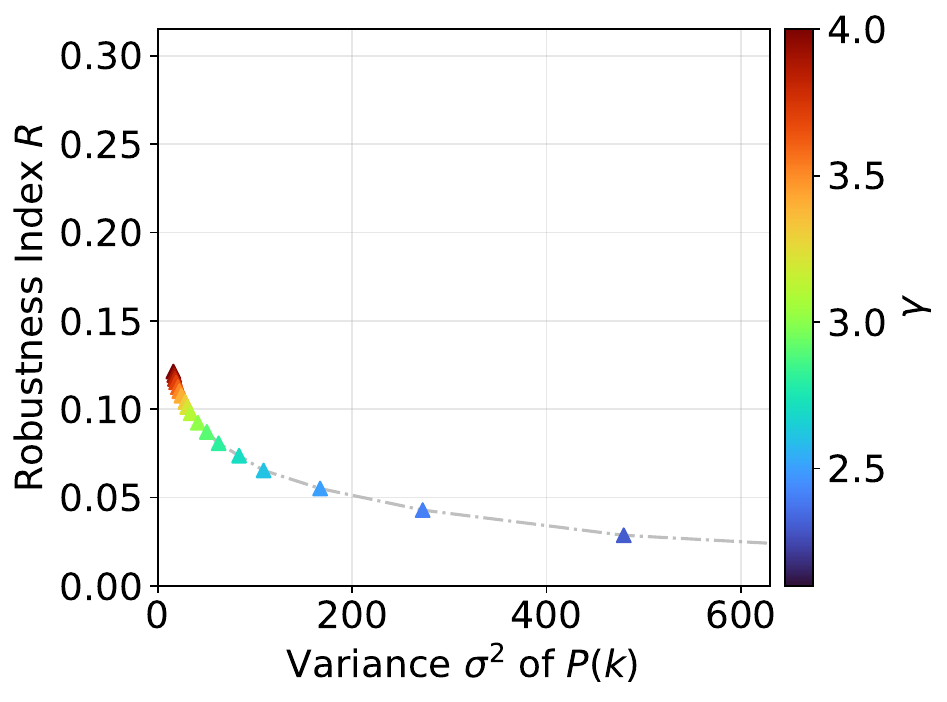}
\caption{Belief propagation attacks}
\end{subfigure}
\hfill
\begin{subfigure}[b]{0.42\textwidth}
\includegraphics[width=\textwidth]{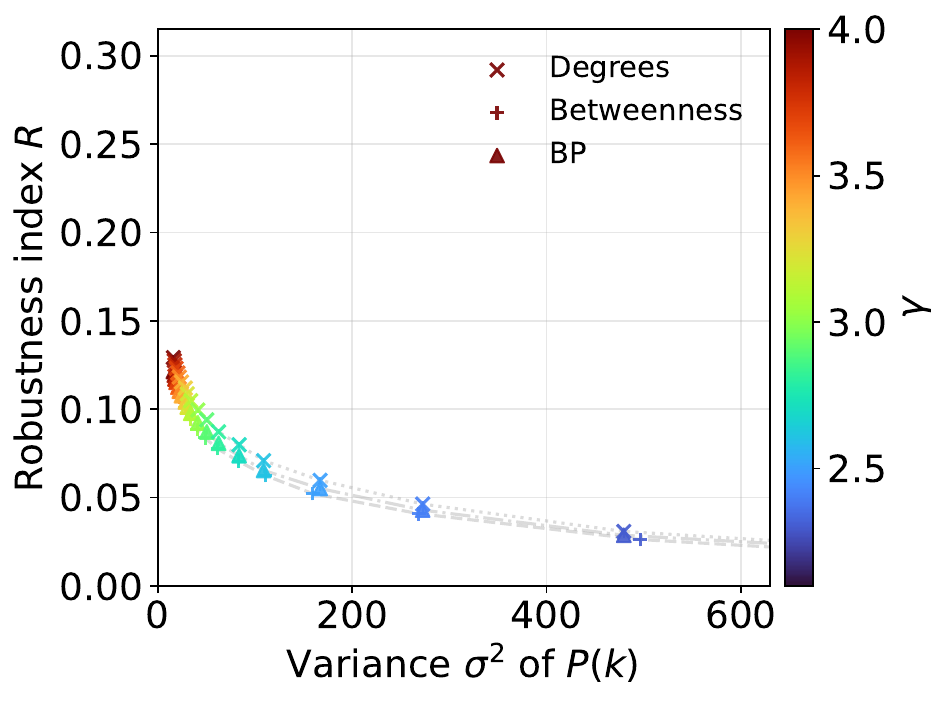}
\caption{Comparison of three attacks}
\end{subfigure}
\caption{Comparison with Figure \ref{fig:r_vs_var} in the case of $N=10^3$ and $m=2$. Robustness index $R$ versus the variance $\sigma^2$ of degree distribution $P(k)$ in randomized SF networks against recalculated (a) degrees, (b) betweenness centralities, (c) belief propagation (BP) attacks, and (d) the comparison of robustness against these attacks for \bm{$N=10^4$} and \bm{$m=2$}. Colored points represent the results for networks with power-law exponents $\gamma$ ranging from $\gamma = 2.1$ (dark purple points) to $\gamma = 4.0$ (red points). It is common that $R$ becomes larger as $\gamma$ increases. However, for $\gamma > 3$ (from green to red points), the improvement of $R$ is bounded.}
\label{fig:r_vs_var_10000_m2}
\end{figure}

\begin{figure}
\centering   
\hfill
\begin{subfigure}[b]{0.42\textwidth}
\includegraphics[width=\textwidth]{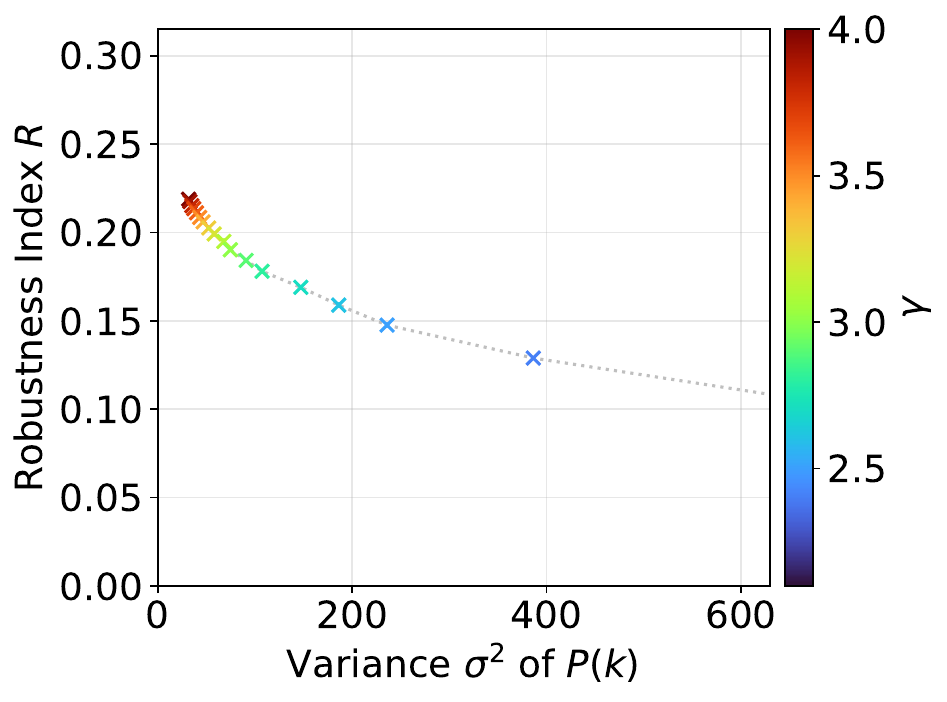}
\caption{Degrees attacks}
\end{subfigure}
\hfill
\begin{subfigure}[b]{0.42\textwidth}
\includegraphics[width=\textwidth]{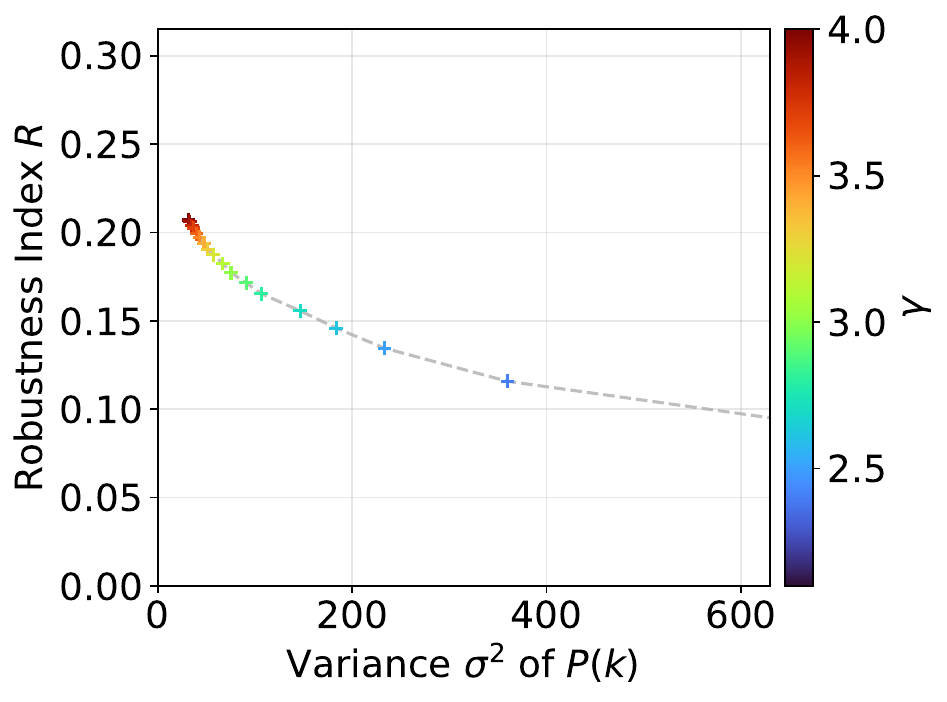}
\caption{Betweenness centralities attacks}
\end{subfigure}

\vspace{1em}
\hfill
\begin{subfigure}[b]{0.42\textwidth}
\includegraphics[width=\textwidth]{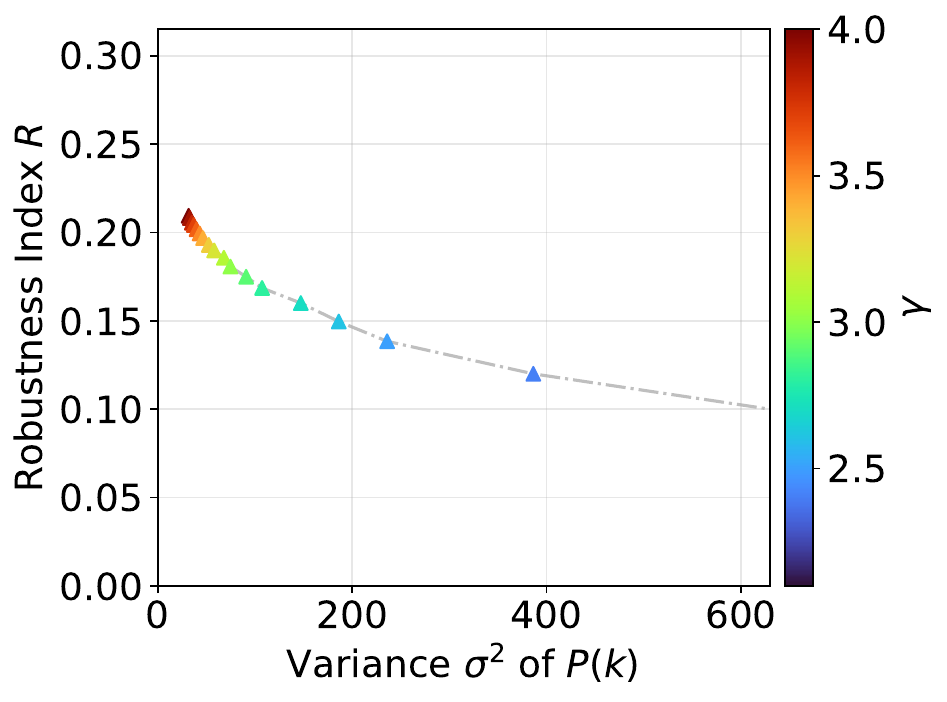}
\caption{Belief propagation attacks}
\end{subfigure}
\hfill
\begin{subfigure}[b]{0.42\textwidth}
\includegraphics[width=\textwidth]{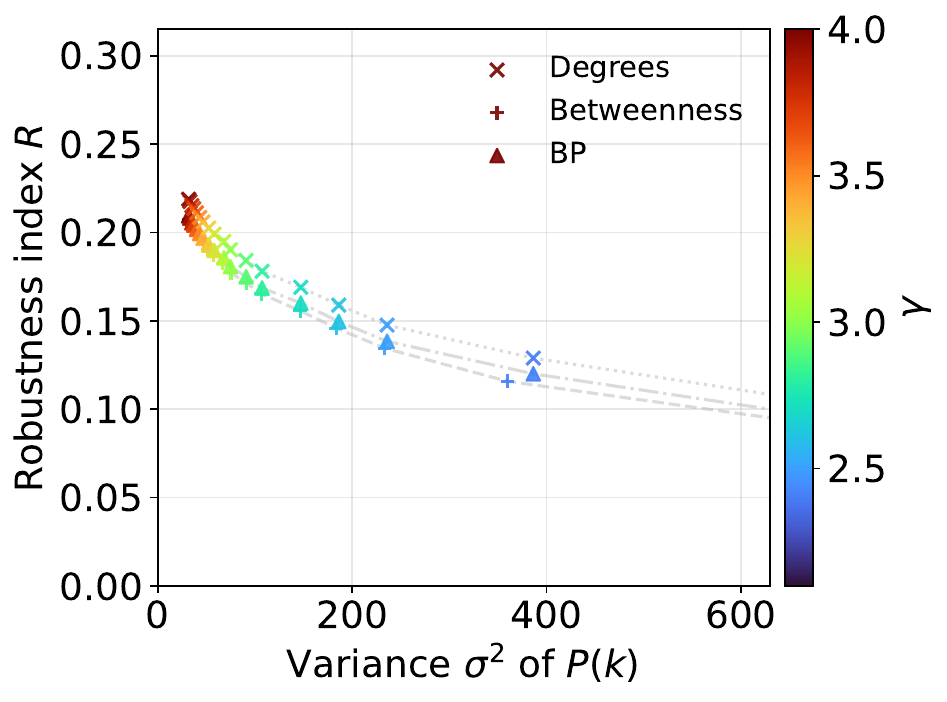}
\caption{Comparison of three attacks}
\end{subfigure}
\caption{Comparison with Figure \ref{fig:r_vs_var} in the case of $N=10^3$ and $m=2$. Robustness index $R$ versus the variance $\sigma^2$ of degree distribution $P(k)$ in randomized SF networks against recalculated (a) degrees, (b) betweenness centralities, (c) belief propagation (BP) attacks, and (d) the comparison of robustness against these attacks for \bm{$N=10^4$} and \bm{$m=3$}. Colored points represent the results for networks with power-law exponents $\gamma$ ranging from $\gamma = 2.1$ (dark purple points) to $\gamma = 4.0$ (red points). It is common that $R$ becomes larger as $\gamma$ increases. However, for $\gamma > 3$ (from green to red points), the improvement of $R$ is bounded.}
\label{fig:r_vs_var_10000_m3}
\end{figure}

\begin{figure}
\centering   
\hfill
\begin{subfigure}[b]{0.42\textwidth}
\includegraphics[width=\textwidth]{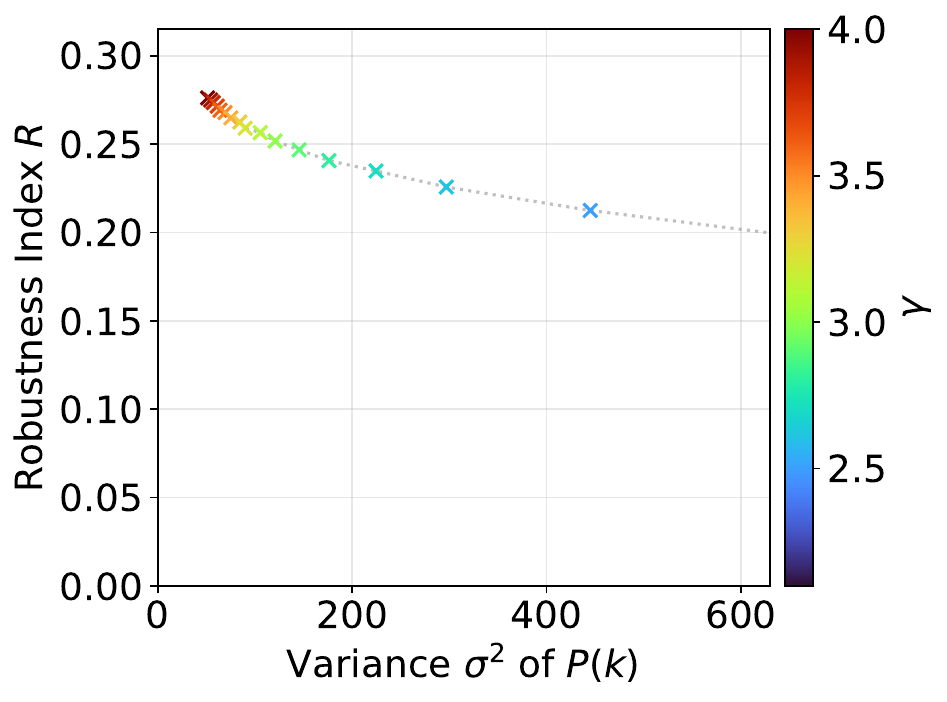}
\caption{Degrees attacks}
\end{subfigure}
\hfill
\begin{subfigure}[b]{0.42\textwidth}
\includegraphics[width=\textwidth]{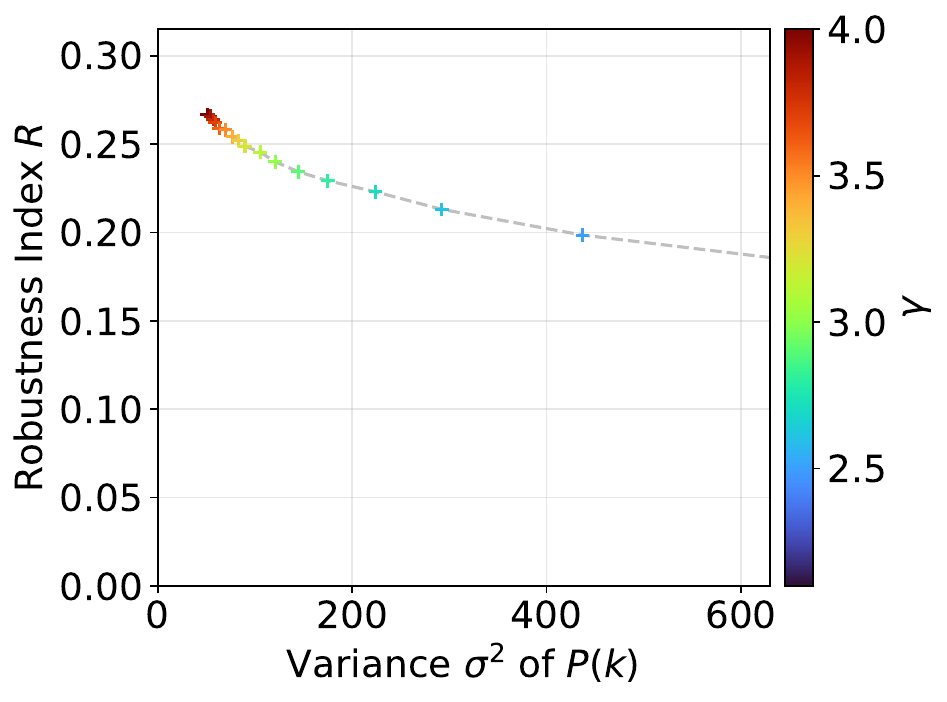}
\caption{Betweenness centralities attacks}
\end{subfigure}

\vspace{1em}
\hfill
\begin{subfigure}[b]{0.42\textwidth}
\includegraphics[width=\textwidth]{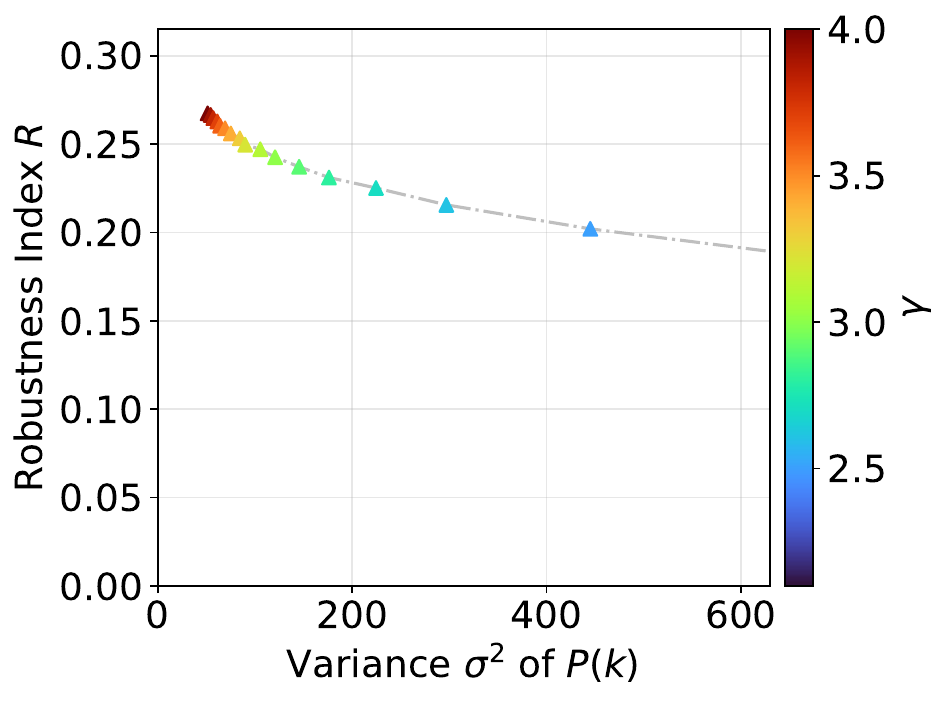}
\caption{Belief propagation attacks}
\end{subfigure}
\hfill
\begin{subfigure}[b]{0.42\textwidth}
\includegraphics[width=\textwidth]{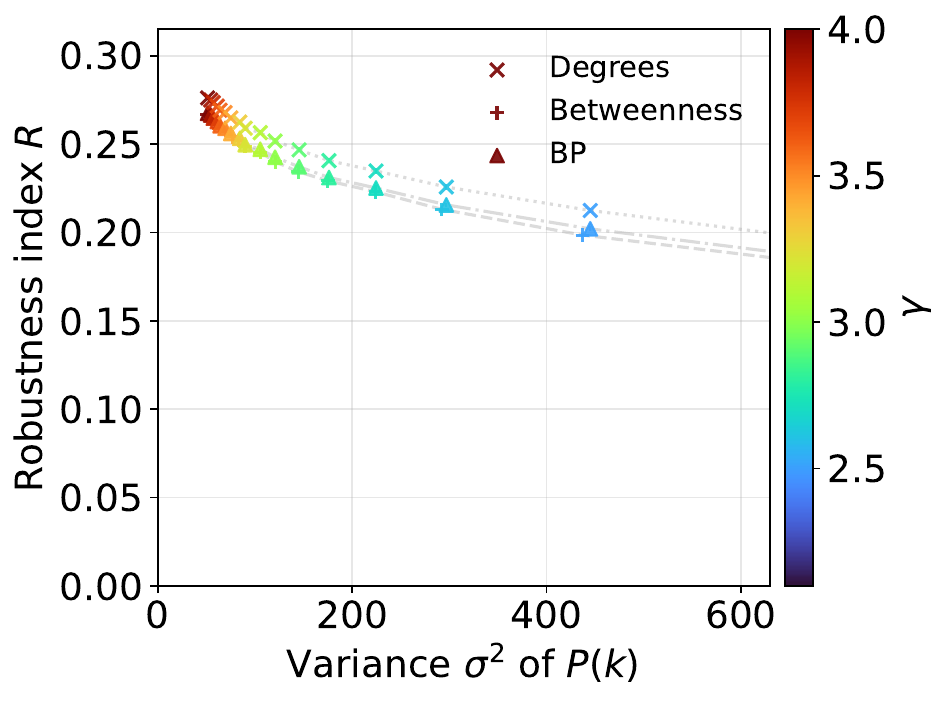}
\caption{Comparison of three attacks}
\end{subfigure}
\caption{Comparison with Figure \ref{fig:r_vs_var} in the case of $N=10^3$ and $m=2$. Robustness index $R$ versus the variance $\sigma^2$ of degree distribution $P(k)$ in randomized SF networks against recalculated (a) degrees, (b) betweenness centralities, (c) belief propagation (BP) attacks, and (d) the comparison of robustness against these attacks for \bm{$N=10^4$} and \bm{$m=4$}. Colored points represent the results for networks with power-law exponents $\gamma$ ranging from $\gamma = 2.1$ (dark purple points) to $\gamma = 4.0$ (red points). It is common that $R$ becomes larger as $\gamma$ increases. However, for $\gamma > 3$ (from green to red points), the improvement of $R$ is bounded.}
\label{fig:r_vs_var_10000_m4}
\end{figure}

\begin{itemize}
    \item[(1)] Figures \ref{fig:degree_distributions_1000_m3} - \ref{fig:degree_distributions_10000_m4} with $N=10^3, ~m=3,4$ and $N=10^4, ~m=2,3,4$ are corresponding to Figure \ref{fig:degree_distributions} with $N=10^3$ and $m=2$ in main body.
    \item[(2)] Figures \ref{fig:kmax_variance_gamma_1000_m3} - \ref{fig:kmax_variance_gamma_10000_m4} with $N=10^3, ~m=3,4$ and $N=10^4, ~m=2,3,4$ are corresponding to Figure \ref{fig:kmax_variance_gamma} with $N=10^3$ and $m=2$ in main body.
    \item[(3)] Figures \ref{fig:r_by_attacks_1000_m3} - \ref{fig:r_by_attacks_10000_m4} with $N=10^3, ~m=3,4$ and $N=10^4, ~m=2,3,4$ are corresponding to Figure \ref{fig:r_by_attacks} with $N=10^3$ and $m=2$ in main body.
    \item[(4)] Figures \ref{fig:r_gamma_1000_m3} - \ref{fig:r_gamma_10000_m4} with $N=10^3, ~m=3,4$ and $N=10^4, ~m=2,3,4$ are corresponding to Figure \ref{fig:r_by_gamma} with $N=10^3$ and $m=2$ in main body.
    \item[(5)] Figures \ref{fig:r_vs_var_1000_m3} - \ref{fig:r_vs_var_10000_m4} with $N=10^3, ~m=3,4$ and $N=10^4, ~m=2,3,4$ are corresponding  to Figure \ref{fig:r_vs_var} with $N=10^3$ and $m=2$ in main body.
\end{itemize}

The supplementary materials associated with this article are available at Zenodo:\\ https://10.5281/zenodo.18372527

\clearpage
\subsection*{Figures}

\clearpage
\subsection*{Tables}
\begin{table}[!p]
\centering
\caption{Variances of $S(q)/N$ for different attacks under various values of the power-law exponent $\gamma$ for $N=10^3$ and $m=2$, corresponding to Figures~\ref{fig:r_by_attacks} and~\ref{fig:r_by_gamma}. All values are scaled by $10^{-3}$. Entries marked with an asterisk (*) indicate values that are nonzero but smaller than the display precision under the current scaling (on the order of $10^{-5}$), and therefore cannot be shown numerically.}
\label{tab:SqN_variance_1000_m2}
\resizebox{\textwidth}{!}{%
\begin{tabular}{c|c|ccccccccccccc}
\hline
\textbf{$\gamma$} & \textbf{fraction q} & \textbf{0.02} & \textbf{0.04} & \textbf{0.06} & \textbf{0.08} & \textbf{0.10} & \textbf{0.12} & \textbf{0.14} & \textbf{0.16} & \textbf{0.18} & \textbf{0.20} & \textbf{0.22} & \textbf{0.24} & \textbf{0.25} \\
\hline
4.0 & degree & 0.01 & 0.03 & 0.09 & 0.12 & 0.31 & 0.73 & 1.14 & 5.47 & 2.76 & 0.40 & 0.05 & 0.01 & 0.01 \\
4.0 & betweenness & 0.01 & 0.03 & 0.08 & 0.28 & 1.17 & 5.94 & 10.62 & 0.02 & * & * & * & * & * \\
4.0 & bp & 0.02 & 0.04 & 0.08 & 0.13 & 0.18 & 0.36 & 21.58 & 2.47 & 0.41 & 0.10 & 0.02 & 0.01 & 0.01 \\
3.9 & degree & 0.02 & 0.08 & 0.10 & 0.19 & 0.44 & 0.82 & 2.12 & 6.97 & 0.82 & 0.18 & 0.11 & 0.02 & * \\
3.9 & betweenness & 0.02 & 0.07 & 0.10 & 0.31 & 1.33 & 10.85 & 6.20 & 0.04 & * & * & * & * & * \\
3.9 & bp & 0.02 & 0.04 & 0.06 & 0.10 & 0.17 & 0.44 & 20.01 & 1.71 & 0.34 & 0.14 & 0.02 & 0.01 & * \\
3.8 & degree & 0.02 & 0.08 & 0.07 & 0.21 & 0.60 & 1.10 & 2.34 & 8.07 & 0.69 & 0.26 & 0.10 & 0.01 & * \\
3.8 & betweenness & 0.02 & 0.10 & 0.12 & 0.48 & 1.63 & 13.97 & 1.25 & 0.02 & * & * & * & * & * \\
3.8 & bp & 0.02 & 0.04 & 0.09 & 0.12 & 0.19 & 1.08 & 21.35 & 1.32 & 0.23 & 0.07 & 0.02 & 0.01 & * \\
3.7 & degree & 0.03 & 0.08 & 0.12 & 0.25 & 0.33 & 0.81 & 3.57 & 6.54 & 0.58 & 0.06 & 0.02 & * & * \\
3.7 & betweenness & 0.03 & 0.06 & 0.15 & 0.38 & 1.91 & 22.50 & 0.61 & 0.01 & * & * & * & * & * \\
3.7 & bp & 0.02 & 0.06 & 0.08 & 0.13 & 0.17 & 4.20 & 12.27 & 1.44 & 0.15 & 0.04 & 0.02 & 0.01 & * \\
3.6 & degree & 0.02 & 0.04 & 0.12 & 0.17 & 0.59 & 1.05 & 5.36 & 4.52 & 0.51 & 0.07 & 0.02 & * & * \\
3.6 & betweenness & 0.02 & 0.03 & 0.21 & 0.42 & 2.42 & 20.39 & 0.21 & 0.01 & * & * & * & * & * \\
3.6 & bp & 0.02 & 0.05 & 0.09 & 0.18 & 0.24 & 10.21 & 8.55 & 1.09 & 0.10 & 0.06 & 0.02 & 0.01 & * \\
3.5 & degree & 0.02 & 0.08 & 0.15 & 0.24 & 0.44 & 1.18 & 3.74 & 3.77 & 0.23 & 0.07 & 0.01 & * & * \\
3.5 & betweenness & 0.02 & 0.07 & 0.20 & 0.47 & 3.27 & 16.96 & 0.07 & 0.01 & * & * & * & * & * \\
3.5 & bp & 0.02 & 0.06 & 0.14 & 0.20 & 0.35 & 23.04 & 4.53 & 0.90 & 0.20 & 0.06 & 0.01 & * & * \\
3.4 & degree & 0.04 & 0.11 & 0.10 & 0.17 & 0.34 & 1.82 & 6.46 & 0.87 & 0.34 & 0.05 & 0.01 & * & * \\
3.4 & betweenness & 0.05 & 0.12 & 0.14 & 0.69 & 4.59 & 7.98 & 0.07 & 0.01 & * & * & * & * & * \\
3.4 & bp & 0.03 & 0.07 & 0.13 & 0.23 & 0.49 & 36.24 & 3.56 & 0.58 & 0.11 & 0.03 & 0.01 & 0.01 & * \\
3.3 & degree & 0.04 & 0.08 & 0.14 & 0.15 & 0.32 & 1.81 & 5.25 & 1.80 & 0.07 & 0.03 & 0.01 & * & * \\
3.3 & betweenness & 0.04 & 0.10 & 0.13 & 0.51 & 6.92 & 3.04 & 0.03 & * & * & * & * & * & * \\
3.3 & bp & 0.04 & 0.07 & 0.15 & 0.23 & 0.40 & 23.42 & 1.13 & 0.23 & 0.05 & 0.03 & 0.01 & * & * \\
3.2 & degree & 0.04 & 0.08 & 0.20 & 0.38 & 0.58 & 2.57 & 6.07 & 1.21 & 0.12 & 0.03 & 0.01 & * & * \\
3.2 & betweenness & 0.04 & 0.09 & 0.32 & 0.89 & 11.69 & 1.79 & 0.02 & * & * & * & * & * & * \\
3.2 & bp & 0.05 & 0.09 & 0.14 & 0.19 & 0.79 & 23.17 & 1.97 & 0.22 & 0.06 & 0.03 & 0.01 & * & * \\
3.1 & degree & 0.06 & 0.17 & 0.32 & 0.59 & 1.41 & 8.16 & 5.31 & 0.98 & 0.09 & 0.01 & * & * & * \\
3.1 & betweenness & 0.06 & 0.17 & 0.33 & 1.10 & 24.92 & 0.29 & 0.01 & * & * & * & * & * & * \\
3.1 & bp & 0.05 & 0.11 & 0.21 & 0.37 & 7.56 & 9.34 & 0.79 & 0.14 & 0.04 & 0.01 & * & * & * \\
3.0 & degree & 0.06 & 0.17 & 0.40 & 0.75 & 3.00 & 14.88 & 2.10 & 0.22 & 0.05 & 0.03 & * & * & * \\
3.0 & betweenness & 0.07 & 0.16 & 0.43 & 2.51 & 26.46 & 0.18 & 0.01 & * & * & * & * & * & * \\
3.0 & bp & 0.06 & 0.14 & 0.23 & 0.48 & 23.75 & 7.05 & 0.57 & 0.10 & 0.04 & 0.01 & * & * & * \\
2.9 & degree & 0.09 & 0.18 & 0.48 & 0.58 & 2.60 & 7.35 & 0.77 & 0.09 & 0.04 & 0.01 & * & * & * \\
2.9 & betweenness & 0.08 & 0.23 & 0.57 & 2.32 & 13.87 & 0.05 & * & * & * & * & * & * & * \\
2.9 & bp & 0.08 & 0.18 & 0.32 & 0.50 & 31.41 & 4.02 & 0.32 & 0.05 & 0.03 & 0.01 & * & * & * \\
2.8 & degree & 0.11 & 0.21 & 0.67 & 1.69 & 8.19 & 6.70 & 0.60 & 0.12 & 0.02 & * & * & * & * \\
2.8 & betweenness & 0.07 & 0.23 & 1.43 & 14.48 & 1.82 & 0.02 & * & * & * & * & * & * & * \\
2.8 & bp & 0.12 & 0.29 & 0.44 & 0.87 & 17.83 & 1.48 & 0.25 & 0.08 & 0.02 & 0.01 & * & * & * \\
2.7 & degree & 0.21 & 0.27 & 0.83 & 3.32 & 8.98 & 3.50 & 0.08 & 0.04 & * & * & * & * & * \\
2.7 & betweenness & 0.19 & 0.22 & 1.66 & 34.17 & 0.16 & 0.01 & * & * & * & * & * & * & * \\
2.7 & bp & 0.19 & 0.32 & 0.63 & 19.60 & 6.01 & 0.59 & 0.09 & 0.03 & 0.01 & * & * & * & * \\
2.6 & degree & 0.16 & 0.58 & 1.20 & 9.13 & 10.02 & 0.61 & 0.12 & 0.02 & * & * & * & * & * \\
2.6 & betweenness & 0.17 & 0.29 & 2.75 & 8.81 & 0.04 & * & * & * & * & * & * & * & * \\
2.6 & bp & 0.28 & 0.56 & 1.02 & 31.79 & 1.46 & 0.17 & 0.06 & 0.02 & * & * & * & * & * \\
2.5 & degree & 0.56 & 1.01 & 3.04 & 11.80 & 0.46 & 0.08 & 0.02 & 0.01 & * & * & * & * & * \\
2.5 & betweenness & 0.48 & 1.10 & 26.51 & 0.12 & * & * & * & * & * & * & * & * & * \\
2.5 & bp & 0.46 & 0.96 & 16.58 & 7.22 & 0.69 & 0.11 & 0.03 & 0.01 & * & * & * & * & * \\
2.4 & degree & 1.24 & 2.99 & 18.06 & 1.28 & 0.09 & 0.01 & 0.01 & 0.01 & * & * & * & * & * \\
2.4 & betweenness & 1.28 & 9.06 & 7.79 & 0.04 & * & * & * & * & * & * & * & * & * \\
2.4 & bp & 1.12 & 3.77 & 24.44 & 0.63 & 0.08 & 0.03 & 0.01 & * & * & * & * & * & * \\
2.3 & degree & 2.21 & 16.19 & 0.98 & 0.03 & 0.01 & 0.01 & * & * & * & * & * & * & * \\
2.3 & betweenness & 2.25 & 19.35 & 0.03 & * & * & * & * & * & * & * & * & * & * \\
2.3 & bp & 1.91 & 27.23 & 0.93 & 0.08 & 0.02 & 0.01 & * & * & * & * & * & * & * \\
2.2 & degree & 11.91 & 8.99 & 0.35 & 0.02 & * & * & * & * & * & * & * & * & * \\
2.2 & betweenness & 21.60 & 0.28 & 0.01 & * & * & * & * & * & * & * & * & * & * \\
2.2 & bp & 8.60 & 3.18 & 0.07 & 0.01 & * & * & * & * & * & * & * & * & * \\
2.1 & degree & 0.41 & 0.01 & * & * & * & * & * & * & * & * & * & * & * \\
2.1 & betweenness & 0.12 & * & * & * & * & * & * & * & * & * & * & * & * \\
2.1 & bp & 0.80 & 0.01 & * & * & * & * & * & * & * & * & * & * & * \\
\hline
\end{tabular}
}
\end{table}

\begin{table}[!ht]
\centering
\caption{Variances of $S(q)/N$ for different attacks under various values of the power-law exponent $\gamma$ for $N=10^3$ and $m=3$, corresponding to Figures~\ref{fig:r_by_attacks_1000_m3} and~\ref{fig:r_gamma_1000_m3}. All values are scaled by $10^{-3}$. Entries marked with an asterisk (*) indicate values that are nonzero but smaller than the display precision under the current scaling (on the order of $10^{-5}$), and therefore cannot be shown numerically.}
\label{tab:SqN_variance_1000_m3}
\resizebox{\textwidth}{!}{%
\begin{tabular}{c|c|ccccccccccccc}
\hline
\textbf{$\gamma$} & \textbf{fraction q} & \textbf{0.02} & \textbf{0.04} & \textbf{0.06} & \textbf{0.08} & \textbf{0.10} & \textbf{0.12} & \textbf{0.14} & \textbf{0.16} & \textbf{0.18} & \textbf{0.20} & \textbf{0.22} & \textbf{0.24} & \textbf{0.25} \\
\hline
4.0 & degree & * & * & 0.01 & 0.01 & 0.02 & 0.03 & 0.03 & 0.07 & 0.13 & 0.19 & 0.35 & 0.92 & 2.22 \\
4.0 & betweenness & * & * & 0.01 & 0.01 & 0.02 & 0.03 & 0.06 & 0.11 & 0.26 & 1.29 & 4.42 & 11.02 & 2.18 \\
4.0 & bp & * & * & * & 0.01 & 0.01 & 0.01 & 0.02 & 0.03 & 0.04 & 0.06 & 0.09 & 24.87 & 13.41 \\
3.9 & degree & * & * & 0.01 & 0.01 & 0.03 & 0.04 & 0.05 & 0.09 & 0.11 & 0.16 & 0.30 & 1.00 & 2.29 \\
3.9 & betweenness & * & * & 0.01 & 0.01 & 0.02 & 0.03 & 0.07 & 0.19 & 0.38 & 1.02 & 4.66 & 12.11 & 1.67 \\
3.9 & bp & * & * & * & 0.01 & 0.01 & 0.02 & 0.02 & 0.03 & 0.04 & 0.06 & 0.10 & 23.36 & 10.43 \\
3.8 & degree & * & * & 0.01 & 0.01 & 0.02 & 0.03 & 0.04 & 0.07 & 0.14 & 0.30 & 0.55 & 1.56 & 3.50 \\
3.8 & betweenness & * & * & 0.01 & 0.01 & 0.02 & 0.03 & 0.05 & 0.20 & 0.44 & 1.57 & 7.99 & 9.33 & 1.59 \\
3.8 & bp & * & * & * & 0.01 & 0.01 & 0.02 & 0.03 & 0.04 & 0.06 & 0.10 & 0.30 & 25.09 & 10.08 \\
3.7 & degree & * & * & 0.01 & 0.01 & 0.02 & 0.03 & 0.05 & 0.10 & 0.16 & 0.25 & 0.50 & 1.60 & 4.04 \\
3.7 & betweenness & * & * & 0.01 & 0.01 & 0.02 & 0.03 & 0.06 & 0.19 & 0.58 & 2.02 & 8.70 & 4.32 & 0.26 \\
3.7 & bp & * & * & * & 0.01 & 0.01 & 0.01 & 0.03 & 0.04 & 0.06 & 0.10 & 2.53 & 17.84 & 5.06 \\
3.6 & degree & * & * & 0.01 & 0.01 & 0.03 & 0.04 & 0.07 & 0.09 & 0.14 & 0.21 & 0.58 & 2.00 & 6.73 \\
3.6 & betweenness & * & * & 0.01 & 0.01 & 0.02 & 0.05 & 0.10 & 0.25 & 0.71 & 2.70 & 12.36 & 4.28 & 0.29 \\
3.6 & bp & * & * & 0.01 & 0.01 & 0.01 & 0.02 & 0.03 & 0.05 & 0.06 & 0.08 & 0.99 & 18.38 & 5.35 \\
3.5 & degree & * & * & 0.01 & 0.02 & 0.03 & 0.05 & 0.07 & 0.13 & 0.22 & 0.34 & 0.68 & 3.81 & 8.45 \\
3.5 & betweenness & * & * & 0.01 & 0.01 & 0.03 & 0.04 & 0.09 & 0.28 & 0.78 & 3.33 & 14.87 & 0.54 & 0.06 \\
3.5 & bp & * & * & 0.01 & 0.01 & 0.02 & 0.03 & 0.04 & 0.05 & 0.07 & 0.13 & 8.68 & 9.27 & 2.69 \\
3.4 & degree & * & * & 0.01 & 0.02 & 0.03 & 0.06 & 0.07 & 0.14 & 0.23 & 0.44 & 0.83 & 5.73 & 7.93 \\
3.4 & betweenness & * & * & 0.01 & 0.02 & 0.03 & 0.05 & 0.09 & 0.52 & 0.87 & 4.71 & 16.75 & 0.51 & 0.06 \\
3.4 & bp & * & * & 0.01 & 0.01 & 0.02 & 0.03 & 0.04 & 0.06 & 0.08 & 0.13 & 22.18 & 5.33 & 1.56 \\
3.3 & degree & * & * & 0.01 & 0.02 & 0.03 & 0.06 & 0.08 & 0.14 & 0.19 & 0.38 & 1.05 & 6.83 & 8.63 \\
3.3 & betweenness & * & * & 0.01 & 0.02 & 0.03 & 0.05 & 0.11 & 0.36 & 1.11 & 5.32 & 16.52 & 0.16 & 0.03 \\
3.3 & bp & * & * & 0.01 & 0.01 & 0.02 & 0.02 & 0.04 & 0.06 & 0.10 & 0.18 & 29.12 & 6.04 & 2.48 \\
3.2 & degree & * & * & 0.01 & 0.02 & 0.04 & 0.06 & 0.10 & 0.17 & 0.31 & 0.54 & 1.69 & 6.86 & 6.11 \\
3.2 & betweenness & * & * & 0.01 & 0.02 & 0.03 & 0.06 & 0.17 & 0.38 & 1.64 & 8.37 & 6.78 & 0.08 & 0.01 \\
3.2 & bp & * & * & 0.01 & 0.02 & 0.02 & 0.03 & 0.04 & 0.06 & 0.12 & 0.37 & 23.47 & 2.60 & 1.10 \\
3.1 & degree & * & 0.01 & 0.01 & 0.03 & 0.04 & 0.07 & 0.09 & 0.17 & 0.37 & 0.78 & 2.59 & 9.35 & 5.18 \\
3.1 & betweenness & * & 0.01 & 0.01 & 0.03 & 0.06 & 0.10 & 0.21 & 0.54 & 2.67 & 14.98 & 4.16 & 0.03 & 0.01 \\
3.1 & bp & * & * & 0.01 & 0.02 & 0.03 & 0.04 & 0.07 & 0.08 & 0.11 & 5.95 & 12.99 & 1.44 & 0.58 \\
3.0 & degree & * & 0.01 & 0.01 & 0.03 & 0.05 & 0.08 & 0.14 & 0.24 & 0.40 & 1.04 & 8.74 & 6.78 & 3.20 \\
3.0 & betweenness & * & 0.01 & 0.01 & 0.03 & 0.05 & 0.10 & 0.24 & 1.01 & 4.23 & 17.61 & 0.57 & 0.01 & * \\
3.0 & bp & * & 0.01 & 0.01 & 0.02 & 0.03 & 0.04 & 0.06 & 0.09 & 0.13 & 16.21 & 7.68 & 0.87 & 0.42 \\
2.9 & degree & * & 0.01 & 0.02 & 0.03 & 0.04 & 0.08 & 0.14 & 0.30 & 0.57 & 2.15 & 12.19 & 4.22 & 1.74 \\
2.9 & betweenness & * & 0.01 & 0.02 & 0.04 & 0.06 & 0.11 & 0.48 & 2.48 & 11.36 & 8.92 & 0.08 & 0.01 & * \\
2.9 & bp & * & 0.01 & 0.01 & 0.02 & 0.02 & 0.04 & 0.06 & 0.10 & 1.67 & 33.19 & 2.95 & 0.61 & 0.31 \\
2.8 & degree & 0.01 & 0.01 & 0.03 & 0.05 & 0.08 & 0.13 & 0.20 & 0.39 & 1.06 & 5.00 & 7.24 & 2.13 & 0.74 \\
2.8 & betweenness & * & 0.01 & 0.03 & 0.04 & 0.07 & 0.17 & 0.60 & 3.41 & 22.27 & 0.77 & 0.02 & * & * \\
2.8 & bp & * & 0.01 & 0.02 & 0.03 & 0.05 & 0.06 & 0.08 & 0.13 & 12.97 & 13.28 & 1.71 & 0.22 & 0.12 \\
2.7 & degree & 0.01 & 0.02 & 0.04 & 0.07 & 0.10 & 0.17 & 0.29 & 0.73 & 3.49 & 11.76 & 5.18 & 0.98 & 0.38 \\
2.7 & betweenness & 0.01 & 0.02 & 0.03 & 0.06 & 0.10 & 0.39 & 1.48 & 10.29 & 18.71 & 0.15 & 0.01 & * & * \\
2.7 & bp & 0.01 & 0.02 & 0.03 & 0.05 & 0.07 & 0.09 & 0.12 & 1.01 & 36.51 & 3.11 & 0.63 & 0.13 & 0.05 \\
2.6 & degree & 0.01 & 0.03 & 0.06 & 0.08 & 0.20 & 0.36 & 0.64 & 1.74 & 9.07 & 7.67 & 1.75 & 0.26 & 0.12 \\
2.6 & betweenness & 0.01 & 0.03 & 0.05 & 0.08 & 0.13 & 0.76 & 4.60 & 26.90 & 1.22 & 0.02 & * & * & * \\
2.6 & bp & 0.01 & 0.03 & 0.04 & 0.06 & 0.08 & 0.14 & 0.26 & 27.74 & 8.66 & 1.19 & 0.23 & 0.10 & 0.06 \\
2.5 & degree & 0.02 & 0.05 & 0.09 & 0.15 & 0.22 & 0.48 & 1.20 & 5.63 & 8.31 & 1.69 & 0.14 & 0.04 & 0.02 \\
2.5 & betweenness & 0.01 & 0.04 & 0.09 & 0.20 & 0.57 & 3.16 & 25.04 & 3.79 & 0.04 & * & * & * & * \\
2.5 & bp & 0.01 & 0.04 & 0.06 & 0.09 & 0.14 & 0.23 & 9.16 & 18.37 & 1.51 & 0.33 & 0.12 & 0.03 & 0.01 \\
2.4 & degree & 0.04 & 0.10 & 0.15 & 0.26 & 0.43 & 1.19 & 8.80 & 8.78 & 1.45 & 0.24 & 0.03 & 0.01 & * \\
2.4 & betweenness & 0.04 & 0.09 & 0.14 & 0.29 & 1.99 & 23.91 & 1.25 & 0.02 & * & * & * & * & * \\
2.4 & bp & 0.03 & 0.09 & 0.12 & 0.16 & 0.29 & 8.03 & 13.26 & 1.77 & 0.34 & 0.10 & 0.04 & 0.01 & * \\
2.3 & degree & 0.12 & 0.32 & 0.60 & 1.25 & 4.32 & 18.76 & 4.02 & 0.49 & 0.07 & 0.03 & * & * & * \\
2.3 & betweenness & 0.12 & 0.31 & 0.65 & 6.62 & 31.27 & 0.31 & 0.01 & * & * & * & * & * & * \\
2.3 & bp & 0.10 & 0.27 & 0.48 & 0.84 & 38.67 & 7.61 & 0.53 & 0.14 & 0.04 & 0.01 & * & * & * \\
2.2 & degree & 0.97 & 2.90 & 11.91 & 18.45 & 2.38 & 0.27 & 0.07 & 0.01 & * & * & * & * & * \\
2.2 & betweenness & 0.91 & 3.73 & 49.02 & 0.52 & 0.02 & * & * & * & * & * & * & * & * \\
2.2 & bp & 0.85 & 2.21 & 28.44 & 10.73 & 0.59 & 0.13 & 0.03 & 0.01 & * & * & * & * & * \\
2.1 & degree & 3.78 & 1.94 & 0.04 & 0.02 & * & * & * & * & * & * & * & * & * \\
2.1 & betweenness & 7.32 & 0.06 & * & * & * & * & * & * & * & * & * & * & * \\
2.1 & bp & 3.36 & 0.74 & 0.05 & 0.01 & * & * & * & * & * & * & * & * & * \\
\hline
\end{tabular}
}
\end{table}

\begin{table}[!ht]
\centering
\caption{Variances of $S(q)/N$ for different attacks under various values of the power-law exponent $\gamma$ for $N=10^3$ and $m=4$, corresponding to Figures~\ref{fig:r_by_attacks_1000_m4} and~\ref{fig:r_gamma_1000_m4}. All values are scaled by $10^{-3}$. Entries marked with an asterisk (*) indicate values that are nonzero but smaller than the display precision under the current scaling (on the order of $10^{-5}$), and therefore cannot be shown numerically.}
\label{tab:SqN_variance_1000_m4}
\resizebox{\textwidth}{!}{%
\begin{tabular}{c|c|ccccccccccccc}
\hline
\textbf{$\gamma$} & \textbf{fraction q} & \textbf{0.02} & \textbf{0.04} & \textbf{0.06} & \textbf{0.08} & \textbf{0.10} & \textbf{0.12} & \textbf{0.14} & \textbf{0.16} & \textbf{0.18} & \textbf{0.20} & \textbf{0.22} & \textbf{0.24} & \textbf{0.25} \\
\hline
4.0 & degree & * & * & * & * & * & 0.01 & 0.01 & 0.01 & 0.02 & 0.03 & 0.05 & 0.07 & 0.06 \\
4.0 & betweenness & * & * & * & * & * & * & 0.01 & 0.01 & 0.02 & 0.03 & 0.07 & 0.13 & 0.20 \\
4.0 & bp & * & * & * & * & * & * & * & 0.01 & 0.01 & 0.01 & 0.02 & 0.03 & 0.03 \\
3.9 & degree & * & * & * & * & * & 0.01 & 0.01 & 0.01 & 0.02 & 0.03 & 0.05 & 0.05 & 0.07 \\
3.9 & betweenness & * & * & * & * & * & * & 0.01 & 0.01 & 0.02 & 0.03 & 0.08 & 0.27 & 0.36 \\
3.9 & bp & * & * & * & * & * & * & * & 0.01 & 0.01 & 0.01 & 0.02 & 0.02 & 0.02 \\
3.8 & degree & * & * & * & * & * & 0.01 & 0.01 & 0.01 & 0.02 & 0.02 & 0.04 & 0.07 & 0.08 \\
3.8 & betweenness & * & * & * & * & * & * & 0.01 & 0.01 & 0.02 & 0.03 & 0.07 & 0.19 & 0.32 \\
3.8 & bp & * & * & * & * & * & * & * & 0.01 & 0.01 & 0.01 & 0.01 & 0.02 & 0.03 \\
3.7 & degree & * & * & * & * & * & 0.01 & 0.01 & 0.01 & 0.02 & 0.03 & 0.05 & 0.08 & 0.10 \\
3.7 & betweenness & * & * & * & * & * & * & 0.01 & 0.01 & 0.02 & 0.05 & 0.09 & 0.25 & 0.40 \\
3.7 & bp & * & * & * & * & * & * & * & 0.01 & 0.01 & 0.01 & 0.02 & 0.03 & 0.03 \\
3.6 & degree & * & * & * & * & * & 0.01 & 0.01 & 0.01 & 0.02 & 0.03 & 0.05 & 0.07 & 0.09 \\
3.6 & betweenness & * & * & * & * & * & 0.01 & 0.01 & 0.01 & 0.02 & 0.04 & 0.09 & 0.17 & 0.37 \\
3.6 & bp & * & * & * & * & * & * & * & 0.01 & 0.01 & 0.01 & 0.01 & 0.02 & 0.02 \\
3.5 & degree & * & * & * & * & * & 0.01 & 0.01 & 0.01 & 0.02 & 0.03 & 0.05 & 0.07 & 0.10 \\
3.5 & betweenness & * & * & * & * & * & 0.01 & 0.01 & 0.01 & 0.02 & 0.05 & 0.12 & 0.28 & 0.46 \\
3.5 & bp & * & * & * & * & * & * & 0.01 & 0.01 & 0.01 & 0.01 & 0.02 & 0.03 & 0.03 \\
3.4 & degree & * & * & * & * & * & 0.01 & 0.01 & 0.02 & 0.03 & 0.04 & 0.07 & 0.09 & 0.10 \\
3.4 & betweenness & * & * & * & * & * & 0.01 & 0.01 & 0.01 & 0.03 & 0.06 & 0.12 & 0.22 & 0.47 \\
3.4 & bp & * & * & * & * & * & * & 0.01 & 0.01 & 0.01 & 0.02 & 0.02 & 0.03 & 0.04 \\
3.3 & degree & * & * & * & * & 0.01 & 0.01 & 0.02 & 0.02 & 0.03 & 0.05 & 0.07 & 0.12 & 0.15 \\
3.3 & betweenness & * & * & * & * & * & 0.01 & 0.01 & 0.02 & 0.02 & 0.05 & 0.15 & 0.42 & 0.65 \\
3.3 & bp & * & * & * & * & * & 0.01 & 0.01 & 0.01 & 0.02 & 0.02 & 0.03 & 0.04 & 0.05 \\
3.2 & degree & * & * & * & * & 0.01 & 0.01 & 0.02 & 0.03 & 0.03 & 0.05 & 0.08 & 0.11 & 0.16 \\
3.2 & betweenness & * & * & * & * & * & 0.01 & 0.01 & 0.02 & 0.04 & 0.08 & 0.23 & 0.55 & 1.11 \\
3.2 & bp & * & * & * & * & * & 0.01 & 0.01 & 0.01 & 0.02 & 0.02 & 0.03 & 0.04 & 0.04 \\
3.1 & degree & * & * & * & * & 0.01 & 0.01 & 0.02 & 0.03 & 0.04 & 0.06 & 0.10 & 0.16 & 0.24 \\
3.1 & betweenness & * & * & * & * & 0.01 & 0.01 & 0.01 & 0.04 & 0.07 & 0.14 & 0.26 & 0.87 & 1.45 \\
3.1 & bp & * & * & * & * & * & 0.01 & 0.01 & 0.01 & 0.02 & 0.02 & 0.04 & 0.04 & 0.05 \\
3.0 & degree & * & * & * & 0.01 & 0.01 & 0.01 & 0.02 & 0.03 & 0.05 & 0.08 & 0.12 & 0.20 & 0.23 \\
3.0 & betweenness & * & * & * & * & 0.01 & 0.01 & 0.02 & 0.02 & 0.06 & 0.13 & 0.34 & 1.06 & 1.76 \\
3.0 & bp & * & * & * & * & 0.01 & 0.01 & 0.01 & 0.01 & 0.02 & 0.02 & 0.03 & 0.04 & 0.04 \\
2.9 & degree & * & * & * & 0.01 & 0.01 & 0.02 & 0.03 & 0.04 & 0.06 & 0.09 & 0.12 & 0.17 & 0.28 \\
2.9 & betweenness & * & * & * & 0.01 & 0.01 & 0.01 & 0.02 & 0.05 & 0.07 & 0.19 & 0.58 & 1.88 & 4.22 \\
2.9 & bp & * & * & * & * & 0.01 & 0.01 & 0.01 & 0.02 & 0.02 & 0.03 & 0.04 & 0.06 & 0.14 \\
2.8 & degree & * & * & * & 0.01 & 0.01 & 0.02 & 0.03 & 0.05 & 0.06 & 0.09 & 0.14 & 0.28 & 0.44 \\
2.8 & betweenness & * & * & * & 0.01 & 0.01 & 0.02 & 0.03 & 0.04 & 0.11 & 0.36 & 0.84 & 3.46 & 7.07 \\
2.8 & bp & * & * & * & 0.01 & 0.01 & 0.01 & 0.02 & 0.02 & 0.02 & 0.03 & 0.05 & 0.09 & 0.31 \\
2.7 & degree & * & * & 0.01 & 0.01 & 0.02 & 0.02 & 0.04 & 0.06 & 0.07 & 0.14 & 0.24 & 0.53 & 0.75 \\
2.7 & betweenness & * & * & * & 0.01 & 0.02 & 0.02 & 0.04 & 0.09 & 0.17 & 0.43 & 1.71 & 8.23 & 21.37 \\
2.7 & bp & * & * & * & 0.01 & 0.01 & 0.02 & 0.03 & 0.03 & 0.04 & 0.05 & 0.09 & 0.40 & 9.43 \\
2.6 & degree & * & * & 0.01 & 0.01 & 0.02 & 0.03 & 0.05 & 0.07 & 0.12 & 0.16 & 0.33 & 0.83 & 1.82 \\
2.6 & betweenness & * & * & 0.01 & 0.01 & 0.02 & 0.03 & 0.04 & 0.08 & 0.25 & 0.95 & 2.83 & 16.59 & 8.37 \\
2.6 & bp & * & * & 0.01 & 0.01 & 0.01 & 0.02 & 0.02 & 0.03 & 0.04 & 0.06 & 0.11 & 22.68 & 26.20 \\
2.5 & degree & * & 0.01 & 0.01 & 0.02 & 0.04 & 0.06 & 0.10 & 0.17 & 0.25 & 0.39 & 0.90 & 3.92 & 8.47 \\
2.5 & betweenness & * & 0.01 & 0.01 & 0.02 & 0.03 & 0.05 & 0.11 & 0.37 & 0.88 & 3.20 & 17.20 & 1.05 & 0.12 \\
2.5 & bp & * & * & 0.01 & 0.02 & 0.02 & 0.03 & 0.04 & 0.05 & 0.08 & 0.11 & 11.48 & 14.80 & 5.60 \\
2.4 & degree & 0.01 & 0.02 & 0.03 & 0.05 & 0.08 & 0.12 & 0.21 & 0.36 & 0.67 & 1.54 & 7.53 & 9.54 & 5.52 \\
2.4 & betweenness & 0.01 & 0.01 & 0.03 & 0.05 & 0.07 & 0.13 & 0.31 & 1.21 & 7.83 & 24.27 & 1.11 & 0.02 & 0.01 \\
2.4 & bp & * & 0.01 & 0.02 & 0.03 & 0.05 & 0.06 & 0.08 & 0.11 & 0.19 & 25.49 & 14.49 & 1.04 & 0.48 \\
2.3 & degree & 0.01 & 0.03 & 0.05 & 0.10 & 0.16 & 0.28 & 0.53 & 1.02 & 3.85 & 11.85 & 4.58 & 0.39 & 0.18 \\
2.3 & betweenness & 0.01 & 0.03 & 0.05 & 0.08 & 0.17 & 0.55 & 2.73 & 21.56 & 7.33 & 0.09 & 0.01 & * & * \\
2.3 & bp & 0.01 & 0.03 & 0.05 & 0.06 & 0.10 & 0.12 & 0.18 & 5.11 & 27.38 & 3.48 & 0.41 & 0.15 & 0.07 \\
2.2 & degree & 0.07 & 0.21 & 0.37 & 0.61 & 1.17 & 3.38 & 14.69 & 8.26 & 1.74 & 0.23 & 0.05 & 0.01 & * \\
2.2 & betweenness & 0.08 & 0.19 & 0.30 & 0.82 & 7.02 & 32.20 & 2.59 & 0.03 & * & * & * & * & * \\
2.2 & bp & 0.06 & 0.16 & 0.26 & 0.37 & 0.51 & 37.56 & 15.70 & 1.60 & 0.26 & 0.09 & 0.02 & 0.01 & * \\
2.1 & degree & 1.41 & 4.11 & 17.43 & 6.14 & 0.44 & 0.06 & 0.02 & * & * & * & * & * & * \\
2.1 & betweenness & 1.33 & 8.34 & 9.97 & 0.05 & * & * & * & * & * & * & * & * & * \\
2.1 & bp & 1.24 & 2.75 & 35.72 & 1.51 & 0.23 & 0.06 & 0.02 & * & * & * & * & * & * \\
\hline
\end{tabular}
}
\end{table}

\begin{table}[!ht]
\centering
\caption{Variances of $S(q)/N$ for different attacks under various values of the power-law exponent $\gamma$ for $N=10^4$ and $m=2$, corresponding to Figures~\ref{fig:r_by_attacks_10000_m2} and~\ref{fig:r_gamma_10000_m2}. All values are scaled by $10^{-3}$. Entries marked with an asterisk (*) indicate values that are nonzero but smaller than the display precision under the current scaling (on the order of $10^{-5}$), and therefore cannot be shown numerically.}
\label{tab:SqN_variance_10000_m2}
\resizebox{\textwidth}{!}{%
\begin{tabular}{c|c|ccccccccccccc}
\hline
\textbf{$\gamma$} & \textbf{fraction q} & \textbf{0.02} & \textbf{0.04} & \textbf{0.06} & \textbf{0.08} & \textbf{0.10} & \textbf{0.12} & \textbf{0.14} & \textbf{0.16} & \textbf{0.18} & \textbf{0.20} & \textbf{0.22} & \textbf{0.24} & \textbf{0.25} \\
\hline
4.0 & degree & * & * & 0.01 & 0.02 & 0.04 & 0.08 & 0.21 & 1.95 & 0.14 & 0.01 & * & * & * \\
4.0 & betweenness & 106.64 & 99.40 & 91.83 & 84.03 & 75.87 & 65.67 & 0.03 & * & * & * & * & * & * \\
4.0 & bp & * & * & 0.01 & 0.01 & 0.01 & 0.02 & 2.00 & 0.05 & * & * & * & * & * \\
3.9 & degree & * & 0.01 & 0.01 & 0.01 & 0.03 & 0.07 & 0.19 & 2.60 & 0.11 & * & * & * & * \\
3.9 & betweenness & 42.64 & 39.63 & 36.51 & 33.32 & 30.00 & 25.39 & 0.02 & * & * & * & * & * & * \\
3.9 & bp & * & * & 0.01 & 0.01 & 0.02 & 0.03 & 1.47 & 0.04 & * & * & * & * & * \\
3.8 & degree & * & 0.01 & 0.01 & 0.02 & 0.05 & 0.11 & 0.35 & 4.35 & 0.07 & * & * & * & * \\
3.8 & betweenness & * & * & 0.01 & 0.02 & 0.03 & 1.28 & * & * & * & * & * & * & * \\
3.8 & bp & * & * & 0.01 & 0.02 & 0.03 & 0.06 & 0.74 & 0.02 & * & * & * & * & * \\
3.7 & degree & * & 0.01 & 0.01 & 0.02 & 0.04 & 0.11 & 0.31 & 3.38 & 0.03 & * & * & * & * \\
3.7 & betweenness & 58.22 & 53.91 & 49.57 & 45.09 & 40.12 & 29.17 & * & * & * & * & * & * & * \\
3.7 & bp & * & * & 0.01 & 0.01 & 0.02 & 0.05 & 0.45 & 0.02 & * & * & * & * & * \\
3.6 & degree & * & 0.01 & 0.02 & 0.02 & 0.04 & 0.12 & 0.40 & 1.89 & 0.02 & * & * & * & * \\
3.6 & betweenness & 40.46 & 37.39 & 34.22 & 31.02 & 27.11 & 18.09 & * & * & * & * & * & * & * \\
3.6 & bp & * & 0.01 & 0.01 & 0.01 & 0.02 & 0.79 & 0.33 & 0.01 & * & * & * & * & * \\
3.5 & degree & * & 0.01 & 0.01 & 0.03 & 0.05 & 0.10 & 0.40 & 0.67 & 0.02 & * & * & * & * \\
3.5 & betweenness & 38.72 & 35.64 & 32.43 & 29.26 & 25.14 & 18.33 & * & * & * & * & * & * & * \\
3.5 & bp & * & 0.01 & 0.01 & 0.02 & 0.03 & 24.39 & 0.18 & 0.01 & * & * & * & * & * \\
3.4 & degree & * & 0.01 & 0.02 & 0.04 & 0.09 & 0.22 & 0.96 & 0.29 & 0.01 & * & * & * & * \\
3.4 & betweenness & 42.04 & 38.59 & 34.97 & 31.49 & 27.75 & 28.27 & * & * & * & * & * & * & * \\
3.4 & bp & * & 0.01 & 0.01 & 0.02 & 0.04 & 28.88 & 0.06 & 0.01 & * & * & * & * & * \\
3.3 & degree & 0.01 & 0.01 & 0.02 & 0.04 & 0.08 & 0.22 & 1.19 & 0.09 & 0.01 & * & * & * & * \\
3.3 & betweenness & 51.20 & 46.78 & 42.50 & 37.85 & 31.86 & 5.17 & * & * & * & * & * & * & * \\
3.3 & bp & * & 0.01 & 0.01 & 0.02 & 0.03 & 4.20 & 0.08 & 0.01 & * & * & * & * & * \\
3.2 & degree & * & 0.01 & 0.02 & 0.02 & 0.05 & 0.19 & 2.78 & 0.05 & * & * & * & * & * \\
3.2 & betweenness & * & 0.01 & 0.01 & 0.03 & 0.22 & 0.01 & * & * & * & * & * & * & * \\
3.2 & bp & * & 0.01 & 0.02 & 0.02 & 0.03 & 1.58 & 0.03 & * & * & * & * & * & * \\
3.1 & degree & * & 0.01 & 0.02 & 0.04 & 0.08 & 0.29 & 2.99 & 0.03 & * & * & * & * & * \\
3.1 & betweenness & 126.84 & 114.71 & 102.85 & 90.49 & 66.03 & * & * & * & * & * & * & * & * \\
3.1 & bp & * & 0.01 & 0.01 & 0.02 & 0.04 & 0.40 & 0.03 & * & * & * & * & * & * \\
3.0 & degree & 0.01 & 0.02 & 0.04 & 0.09 & 0.18 & 0.70 & 0.49 & 0.01 & * & * & * & * & * \\
3.0 & betweenness & 71.83 & 64.32 & 56.95 & 49.10 & 34.98 & * & * & * & * & * & * & * & * \\
3.0 & bp & 0.01 & 0.02 & 0.03 & 0.04 & 7.95 & 0.16 & 0.01 & * & * & * & * & * & * \\
2.9 & degree & 0.01 & 0.02 & 0.05 & 0.10 & 0.26 & 1.48 & 0.07 & 0.01 & * & * & * & * & * \\
2.9 & betweenness & 46.95 & 41.48 & 36.42 & 31.17 & 4.03 & * & * & * & * & * & * & * & * \\
2.9 & bp & 0.01 & 0.02 & 0.03 & 0.05 & 10.32 & 0.06 & 0.01 & * & * & * & * & * & * \\
2.8 & degree & 0.01 & 0.03 & 0.05 & 0.13 & 0.43 & 3.10 & 0.01 & * & * & * & * & * & * \\
2.8 & betweenness & 69.58 & 60.83 & 52.47 & 41.71 & * & * & * & * & * & * & * & * & * \\
2.8 & bp & 0.01 & 0.02 & 0.03 & 0.05 & 0.61 & 0.02 & * & * & * & * & * & * & * \\
2.7 & degree & 0.01 & 0.03 & 0.10 & 0.31 & 1.52 & 0.24 & 0.01 & * & * & * & * & * & * \\
2.7 & betweenness & 70.52 & 60.67 & 51.20 & 51.30 & * & * & * & * & * & * & * & * & * \\
2.7 & bp & 0.01 & 0.02 & 0.04 & 9.74 & 0.14 & 0.01 & * & * & * & * & * & * & * \\
2.6 & degree & 0.03 & 0.05 & 0.13 & 0.55 & 3.71 & 0.01 & * & * & * & * & * & * & * \\
2.6 & betweenness & 48.20 & 40.02 & 32.31 & 0.24 & * & * & * & * & * & * & * & * & * \\
2.6 & bp & 0.02 & 0.05 & 0.08 & 3.49 & 0.02 & * & * & * & * & * & * & * & * \\
2.5 & degree & 0.06 & 0.13 & 0.35 & 2.91 & 0.07 & * & * & * & * & * & * & * & * \\
2.5 & betweenness & 65.60 & 51.54 & 31.45 & * & * & * & * & * & * & * & * & * & * \\
2.5 & bp & 0.05 & 0.10 & 0.25 & 0.09 & 0.01 & * & * & * & * & * & * & * & * \\
2.4 & degree & 0.11 & 0.25 & 1.58 & 0.21 & * & * & * & * & * & * & * & * & * \\
2.4 & betweenness & 0.06 & 0.13 & * & * & * & * & * & * & * & * & * & * & * \\
2.4 & bp & 0.08 & 0.19 & 1.21 & 0.01 & * & * & * & * & * & * & * & * & * \\
2.3 & degree & 0.28 & 1.25 & 0.14 & * & * & * & * & * & * & * & * & * & * \\
2.3 & betweenness & 26.39 & 0.89 & * & * & * & * & * & * & * & * & * & * & * \\
2.3 & bp & 0.25 & 6.17 & 0.01 & * & * & * & * & * & * & * & * & * & * \\
2.2 & degree & 0.87 & 0.01 & * & * & * & * & * & * & * & * & * & * & * \\
2.2 & betweenness & 18.76 & * & * & * & * & * & * & * & * & * & * & * & * \\
2.2 & bp & 12.62 & 0.01 & * & * & * & * & * & * & * & * & * & * & * \\
2.1 & degree & * & * & * & * & * & * & * & * & * & * & * & * & * \\
2.1 & betweenness & * & * & * & * & * & * & * & * & * & * & * & * & * \\
2.1 & bp & * & * & * & * & * & * & * & * & * & * & * & * & * \\
\hline
\end{tabular}
}
\end{table}

\begin{table}[!ht]
\centering
\caption{Variances of $S(q)/N$ for different attacks under various values of the power-law exponent $\gamma$ for $N=10^4$ and $m=3$, corresponding to Figures~\ref{fig:r_by_attacks_10000_m3} and~\ref{fig:r_gamma_10000_m3}. All values are scaled by $10^{-3}$. Entries marked with an asterisk (*) indicate values that are nonzero but smaller than the display precision under the current scaling (on the order of $10^{-5}$), and therefore cannot be shown numerically.}
\label{tab:SqN_variance_10000_m3}
\resizebox{\textwidth}{!}{%
\begin{tabular}{c|c|ccccccccccccc}
\hline
\textbf{$\gamma$} & \textbf{fraction q} & \textbf{0.02} & \textbf{0.04} & \textbf{0.06} & \textbf{0.08} & \textbf{0.10} & \textbf{0.12} & \textbf{0.14} & \textbf{0.16} & \textbf{0.18} & \textbf{0.20} & \textbf{0.22} & \textbf{0.24} & \textbf{0.25} \\
\hline
4.0 & degree & * & * & * & * & * & * & * & 0.01 & 0.01 & 0.01 & 0.02 & 0.06 & 0.14 \\
4.0 & betweenness & * & * & * & * & * & * & * & 0.01 & 0.01 & 0.02 & 0.15 & 12.22 & 0.01 \\
4.0 & bp & * & * & * & * & * & * & * & * & * & 0.01 & 0.01 & 25.65 & 0.86 \\
3.9 & degree & * & * & * & * & * & * & * & 0.01 & 0.01 & 0.02 & 0.02 & 0.06 & 0.15 \\
3.9 & betweenness & * & * & * & * & * & * & * & * & * & * & 0.01 & 16.08 & 0.01 \\
3.9 & bp & * & * & * & * & * & * & * & * & * & 0.01 & 0.01 & 11.82 & 0.97 \\
3.8 & degree & * & * & * & * & * & * & 0.01 & 0.01 & 0.01 & 0.02 & 0.04 & 0.08 & 0.18 \\
3.8 & betweenness & * & * & * & * & * & * & * & * & * & * & 0.24 & 17.93 & 0.01 \\
3.8 & bp & * & * & * & * & * & * & * & * & 0.01 & 0.01 & 0.01 & 3.81 & 0.35 \\
3.7 & degree & * & * & * & * & * & * & 0.01 & 0.01 & 0.01 & 0.02 & 0.04 & 0.14 & 0.34 \\
3.7 & betweenness & * & * & * & * & * & * & * & * & 0.01 & 0.01 & 0.14 & 0.32 & * \\
3.7 & bp & * & * & * & * & * & * & * & * & * & * & 0.01 & 1.97 & 0.17 \\
3.6 & degree & * & * & * & * & * & 0.01 & 0.01 & 0.01 & 0.02 & 0.03 & 0.07 & 0.19 & 0.40 \\
3.6 & betweenness & * & * & * & * & * & * & * & 0.01 & 0.01 & 0.02 & 0.66 & 0.01 & * \\
3.6 & bp & * & * & * & * & * & * & * & * & 0.01 & 0.01 & 0.02 & 0.89 & 0.18 \\
3.5 & degree & * & * & * & * & * & * & * & 0.01 & 0.01 & 0.02 & 0.06 & 0.18 & 0.42 \\
3.5 & betweenness & * & * & * & * & * & * & * & * & 0.01 & 0.03 & 1.33 & * & * \\
3.5 & bp & * & * & * & * & * & * & * & * & * & 0.01 & 0.03 & 0.58 & 0.10 \\
3.4 & degree & * & * & * & * & * & 0.01 & 0.01 & 0.01 & 0.02 & 0.04 & 0.08 & 0.23 & 0.61 \\
3.4 & betweenness & * & * & * & * & * & * & * & 0.01 & 0.01 & 0.03 & 1.77 & * & * \\
3.4 & bp & * & * & * & * & * & * & * & * & 0.01 & 0.01 & 3.46 & 0.47 & 0.04 \\
3.3 & degree & * & * & * & * & * & 0.01 & 0.01 & 0.01 & 0.02 & 0.04 & 0.08 & 0.28 & 1.08 \\
3.3 & betweenness & * & * & * & * & * & * & 0.01 & 0.01 & 0.02 & 0.05 & 20.35 & * & * \\
3.3 & bp & * & * & * & * & * & * & * & 0.01 & 0.01 & 0.01 & 16.50 & 0.16 & 0.03 \\
3.2 & degree & * & * & * & * & * & 0.01 & 0.01 & 0.02 & 0.03 & 0.05 & 0.11 & 0.73 & 2.32 \\
3.2 & betweenness & * & * & * & * & * & * & 0.01 & 0.02 & 0.03 & 0.06 & 1.82 & * & * \\
3.2 & bp & * & * & * & * & * & * & 0.01 & 0.01 & 0.01 & 0.02 & 7.65 & 0.09 & 0.02 \\
3.1 & degree & * & * & * & * & * & 0.01 & 0.01 & 0.02 & 0.04 & 0.06 & 0.22 & 1.74 & 3.83 \\
3.1 & betweenness & * & * & * & * & * & * & * & * & 0.01 & 0.85 & 0.01 & * & * \\
3.1 & bp & * & * & * & * & * & * & * & 0.01 & 0.01 & 0.02 & 1.35 & 0.02 & 0.01 \\
3.0 & degree & * & * & * & * & 0.01 & 0.01 & 0.02 & 0.02 & 0.04 & 0.07 & 0.23 & 2.99 & 1.23 \\
3.0 & betweenness & * & * & * & * & * & * & * & 0.03 & 0.14 & 3.41 & * & * & * \\
3.0 & bp & * & * & * & * & * & 0.01 & 0.01 & 0.01 & 0.01 & 8.15 & 0.26 & 0.03 & 0.01 \\
2.9 & degree & * & * & * & * & * & 0.01 & 0.01 & 0.03 & 0.05 & 0.15 & 0.70 & 2.44 & 0.11 \\
2.9 & betweenness & * & * & * & * & 0.01 & 0.01 & 0.02 & 0.03 & 0.08 & 3.19 & * & * & * \\
2.9 & bp & * & * & * & * & * & 0.01 & 0.01 & 0.01 & 0.01 & 10.61 & 0.11 & 0.01 & * \\
2.8 & degree & * & * & * & * & 0.01 & 0.01 & 0.01 & 0.03 & 0.03 & 0.12 & 2.01 & 0.34 & 0.03 \\
2.8 & betweenness & * & * & * & * & * & * & * & 0.01 & 0.15 & * & * & * & * \\
2.8 & bp & * & * & * & * & * & * & 0.01 & 0.01 & 0.01 & 1.14 & 0.03 & 0.01 & * \\
2.7 & degree & * & * & * & 0.01 & 0.01 & 0.01 & 0.02 & 0.04 & 0.08 & 0.65 & 2.56 & 0.06 & 0.01 \\
2.7 & betweenness & * & * & * & * & * & 0.01 & 0.01 & 0.02 & 13.13 & * & * & * & * \\
2.7 & bp & * & * & * & * & * & 0.01 & 0.01 & 0.01 & 20.29 & 0.17 & 0.02 & * & * \\
2.6 & degree & * & * & * & 0.01 & 0.01 & 0.02 & 0.04 & 0.10 & 0.31 & 3.32 & 0.35 & 0.01 & * \\
2.6 & betweenness & * & * & * & * & * & 0.01 & 0.01 & 2.23 & * & * & * & * & * \\
2.6 & bp & * & * & * & 0.01 & 0.01 & 0.01 & 0.02 & 0.06 & 0.76 & 0.04 & 0.01 & * & * \\
2.5 & degree & * & * & 0.01 & 0.01 & 0.02 & 0.03 & 0.07 & 0.29 & 1.39 & 1.06 & 0.03 & * & * \\
2.5 & betweenness & * & * & * & 0.01 & 0.02 & 0.02 & 0.34 & 0.09 & * & * & * & * & * \\
2.5 & bp & * & * & 0.01 & 0.01 & 0.01 & 0.01 & 0.02 & 3.71 & 0.08 & 0.01 & * & * & * \\
2.4 & degree & * & 0.01 & 0.01 & 0.02 & 0.03 & 0.05 & 0.08 & 1.50 & 0.54 & 0.02 & * & * & * \\
2.4 & betweenness & * & 0.01 & * & 0.01 & 0.03 & 0.74 & * & * & * & * & * & * & * \\
2.4 & bp & * & 0.01 & 0.01 & 0.01 & 0.02 & 0.03 & 3.56 & 0.09 & 0.01 & * & * & * & * \\
2.3 & degree & 0.01 & 0.01 & 0.04 & 0.08 & 0.19 & 0.70 & 2.23 & 0.05 & * & * & * & * & * \\
2.3 & betweenness & 0.01 & 0.01 & 0.03 & 0.08 & 8.44 & * & * & * & * & * & * & * & * \\
2.3 & bp & 0.01 & 0.01 & 0.02 & 0.04 & 4.00 & 0.16 & 0.02 & * & * & * & * & * & * \\
2.2 & degree & 0.01 & 0.04 & 0.14 & 0.74 & 0.19 & 0.02 & * & * & * & * & * & * & * \\
2.2 & betweenness & * & 0.03 & 0.49 & * & * & * & * & * & * & * & * & * & * \\
2.2 & bp & 0.01 & 0.04 & 0.09 & 0.42 & 0.01 & * & * & * & * & * & * & * & * \\
2.1 & degree & 0.09 & 0.15 & * & * & * & * & * & * & * & * & * & * & * \\
2.1 & betweenness & 0.53 & * & * & * & * & * & * & * & * & * & * & * & * \\
2.1 & bp & 0.11 & 0.01 & * & * & * & * & * & * & * & * & * & * & * \\
\hline
\end{tabular}}
\end{table}

\begin{table}[!ht]
\centering
\caption{Variances of $S(q)/N$ for different attacks under various values of the power-law exponent $\gamma$ for $N=10^4$ and $m=4$, corresponding to Figures~\ref{fig:r_by_attacks_10000_m4} and~\ref{fig:r_gamma_10000_m4}. All values are scaled by $10^{-3}$. Entries marked with an asterisk (*) indicate values that are nonzero but smaller than the display precision under the current scaling (on the order of $10^{-5}$), and therefore cannot be shown numerically.}
\label{tab:SqN_variance_10000_m4}
\resizebox{\textwidth}{!}{%
\begin{tabular}{c|c|ccccccccccccc}
\hline
\textbf{$\gamma$} & \textbf{fraction q} & \textbf{0.02} & \textbf{0.04} & \textbf{0.06} & \textbf{0.08} & \textbf{0.10} & \textbf{0.12} & \textbf{0.14} & \textbf{0.16} & \textbf{0.18} & \textbf{0.20} & \textbf{0.22} & \textbf{0.24} & \textbf{0.25} \\
\hline
4.0 & degree & * & * & * & * & * & * & * & * & * & * & * & 0.01 & 0.01 \\
4.0 & betweenness & * & * & * & * & * & * & * & * & * & * & * & * & * \\
4.0 & bp & * & * & * & * & * & * & * & * & * & * & * & * & * \\
3.9 & degree & * & * & * & * & * & * & * & * & * & * & * & 0.01 & 0.01 \\
3.9 & betweenness & * & * & * & * & * & * & * & * & * & * & * & * & * \\
3.9 & bp & * & * & * & * & * & * & * & * & * & * & * & * & * \\
3.8 & degree & * & * & * & * & * & * & * & * & * & * & 0.01 & 0.01 & 0.01 \\
3.8 & betweenness & * & * & * & * & * & * & * & * & * & * & * & * & 0.01 \\
3.8 & bp & * & * & * & * & * & * & * & * & * & * & * & * & * \\
3.7 & degree & * & * & * & * & * & * & * & * & * & * & 0.01 & 0.01 & 0.01 \\
3.7 & betweenness & * & * & * & * & * & * & * & * & * & * & 0.01 & 0.01 & 0.01 \\
3.7 & bp & * & * & * & * & * & * & * & * & * & * & * & * & * \\
3.6 & degree & * & * & * & * & * & * & * & * & * & * & 0.01 & 0.01 & 0.01 \\
3.6 & betweenness & * & * & * & * & * & * & * & * & * & * & * & * & * \\
3.6 & bp & * & * & * & * & * & * & * & * & * & * & * & * & * \\
3.5 & degree & * & * & * & * & * & * & * & * & * & * & 0.01 & 0.01 & 0.01 \\
3.5 & betweenness & * & * & * & * & * & * & * & * & * & * & * & * & * \\
3.5 & bp & * & * & * & * & * & * & * & * & * & * & * & * & * \\
3.4 & degree & * & * & * & * & * & * & * & * & * & * & 0.01 & 0.01 & 0.01 \\
3.4 & betweenness & * & * & * & * & * & * & * & * & * & * & * & * & * \\
3.4 & bp & * & * & * & * & * & * & * & * & * & * & * & * & * \\
3.3 & degree & * & * & * & * & * & * & * & * & * & 0.01 & 0.01 & 0.01 & 0.01 \\
3.3 & betweenness & * & * & * & * & * & * & * & * & * & * & * & * & 0.01 \\
3.3 & bp & * & * & * & * & * & * & * & * & * & * & * & * & * \\
3.2 & degree & * & * & * & * & * & * & * & * & * & * & 0.01 & 0.01 & 0.01 \\
3.2 & betweenness & * & * & * & * & * & * & * & * & * & * & * & * & * \\
3.2 & bp & * & * & * & * & * & * & * & * & * & * & * & * & * \\
3.1 & degree & * & * & * & * & * & * & * & * & * & * & 0.01 & 0.01 & 0.01 \\
3.1 & betweenness & * & * & * & * & * & * & * & * & * & * & 0.01 & 0.03 & 0.04 \\
3.1 & bp & * & * & * & * & * & * & * & * & * & * & * & * & * \\
3.0 & degree & * & * & * & * & * & * & * & * & * & 0.01 & 0.01 & 0.02 & 0.03 \\
3.0 & betweenness & * & * & * & * & * & * & * & * & * & * & * & 0.01 & 0.01 \\
3.0 & bp & * & * & * & * & * & * & * & * & * & * & * & * & * \\
2.9 & degree & * & * & * & * & * & * & * & * & * & 0.01 & 0.01 & 0.03 & 0.03 \\
2.9 & betweenness & * & * & * & * & * & * & * & * & 0.01 & * & * & 0.01 & 0.03 \\
2.9 & bp & * & * & * & * & * & * & * & * & * & * & * & * & 0.01 \\
2.8 & degree & * & * & * & * & * & * & * & * & * & * & 0.01 & 0.01 & 0.03 \\
2.8 & betweenness & * & * & * & * & * & * & * & * & * & * & 0.01 & 0.02 & 0.09 \\
2.8 & bp & * & * & * & * & * & * & * & * & * & * & * & * & * \\
2.7 & degree & * & * & * & * & * & * & * & * & 0.01 & 0.01 & 0.02 & 0.04 & 0.05 \\
2.7 & betweenness & * & * & * & * & * & * & * & * & * & * & 0.01 & 0.01 & 0.16 \\
2.7 & bp & * & * & * & * & * & * & * & * & * & * & 0.01 & 0.01 & 0.02 \\
2.6 & degree & * & * & * & * & * & * & * & 0.01 & 0.01 & 0.01 & 0.02 & 0.03 & 0.08 \\
2.6 & betweenness & * & * & * & * & * & * & * & * & 0.01 & 0.01 & 0.03 & 0.43 & 21.55 \\
2.6 & bp & * & * & * & * & * & * & * & * & * & * & 0.01 & 0.01 & 11.20 \\
2.5 & degree & * & * & * & * & * & * & 0.01 & 0.01 & 0.02 & 0.03 & 0.05 & 0.14 & 0.32 \\
2.5 & betweenness & * & * & * & * & * & * & * & * & * & 0.04 & 0.36 & 0.01 & * \\
2.5 & bp & * & * & * & * & * & * & * & * & * & 0.01 & 0.01 & 1.37 & 0.14 \\
2.4 & degree & * & * & * & 0.01 & 0.01 & 0.01 & 0.02 & 0.04 & 0.06 & 0.17 & 0.69 & 2.86 & 4.57 \\
2.4 & betweenness & * & * & * & * & * & * & * & 0.01 & 0.02 & 0.78 & 0.03 & * & * \\
2.4 & bp & * & * & * & * & * & 0.01 & 0.01 & 0.01 & 0.01 & 0.04 & 0.93 & 0.04 & 0.01 \\
2.3 & degree & * & * & * & 0.01 & 0.01 & 0.01 & 0.02 & 0.03 & 0.10 & 0.38 & 2.43 & 0.18 & 0.02 \\
2.3 & betweenness & * & * & * & * & 0.01 & 0.01 & 0.01 & 0.02 & 26.44 & * & * & * & * \\
2.3 & bp & * & * & * & * & 0.01 & 0.01 & 0.01 & 0.01 & 23.88 & 0.29 & 0.03 & * & * \\
2.2 & degree & * & 0.01 & 0.01 & 0.02 & 0.03 & 0.06 & 0.17 & 1.40 & 1.12 & 0.02 & * & * & * \\
2.2 & betweenness & * & 0.01 & 0.01 & 0.01 & 0.02 & 0.21 & 0.01 & * & * & * & * & * & * \\
2.2 & bp & * & 0.01 & 0.01 & 0.01 & 0.01 & 0.02 & 3.52 & 0.07 & 0.01 & * & * & * & * \\
2.1 & degree & 0.04 & 0.11 & 0.58 & 0.32 & 0.01 & * & * & * & * & * & * & * & * \\
2.1 & betweenness & 0.02 & 0.93 & * & * & * & * & * & * & * & * & * & * & * \\
2.1 & bp & 0.04 & 0.09 & 0.38 & 0.01 & * & * & * & * & * & * & * & * & * \\
\hline
\end{tabular}
}
\end{table}

\begin{table}[!ht]
\centering
\caption{Variances of the robustness indexes $R$ against different attacks and the average lengths $\langle l\rangle$ of the shortest loops for the various values of the power-law exponent $\gamma$ for the case $(N=10^3 \text{ and } m=2)$. All values are scaled by $10^{-3}$. Entries marked with an asterisk (*) indicate values that are nonzero but smaller than the display precision under the current scaling (on the order of $10^{-5}$), and therefore cannot be shown numerically.}
\label{tab:variance_1000_m2}
\resizebox{\textwidth}{!}{%
\begin{tabular}{c|cccccccccccccccccccc}
\hline
$\gamma$ & 2.1 & 2.2 & 2.3 & 2.4 & 2.5 & 2.6 & 2.7 & 2.8 & 2.9 & 3.0 & 3.1 & 3.2 & 3.3 & 3.4 & 3.5 & 3.6 & 3.7 & 3.8 & 3.9 & 4.0 \\
\hline
$R^{\mathrm{degree}}$ & * & * & * & * & * & * & * & * & * & * & * & * & * & * & * & * & * & * & * & * \\
$R^{\mathrm{betweenness}}$ & * & * & * & * & * & * & * & * & * & * & * & * & * & * & * & * & * & * & * & * \\
$R^{\mathrm{BP}}$ & * & * & * & * & * & * & * & * & * & * & * & * & * & * & * & * & * & * & * & * \\
\hline
$\langle l \rangle$ & 3.6 & 7.9 & 7.4 & 9.1 & 7.2 & 5.9 & 6.2 & 4.3 & 4.9 & 5.3 & 4.1 & 3.8 & 3.4 & 5.2 & 4.0 & 3.4 & 3.0 & 3.5 & 2.8 & 2.4 \\
\hline
\end{tabular}
}
\end{table}

\begin{table}[!ht]
\centering
\caption{Variances of the robustness indexes $R$ against different attacks and the average lengths $\langle l\rangle$ of the shortest loops for the various values of the power-law exponent $\gamma$ for the case $(N=10^3 \text{ and } m=3)$. All values are scaled by $10^{-3}$. Entries marked with an asterisk (*) indicate values that are nonzero but smaller than the display precision under the current scaling (on the order of $10^{-5}$), and therefore cannot be shown numerically.}
\label{tab:variance_1000_m3}
\resizebox{\textwidth}{!}{%
\begin{tabular}{c|cccccccccccccccccccc}
\hline
$\gamma$ & 2.1 & 2.2 & 2.3 & 2.4 & 2.5 & 2.6 & 2.7 & 2.8 & 2.9 & 3.0 & 3.1 & 3.2 & 3.3 & 3.4 & 3.5 & 3.6 & 3.7 & 3.8 & 3.9 & 4.0 \\
\hline
$R^{\mathrm{degree}}$ 
& * & 0.1 & * & * & * & * & * & * & * & * & * & * & * & * & * & * & * & * & * & * \\
$R^{\mathrm{betweenness}}$ 
& * & * & * & * & * & * & * & * & * & * & * & * & * & * & * & * & * & * & * & * \\
$R^{\mathrm{BP}}$ 
& * & * & * & * & * & * & * & * & * & * & * & * & * & * & * & * & * & * & * & * \\
\hline
$\langle l \rangle$ 
& 0.3 & 2.1 & 1.4 & 2.3 & 2.2 & 3.0 & 2.1 & 1.7 & 2.1 & 1.4 & 1.5 & 0.9 & 1.0 & 1.3 & 1.2 & 0.9 & 1.0 & 0.7 & 0.9 & 0.9 \\
\hline
\end{tabular}
}
\end{table}

\begin{table}[!ht]
\centering
\caption{Variances of the robustness indexes $R$ against different attacks and the average lengths $\langle l\rangle$ of the shortest loops for the various values of the power-law exponent $\gamma$ for the case $(N=10^3 \text{ and } m=4)$. 
All values are scaled by $10^{-3}$. 
Entries marked with an asterisk (*) indicate values that are nonzero but smaller than the display precision under the current scaling (on the order of $10^{-5}$), and therefore cannot be shown numerically.}
\label{tab:variance_1000_m4}
\resizebox{\textwidth}{!}{%
\begin{tabular}{c|cccccccccccccccccccc}
\hline
$\gamma$ & 2.1 & 2.2 & 2.3 & 2.4 & 2.5 & 2.6 & 2.7 & 2.8 & 2.9 & 3.0 & 3.1 & 3.2 & 3.3 & 3.4 & 3.5 & 3.6 & 3.7 & 3.8 & 3.9 & 4.0 \\
\hline
$R^{\mathrm{degree}}$ 
& * & 0.1 & * & * & * & * & * & * & * & * & * & * & * & * & * & * & * & * & * & * \\
$R^{\mathrm{betweenness}}$ 
& * & * & * & * & * & * & * & * & * & * & * & * & * & * & * & * & * & * & * & * \\
$R^{\mathrm{BP}}$ 
& * & 0.1 & * & * & * & * & * & * & * & * & * & * & * & * & * & * & * & * & * & * \\
\hline
$\langle l \rangle$ 
& 0.2 & 0.5 & 0.8 & 1.0 & 1.0 & 1.0 & 1.2 & 0.9 & 0.9 & 0.6 & 0.6 & 0.5 & 0.5 & 0.6 & 0.5 & 0.5 & 0.4 & 0.4 & 0.4 & 0.4 \\
\hline
\end{tabular}
}
\end{table}

\begin{table}[!ht]
\centering
\caption{Variances of the robustness indexes $R$ against different attacks and the average lengths $\langle l\rangle$ of the shortest loops for the various values of the power-law exponent $\gamma$ for the case $(N=10^4 \text{ and } m=2)$. All values are scaled by $10^{-4}$. Entries marked with an asterisk (*) indicate values that are nonzero but smaller than the display precision under the current scaling (on the order of $10^{-6}$), and therefore cannot be shown numerically.}
\label{tab:variance_10000_m2}
\resizebox{\textwidth}{!}{%
\begin{tabular}{c|cccccccccccccccccccc}
\hline
$\gamma$ & 2.1 & 2.2 & 2.3 & 2.4 & 2.5 & 2.6 & 2.7 & 2.8 & 2.9 & 3.0 & 3.1 & 3.2 & 3.3 & 3.4 & 3.5 & 3.6 & 3.7 & 3.8 & 3.9 & 4.0 \\
\hline
$R^{\mathrm{degree}}$ & * & * & * & * & * & * & * & * & * & * & * & * & * & * & * & * & * & * & * & * \\
$R^{\mathrm{betweenness}}$ & * & * & * & * & * & * & * & * & * & * & * & * & * & * & * & * & * & * & * & * \\
$R^{\mathrm{BP}}$ & * & * & * & * & * & * & * & * & * & * & * & * & * & * & * & * & * & * & * & * \\
$\langle l \rangle$ & 20.3 & 21.2 & 34.1 & 23.8 & 20.6 & 25.5 & 29.1 & 30.7 & 22.3 & 18.3 & 7.2 & 11.6 & 18.2 & 17.1 & 11.0 & 13.3 & 12.3 & 16.0 & 7.5 & 7.4 \\
\hline
\end{tabular}
}
\end{table}

\begin{table}[!ht]
\centering
\caption{Variances of the robustness indexes $R$ against different attacks and the average lengths $\langle l\rangle$ of the shortest loops for the various values of the power-law exponent $\gamma$ for the case $(N=10^4 \text{ and } m=3)$. All values are scaled by $10^{-4}$. Entries marked with an asterisk (*) indicate values that are nonzero but smaller than the display precision under the current scaling (on the order of $10^{-6}$), and therefore cannot be shown numerically.}
\label{tab:variance_10000_m3}
\resizebox{\textwidth}{!}{%
\begin{tabular}{c|cccccccccccccccccccc}
\hline
$\gamma$ & 2.1 & 2.2 & 2.3 & 2.4 & 2.5 & 2.6 & 2.7 & 2.8 & 2.9 & 3.0 & 3.1 & 3.2 & 3.3 & 3.4 & 3.5 & 3.6 & 3.7 & 3.8 & 3.9 & 4.0 \\
\hline
$R^{\mathrm{degree}}$ & * & * & * & * & * & * & * & * & * & * & * & * & * & * & * & * & * & * & * & * \\
$R^{\mathrm{betweenness}}$ & * & * & * & * & * & * & * & * & * & * & * & * & * & * & * & * & * & * & * & * \\
$R^{\mathrm{BP}}$ & * & * & * & * & * & * & * & * & * & * & * & * & * & * & * & * & * & * & * & * \\
$\langle l \rangle$ & 0.1 & 0.6 & 1.8 & 5.7 & 4.0 & 3.6 & 6.8 & 12.9 & 5.6 & 3.5 & 3.9 & 2.4 & 2.9 & 2.3 & 3.5 & 2.0 & 2.0 & 3.3 & 0.9 & 2.1 \\
\hline
\end{tabular}
}
\end{table}

\begin{table}[!ht]
\centering
\caption{Variances of the robustness indexes $R$ against different attacks and the average lengths $\langle l\rangle$ of the shortest loops for the various values of the power-law exponent $\gamma$ for the case $(N=10^4 \text{ and } m=4)$. All values are scaled by $10^{-4}$. Entries marked with an asterisk (*) indicate values that are nonzero but smaller than the display precision under the current scaling (on the order of $10^{-6}$), and therefore cannot be shown numerically.}
\label{tab:variance_10000_m4}
\resizebox{\textwidth}{!}{%
\begin{tabular}{c|cccccccccccccccccccc}
\hline
$\gamma$ & 2.1 & 2.2 & 2.3 & 2.4 & 2.5 & 2.6 & 2.7 & 2.8 & 2.9 & 3.0 & 3.1 & 3.2 & 3.3 & 3.4 & 3.5 & 3.6 & 3.7 & 3.8 & 3.9 & 4.0 \\
\hline
$R^{\mathrm{degree}}$ & * & * & * & 0.1 & * & * & * & * & * & * & * & * & * & * & * & * & * & * & * & * \\
$R^{\mathrm{betweenness}}$ & * & * & * & 0.1 & * & * & * & * & * & * & * & * & * & * & * & * & * & * & * & * \\
$R^{\mathrm{BP}}$ & * & * & * & 0.1 & * & * & * & * & * & * & * & * & * & * & * & * & * & * & * & * \\
$\langle l \rangle$ & * & 0.3 & 0.2 & 12.5 & 0.7 & 1.3 & 1.3 & 2.3 & 2.8 & 5.4 & 0.6 & 1.0 & 1.4 & 1.4 & 0.9 & 1.1 & 1.2 & 1.7 & 0.8 & 0.7 \\
\hline
\end{tabular}
}
\end{table}

\clearpage
\funding
\textcolor{red}{(required)}
Not applicable

\conflictsofinterest
\textcolor{red}{(required)}
The authors have declared that no competing interests exist.

\authorcontribution
\begin{tabular}{ll}
Conceptualization: & Yukio Hayashi. \\
Funding acquisition: & Yukio Hayashi. \\
Investigation: & Yingzhou Mou, Yukio Hayashi. \\
Methodology: & Yingzhou Mou, Yukio Hayashi. \\
Supervision: & Yukio Hayashi. \\
Visualization: & Yingzhou Mou. \\
Writing--original draft: & Yingzhou Mou. \\
Writing--review \& editing: & Yingzhou Mou, Yukio Hayashi.
\end{tabular}

\bibliographystyle{ieeetr}
\bibliography{reference}

\end{document}